\documentclass[11pt, a4paper]{article}
\pdfoutput=1

\usepackage[height=8.85in,width=6.55in]{geometry}

\usepackage{graphicx,rotating}
\usepackage[bookmarksopen,colorlinks=true,linkcolor=steelblue,
citecolor=darkred,urlcolor=darkred,linktoc=all]{hyperref}

\usepackage{amsmath,amssymb,tensor,subdepth}
\usepackage{mathtools}
\usepackage{slashed}
\usepackage{bm}
\usepackage{bbm}
\usepackage{cite}
\usepackage{comment}
\usepackage[utf8]{inputenc}
\usepackage{accents}
\usepackage[footnotesize]{caption}
\usepackage{cancel}
\usepackage{physics}
\usepackage{here}

\usepackage{libertine}
\usepackage[ISO, upint]{libertinust1math}
\usepackage[scr=boondoxo,bb=boondox]{mathalpha}
\usepackage[T1]{fontenc}

\usepackage{stmaryrd}
\usepackage{trimclip}
\usepackage[framemethod=default]{mdframed}
\newmdenv[skipabove=7pt,
skipbelow=7pt,
rightline=false,
leftline=false,
topline=false,
bottomline=false,
backgroundcolor=gray!10,
linecolor=gray,
innerleftmargin=5pt,
innerrightmargin=5pt,
innertopmargin=5pt,
innerbottommargin=5pt,
leftmargin=0cm,
rightmargin=0cm,
linewidth=4pt]{eBox}
\newmdenv[skipabove=7pt,
skipbelow=7pt,
rightline=true,
leftline=true,
topline=true,
bottomline=true,
backgroundcolor=white,
linecolor=gray,
innerleftmargin=5pt,
innerrightmargin=5pt,
innertopmargin=5pt,
innerbottommargin=5pt,
leftmargin=0cm,
rightmargin=0cm,
linewidth=1pt]{eBox2}

\usepackage{chngcntr}
\counterwithin{equation}{section}

\usepackage{xcolor}
\definecolor{darkred}{rgb}{0.7, 0., 0.}
\definecolor{orangered}{rgb}{1,0.27,0.}
\definecolor{steelblue}{rgb}{0.275,0.51, 0.706}
\definecolor{forestgreen}{rgb}{0.13,0.55,0.13}
\definecolor{violet}{cmyk}{.32,.95,.17,.00}
\definecolor{sagegreen}{rgb}{0.5, 0.65, 0.4}
\definecolor{sepia}{rgb}{0.55, 0.45, 0.3}

\usepackage{tikz}
\usepackage{tikz-feynman}
\tikzfeynmanset{compat=1.0.0}
\pgfdeclarelayer{bg}
\pgfsetlayers{bg,main}
\usetikzlibrary{decorations.shapes}
\allowdisplaybreaks

\pgfdeclaredecoration{dashsoliddouble}{initial}{
    \state{initial}[width=\pgfdecoratedinputsegmentlength]{
    \pgfmathsetlengthmacro\lw{.3pt+.5\pgflinewidth}
    \begin{pgfscope}
    \pgfpathmoveto{\pgfpoint{0pt}{\lw}}%
    \pgfpathlineto{\pgfpoint{\pgfdecoratedinputsegmentlength}{\lw}}%
    \pgfmathtruncatemacro\dashnum{%
        round((\pgfdecoratedinputsegmentlength-3pt)/6pt)
    }
    \pgfmathsetmacro\dashscale{%
        \pgfdecoratedinputsegmentlength/(\dashnum*6pt + 3pt)
    }
    \pgfmathsetlengthmacro\dashunit{3pt*\dashscale}
    \pgfsetdash{{\dashunit}{\dashunit}}{0pt}
    \pgfusepath{stroke}
    \pgfsetdash{}{0pt}
    \pgfpathmoveto{\pgfpoint{0pt}{-\lw}}%
    \pgfpathlineto{\pgfpoint{\pgfdecoratedinputsegmentlength}{-\lw}}%
    \pgfusepath{stroke}
    \end{pgfscope}
}
}

\begin{document}

\hypersetup{pageanchor=false}
\begin{titlepage}

\begin{center}

\hfill KEK-TH-2737\\
\hfill IPMU25-0035

\vskip 1in

{\Huge \bfseries
Symmetries of Hot SM, Magnetic Flux
\&\\[.5em]
Baryogenesis from Helicity Decay
} \\
\vskip .8in

{\Large Yuta Hamada$^{a,b}$, Kyohei Mukaida$^{a,b}$, Fumio Uchida$^{c,a,d}$}

\vskip .3in
\begin{tabular}{ll}
$^a$
& \!\!\!\!\!\emph{Theory Center, IPNS, KEK, 1-1 Oho, Tsukuba, Ibaraki 305-0801, Japan}\\
$^b$
& \!\!\!\!\!\emph{Graduate University for Advanced Studies (Sokendai), }\\[-.15em]
& \!\!\!\!\!\emph{1-1 Oho, Tsukuba, Ibaraki 305-0801, Japan}\\
$^c$
& \!\!\!\!\!\emph{Kavli IPMU (WPI), UTIAS, University of Tokyo, }\\[-.15em]
& \!\!\!\!\!\emph{5-1-5 Kashiwanoha, Kashiwa, Chiba 277-8583, Japan}\\
$^d$
& \!\!\!\!\!\emph{CTPU-CGA, Institute for Basic Science (IBS), Daejeon, 34126, Korea}
\end{tabular}

\end{center}
\vskip .6in

\begin{abstract}
\noindent
We revisit the electroweak crossover of the Standard Model (SM) in the early Universe, focusing on the interplay between generalized global symmetries, magnetic flux dynamics, and baryogenesis.
Employing the dimensionally reduced 3d effective field theory of the SM at high temperature, we identify the symmetry structure—including higher-form and magnetic symmetries—and analyze their spontaneous breaking patterns across the crossover.
We further define a gauge-invariant mixing angle that interpolates between $\mathrm U(1)_Y$ and $\mathrm U(1)_\text{em}$ magnetic fields.
Based on this framework, we examine baryogenesis via decaying magnetic helicity and identify three key effects: the baryon asymmetry is modified by an $\mathcal{O}(1)$ factor due to (1) the gauge-invariant definition of the mixing angle and (2) the approximate conservation of the unconfined magnetic flux;
(3) a novel non-perturbative process in the presence of magnetic flux, which has been overlooked in previous analyses.
Our findings suggest that the previous estimation of baryon asymmetry from the magnetic helicity decay may have sizable uncertainties, and we caution against relying on it, calling for further investigation.
\end{abstract}

\end{titlepage}

\tableofcontents
\renewcommand{\thepage}{\arabic{page}}
\renewcommand{\thefootnote}{$\natural$\arabic{footnote}}
\setcounter{footnote}{0}
\newpage
\hypersetup{pageanchor=true}

\section{Introduction}
\label{sec:introduction}

After the Higgs discovery at the LHC in $2012$~\cite{ATLAS:2012yve,CMS:2012qbp}, the Standard Model (SM) of particle physics has been a great success in describing a number of observations, established as the well-tested theory of particle physics.
If the electroweak symmetry breaking (EWSB) is solely described within the SM, we now have a complete handle of EWSB in the high-temperature plasma of the early Universe.
The high-temperature phase can be studied by the $3$d effective field theory (EFT), which is obtained by the dimensional reduction of the original $1$$+$$3$d theory along the thermal circle associated with the trace over the Gibbs state (see \textit{e.g.,} \cite{Laine:2016hma}).
The phase diagram of the SM was already established a few decades ago by the lattice simulations based on the $3$d EFT of the SM~\cite{Kajantie:1996mn,Gurtler:1997hr,Rummukainen:1998as,Csikor:1998eu,Aoki:1999fi,DOnofrio:2015gop}.
The observed Higgs mass of $125.11(11)\,\mathrm{GeV}$~\cite{ATLAS:2023oaq} indicates that the EWSB is a crossover, \textit{i.e.,} no clear distinction between the symmetric and broken phases.

Recently, there are renewed interests on the behavior of the primordial magnetic fields during the electroweak crossover.
The lack of the secondary $\mathrm{GeV}$ gamma rays in blazar spectra suggests the existence of intergalactic magnetic fields (IGMFs) with the field strength of $\gtrsim 10^{-17}\,{\rm G}$ \cite{Neronov:2010gir,Tavecchio:2010mk,Dolag:2010ni,Fermi-LAT:2018jdy,MAGIC:2022piy}.
The electroweak crossover implies that the primordial $\mathrm U(1)_Y$ magnetic fields are smoothly converted to the $\mathrm U(1)_\text{em}$ ones during EWSB, which can be a source of the IGMFs.
If the primordial $\mathrm U(1)_Y$ magnetic fields carry a net magnetic helicity, the conversion process yields the baryon asymmetry via Adler--Bell--Jackiw (ABJ) anomaly~\cite{Adler:1969gk,Bell:1969ts}.
Previous works have shown that the maximally helical $\mathrm U(1)_Y$ magnetic fields cannot explain the IGMFs because of the baryon overproduction~\cite{Joyce:1997uy,Vachaspati:2001nb,Fujita:2016igl,Kamada:2016cnb,Kamada:2016eeb}.
Moreover, non-helical $\mathrm U(1)_Y$ magnetic fields are associated with the spatial fluctuations and thereby constrained by the baryon isocurvature perturbations~\cite{Giovannini:1997eg,Giovannini:1997gp,Kamada:2020bmb}.
These results essentially rule out any magnetogenesis before the EWSB.
However, the apparent gauge dependent relation between the baryon charge and the helicity density makes it difficult to draw a definite conclusion~\cite{Boyer:2025jno}.

The Landau paradigm states that the phase structure of a system is determined by the symmetry breaking patterns.
In the last decade, the notion of the symmetry has been greatly generalized by recognizing that the symmetry is equivalent to the presence of the topological operators~\cite{Gaiotto:2014kfa,Gaiotto:2017yup}.
Various new notions of symmetry such as the higher-form symmetry, non-invertible symmetry, and so forth, are introduced.
Therefore, it is natural to extend the Landau paradigm including the generalized symmetry (see \textit{e.g.,} \cite{Schafer-Nameki:2023jdn,Brennan:2023mmt} and the references therein for developments).
In our context, the presence of the magnetic field as a macroscopic variable in magnetohydrodynamics (MHD) is recognized as a spontaneous symmetry breaking (SSB) of the magnetic $0$-form symmetry originated from the original $1$$+$$3$d magnetic $1$-form symmetry in the $\mathrm U(1)_\text{em}$ gauge theory~\cite{Grozdanov:2016tdf,Liu:2018kfw}.
Since the $3$d SM also has the magnetic $0$-form symmetry associated with the $\mathrm U(1)_Y$ gauge symmetry, we expect that the conversion from $\mathrm U(1)_Y$ to $\mathrm U(1)_\text{em}$ magnetic field can be understood as the behavior of the magnetic $0$-form symmetry breaking.

In this paper, we take a fresh look at the electroweak crossover from the perspective of the generalized symmetry of the $3$d SM.
After reviewing the $3$d SM (Sec.~\ref{sec:hotSM}), we identify its generalized symmetries and show that their order parameters are consistent with the crossover nature of the EWSB (Sec.~\ref{sec:phases}).
In particular, we clarify how the magnetic $0$-form symmetry of ${\mathrm U}(1)_Y$ is kept spontaneously broken during the electroweak crossover even when the $Z$-boson magnetic flux is confined.
We then identify the associated Nambu--Goldstone mode in a gauge invariant way, clarifying the problems of the previous literature (Sec.~\ref{sec:magnetic}).
Finally, we point out a potential issue in the previous analysis of the baryogenesis from the ${\mathrm U}(1)_Y$ helicity decay by providing an existence proof of a novel non-perturbative process overlooked in the literature (Sec.~\ref{sec:baryon}).
The final section is devoted to discussion (Sec.~\ref{sec:disc}).

\section{Lighting review of \texorpdfstring{\boldmath{$3$}}{3}d SM}
\label{sec:hotSM}
Here, we provide a brief overview of the dimensional reduction of the SM, which forms the foundation for the subsequent discussion.
See Refs.\cite{Kajantie:1995kf,Kajantie:1995dw,Kajantie:1997ky,Gynther:2005dj,Gynther:2005av} for more detailed discussions.

\subsection{Thermal equilibrium and dimensional reduction}
\label{sec:thermodyn}
The thermodynamics of a certain theory is encoded in the partition function and its response to the external sources.
For a given theory with $\mathscr{L}_{4}$ being its $1$$+$$3$d Lagrangian, the partition function at temperature $T$ can be expressed as
\begin{equation}
    Z_T = \text{Tr}\left[e^{-\beta H}\right]
    = \int_\text{(anti-)periodic} \mathscr{D} \phi_4 \, \exp( - \int^\beta_0 \dd \tau \dd^3 x\, \mathscr{L}_{4 \text{E}} ),
    \label{eq:partition_function}
\end{equation}
where $\phi_4$ represents fields on $4$d Euclideanized space, the Euclideanized Lagrangian is $\mathscr{L}_{4 \text{E}}$, the inverse temperature is denoted as $\beta \coloneq 1/T$, and the path integral is taken over the fields with the (anti-)periodic boundary condition in the Euclidean time $\tau$ for bosons (fermions) respectively.
This path-integral expression makes it clear that the partition function is obtained by the path integral on the Euclidean spacetime of $\mathbb{S}_\beta^1 \times \mathbb{R}^3$ with $\mathbb{S}^1_\beta\coloneq \{\tau | \tau \sim \tau + \beta \}$.

If we are interested in the thermodynamic functions (\textit{e.g.,} pressure) or the response to the soft external sources (\textit{i.e.,} $\text{scale} < \pi T$), we can integrate out the higher Kaluza--Klein modes with $\omega > \pi T$ and obtain the EFT in $3$d space, which is known as the dimensional reduction.
The periodic boundary condition with respect to $\mathbb{S}^1_\beta$ makes the boson fields have the Kaluza--Klein modes with the Matsubara frequency $\omega_n = 2\pi n T$, including the zero mode ($n = 0$).
On the other hand, the fermion fields have the Matsubara frequency $\omega_n = (2n+1)\pi T$ owing to the anti-periodic boundary condition, where all the modes are massive.
Hence, the dimensionally reduced $3$d EFT only involves the zero modes of the boson fields, which leads to the following expression of the partition function:
\begin{equation}
    Z_T \simeq \int \mathscr{D}\phi \exp( - \int \dd^3 x\, \mathscr{L}_{3} ),
\end{equation}
where $\phi$ represents the fields in $3$d space corresponding to the zero modes of the original fields in $4$d space, and $\mathscr{L}_{3}$ is the $3$d effective Lagrangian.

The $3$d EFT can be constructed as follows.
First, we write down (lower-dimensional) operators of the $3$d fields $\phi$ that are consistent with the underlying symmetry of the original theory.
The coefficients of these operators are determined by the matching at the scale $\pi T$, \textit{i.e.,} compute certain physical quantities at the scale $\pi T$ in both the $4$d and $3$d theories, and then determine the coefficients of the operators in the $3$d theory so that the physical quantities match.
Once the $3$d EFT is obtained, we can compute the thermodynamic functions and the response to the soft external sources solely by means of $3$d EFT not referring to the original $4$d theory.
Note that if the $3$d EFT involves some interacting fields that are not gapped at the ultra-soft scale $\sim g^2 T$ with $g$ being a $4$d coupling constant, the $3$d EFT becomes non-perturbative and calls for lattice simulations~\cite{Linde:1980ts}.\footnote{
    For instance, a $3$d pure Yang--Mills theory is solely characterized by a dimensionful coupling $g_3^2$, which exhibits confinement below $g_3^2$.
    As we see in the following, this is related to the $4$d coupling by $g_3^2 \simeq g^2 (T)\, T$ with $g(T)$ being the $4$d coupling at the scale $T$.
    Hence, perturbative calculations are invalid for $\text{scale} < g_3^2 \simeq g^2 (T)\, T$.
}
In the case of our interest that will be discussed in the later sections, this is not the case because such dangerous fields are Higgsed whose scale is larger than $g^2 T$ (see {\it Top} panel in Fig.~\ref{fig:theta_eff}).

Let us stop here the abstract discussion and move on to the concrete example of the SM in the next section.

\subsection{\texorpdfstring{\boldmath{$3$}}{3}d SM at the soft scale}
\label{sec:3dSM}

The SM Lagrangian on the $4$d Euclideanized space of $\mathbb{S}_\beta^1 \times \mathbb{R}^3$ is given by
\begin{align}
    \mathscr{L}_{\text{SM},\text{E}} =&
    \qty( D_\mu \Phi_4 )^\dagger D_\mu \Phi_4 + m_\Phi^2 \Phi_4^{\dagger} \Phi_4^{} + \lambda \qty(\Phi_4^\dag \Phi_4^{})^2 + \frac{1}{4} Y_{4\mu\nu} Y_{4\mu\nu} + \frac{1}{4} W_{4\mu\nu}^a W^{a}_{4\mu\nu} + \frac{1}{4} G_{4\mu\nu}^A G^{A}_{4\mu\nu} \\[.3em]
    & + \bar{e}_{\mathrm Rf} \slashed{D} e_{\mathrm Rf}
    + \bar{\ell}_{\mathrm Lf} \slashed{D} \ell_{\mathrm Lf}
    + \bar{u}_{\mathrm Rf} \slashed{D} u_{\mathrm Rf} + \bar{d}_{\mathrm Rf} \slashed{D} d_{\mathrm Rf}
    + \bar{q}_{\mathrm Lf} \slashed{D} q_{\mathrm Lf}
    + y_t \qty( \bar{q}_{\mathrm L3} \tilde\Phi_4 t_{\mathrm R} + \text{H.c.}),
\end{align}
where $\Phi_4$ is the Higgs doublet, and $Y_{4\mu\nu}$, $W_{4\mu\nu}^a$, and $G_{4\mu\nu}^A$ are the field strengths of the $\mathrm{U}(1)_Y$, $\mathrm{SU}(2)_{\mathrm L}$, and $\mathrm{SU}(3)_{\mathrm c}$ gauge fields, respectively.
The covariant derivative is defined by $D_\mu \coloneq \partial_\mu + i g' n_Y Y_{4\mu} - i g W_{4\mu}^a \sigma^a/2 - i g_{\mathrm s} G_{4\mu}^A \lambda^A/2$, 
where $Y_{4\mu}$, $W_{4\mu}^a$, and $G_{4\mu}^A$ are the $\mathrm{U}(1)_Y$, $\mathrm{SU}(2)_{\mathrm L}$, and $\mathrm{SU}(3)_{\mathrm c}$ gauge fields, and the hypercharge is denoted as $n_Y$.
The fields $e_{\mathrm Rf}$, $\ell_{\mathrm Lf}$, $u_{\mathrm Rf}$, $d_{\mathrm Rf}$, and $q_{\mathrm Lf}$ are the right-handed charged lepton, left-handed lepton, right-handed up-type quark, right-handed down-type quark, and left-handed quark, respectively.
We have added the subscript $4$ to the boson fields so that we can distinguish them from the $3$d fields.
The superscript of the field strengths run through $a = 1,2, 3$ and $A = 1,2,\cdots,8$ respectively, and the subscript $f$ denotes the flavor index.
For the fields that transform under the fundamental representation, $\sigma^a$ and $\lambda^A$ are the Pauli and the Gell-Mann matrices, respectively, while these generators are replaced by $0$ for the fields that transform trivially.
We only consider the top Yukawa interaction for simplicity as the other Yukawa couplings are much smaller than the gauge couplings.
In the following discussion, we will adopt the following counting rule for the coupling constants:
\begin{equation}
    \lambda \sim g^2 \sim y_t^2, \qquad g'^2 \sim g^3.
    \label{eq:counting_rule}
\end{equation}

Let us first identify the zero modes of the boson fields, which are the fundamental building blocks of the $3$d EFT.
The zero mode of the Higgs doublet is denoted by a static field as
\begin{equation}
    \Phi_4 (\tau,{\bm x}) \xrightarrow{\text{zero mode}} \Phi ({\bm x}). \label{eq:Higgs_zero_mode}
\end{equation}
The zero modes of the gauge fields are split into the electric and magnetic ones.
The magnetic component is nothing but the $3$d gauge fields, which is denoted as
\begin{equation}
    Y_{4i} (\tau,{\bm x}) \xrightarrow{\text{zero mode}} Y_i ({\bm x}), \qquad
    W_{4i}^a (\tau,{\bm x}) \xrightarrow{\text{zero mode}} W_{i}^a ({\bm x}), \qquad G_{4i}^A (\tau,{\bm x}) \xrightarrow{\text{zero mode}} G_i^A ({\bm x}), \label{eq:3dgauge_zero_mode}
\end{equation}
where the subscript $i$ runs through $i = 1,2,3$.
Note that the $3$d EFT must respect the gauge symmetry associated with these $3$d gauge fields.
On the other hand, the electric component is originated from the temporal component of the gauge fields.
The Wilson lines along the compactified direction $\mathbb{S}^1_\beta$ are the appropriate quantities that transform as the adjoint representation of the gauge group (up to its center symmetry):
\begin{equation}
    \mathcal{W}_Y ({\bm x}) \coloneq e^{ i \int_0^\beta \dd \tau\, g' Y_{4\tau} }, \qquad
    \mathcal{W}_{\mathrm L} ({\bm x}) \coloneq \mathcal{P} e^{ i \int_0^\beta \dd \tau\, g W_{4\tau}^a \sigma^a/2 }, \qquad
    \mathcal{W}_c ({\bm x}) \coloneq \mathcal{P} e^{i \int_0^\beta \dd \tau\, g_s G_{4\tau}^A \lambda^A/2 }. \label{eq:Polyakov_line}
\end{equation}
Without loss of generality, we can expand the temporal components around zero since the center symmetry is spontaneously broken in the deconfined phase.
Then, the Wilson lines can be expressed as the zero modes of the temporal component of the gauge fields: $\mathcal{W}_Y({\bm x}) \simeq 1+i g’ Y_\tau ({\bm x})$, $\mathcal{W}_{\mathrm L}({\bm x}) \simeq 1+i g W_\tau^a ({\bm x}) \sigma^a/2$, and $\mathcal{W}_{\mathrm c}({\bm x}) \simeq 1+i g_s G_\tau^A ({\bm x}) \lambda^A/2$.
For this reason, the electric component of the gauge fields is practically the zero mode of the temporal component as far as the perturbative calculations are concerned:
\begin{equation}
    Y_{4\tau} (\tau,{\bm x}) \xrightarrow{\text{zero mode}} Y_\tau ({\bm x}), \qquad
    W_{4\tau}^a (\tau,{\bm x}) \xrightarrow{\text{zero mode}} W_{\tau}^a ({\bm x}), \qquad 
    G_{4\tau}^A (\tau,{\bm x}) \xrightarrow{\text{zero mode}} G_{\tau}^A ({\bm x}). \label{eq:3dWilson_zero_mode}
\end{equation}

\begin{table}
    \centering
    \begin{tabular}{c|c|c|c}
        & $\mathrm{U}(1)_Y$ & $\mathrm{SU}(2)_{\mathrm L}$ & $\mathrm{SU}(3)_{\mathrm c}$ \\
        \hline
        \hline
        $\Phi$ & $1/2$ & $\mathbf{2}$ & $\mathbf{1}$ \\
        $Y_\tau$ & $0$ & $\mathbf{1}$ & $\mathbf{1}$ \\
        $W^a_\tau$ & $0$ & $\mathbf{3}$ & $\mathbf{1}$ \\
        $G_\tau^A$ & $0$ & $\mathbf{1}$ & $\mathbf{8}$ \\
    \end{tabular}
    \caption{Transformation of the $3$d matter fields, $\Phi$, $Y_\tau$, $W^a_\tau$, and $G_\tau^A$, under the $3$d gauge symmetry of $\mathrm{U}(1)_Y \times \mathrm{SU}(2)_{\mathrm L} \times \mathrm{SU}(3)_{\mathrm c}$.}
    \label{tab:3d_matter}
\end{table}

These $3$d fields given in Eqs.~\eqref{eq:Higgs_zero_mode}, \eqref{eq:3dgauge_zero_mode}, and \eqref{eq:3dWilson_zero_mode} are the building block of EFT.
Let us identify the symmetries of the $3$d SM.
The transformation of the $3$d matter fields
under the $3$d gauge symmetry of $\mathrm{U}(1)_Y \times \mathrm{SU}(2)_{\mathrm L} \times \mathrm{SU}(3)_{\mathrm c}$ is summarized in Tab.~\ref{tab:3d_matter}.
The $3$d Lagrangian must respect the symmetries of the $4$d Euclidean theory.
Any relativistic quantum field theories on $1$\,$+$\,$3$d Minkowski spacetime must fulfill the $CPT$ symmetry, whose corresponding transformation in Euclideanized theory is the parity transformation of $4$d Euclidean space (see \textit{e.g.,} \cite{Harlow:2023hjb}), \textit{i.e.,} $(\tau, {\bm x}) \mapsto (-\tau, -{\bm x})$.
The $4$d parity induces the following transformation of the $3$d fields after the dimensional reduction:
\begin{equation}
    \label{eq:cpt}
    \Phi({\bm x}) \mapsto \Phi (-{\bm x}), \quad
    Y_{\mu} ({\bm x}) \mapsto - Y_{\mu} (-{\bm x}), \quad
    W_{\mu} ({\bm x}) \mapsto - W_{\mu} (-{\bm x}), \quad
    G_{\mu} ({\bm x}) \mapsto - G_{\mu} (-{\bm x}),
\end{equation}
here we collectively denote $\mu = \tau, i$ and utilize the matrix notation $W_{\mu} = W_{\mu}^a \sigma^a/2$ and $G_{\mu} = G_{\mu}^A \lambda^A/2$.\footnote{
        Non-zero chemical potentials are regarded as spurion that transform as $\mu_\alpha \mapsto -\mu_\alpha$ under the $CPT$ transformation, which corresponds to the pure imaginary background of the Wilson lines along $\mathbb{S}_\beta^1$.
        Since the observed baryon asymmetry is tiny, we neglect the chemical potentials in the following discussion.
    }
On top of this, the unitarity of the $1$\,$+$\,$3$d theory imposes the reflection positivity on the $4$d Euclidean theory after the Wick rotation, which implies (see \textit{e.g.,} Ref.~\cite{Simmons-Duffin:2016gjk})
\begin{alignat}{4}
    \label{eq:unitarity1}
    \Phi({\bm x})^\dag &= \Phi ({\bm x}), &\quad
    Y_{\tau} ({\bm x})^\dag &= - Y_{\tau} ({\bm x}), &\quad
    Y_{i} ({\bm x})^\dag &= Y_{i} ({\bm x}), && \\
    \label{eq:unitarity2}
    W_{\tau} ({\bm x})^\dag &= - W_{\tau} ({\bm x}), &\quad
    W_{i} ({\bm x})^\dag &= W_{i} ({\bm x}), &\quad
    G_{\tau} ({\bm x})^\dag &= - G_{\tau} ({\bm x}), &\quad
    G_{i} ({\bm x})^\dag &= G_{i} ({\bm x}).
\end{alignat}
The $3$d Euclidean action should be real in this sense.

The electrostatic $3$d SM Lagrangian consistent with the $3$d gauge symmetry, the $CPT$ symmetry [Eq.~\eqref{eq:cpt}], and the unitarity [Eqs.~\eqref{eq:unitarity1} and \eqref{eq:unitarity2}] is given by
\begin{equation} 
    \label{eq:SM3}
    \begin{split}
    \mathscr{L}_{\text{ESM}_3}
    =&
    \qty( D_i  \Phi )^{\dagger} D_i \Phi
    + m_{3}^{2}\Phi^{\dagger}\Phi
    + \lambda_{3}\qty(\Phi^{\dagger}\Phi )^{2} 
    + \frac{1}{4}Y_{ij}Y_{ij}
    + \frac{1}{4}W^ a_{ij}W ^ a_{ij}
    \\
    &
    + \frac{1}{2} \qty(\partial _ i Y_{\tau})^{2}
    + \frac{1}{2} m_{\rm D}'^{2}Y_{\tau}^{2}
    + \frac{1}{2} \qty( D _ i W^ a _{\tau})^{2}
    + \frac{1}{2}m_{\rm D}^{2}W_\tau ^ a W_\tau^a \\[.5em]
    &
    + h_{3}' \Phi^{\dagger}\Phi Y_{\tau}^{2}
    + h_{3}'' Y_\tau \Phi ^  \dagger W_\tau ^ a \sigma^a  \Phi
    + h_{3}  \Phi^{\dagger}\Phi W^a_{\tau} W^a_\tau \\[.5em]
    &
    +\mathcal{O} (g^4)
    + \Lambda_T + \mathscr{L}_{\mathrm{SU}(3)_{\mathrm c}},
    \end{split}
\end{equation}
where we omit the higher order terms of $\mathcal{O}(g^4)$ in the counting rule of Eq.~\eqref{eq:counting_rule}.
We have not written down the $\mathrm{SU}(3)_{\mathrm c}$ part explicitly since it is irrelevant to our following discussion.
The absence of the odd number in $Y_\tau$ and $W_\tau^a$ terms is due to the $CPT$ symmetry without chemical potentials \eqref{eq:cpt}. 
The same condition also forbids the $3$d Chern--Simons terms.\footnote{
    As the $3$d Chern--Simons terms are consistent with the unitarity condition \eqref{eq:unitarity1} and \eqref{eq:unitarity2}, they can be generated under a certain non-trivial background of the Wilson line along $\mathbb{S}_\beta^1$~\cite{Poppitz:2008hr} (Sec.~5).
}
The total derivative terms, such as $i \partial_i Y_\tau \epsilon_{ijk} Y_{jk}$, vanish as long as we consider $\mathbb{R}^3 + \{\infty\} \simeq \mathbb{S}^3_\infty$.
The reality condition of the $3$d action with respect to Eqs.~\eqref{eq:unitarity1} and \eqref{eq:unitarity2} implies that all the coefficients in Eq.~\eqref{eq:SM3}, $(m_3^2, \lambda_3, m_{\rm D}'^2, m_{\rm D}^2, h_3', h_3'',h_3, \Lambda_T)$, are real.
Note that this lower order $3$d Lagrangian also enjoys the $P$ symmetry of the $1$\,$+$\,$3$d spacetime accidentally, which can be broken by higher order terms~\cite{Kajantie:1997ky}.
The Casimir energy with respect to the compactified $\mathbb{S}_\beta^1$ is denoted by $\Lambda_T$ that depends on the radius of $\mathbb{S}_\beta^1$ and hence $T$, which corresponds to the free energy of relativistic particles of the original $4$d Lagrangian.
All the $3$d matter fields are gapped since the appearance of the mass term is not protected by some symmetries, while the magnetic component of the gauge fields are massless due to the $3$d gauge symmetry unless the Higgs mechanism is triggered.

The coefficients of the $3$d Lagrangian are determined by the matching at the scale $\pi T$.
The $3$d fields are related to the renormalized $4$d fields through the wave function renormalization:
\begin{alignat}{3}
    \Phi^\dag \Phi &= \frac{1}{T}\, Z^\Phi_\mu \big[\Phi_4^\dag \Phi_4 \big]_\mu, &\qquad
    Y_{\tau}^2 &= \frac{1}{T}\, Z^{Y_\tau}_\mu \big[Y_{4\tau}^2 \big]_\mu, &\qquad
    W_{\tau}^a W_{\tau}^a &= \frac{1}{T}\, Z^{W_\tau}_\mu \big[W_{4\tau}^a W_{4\tau}^a \big]_\mu, 
    \label{eq:matching_fields1}\\
    Y_{i} Y_{i} &= \frac{1}{T}\, Z^{Y_i}_\mu \big[Y_{4i} Y_{4i} \big]_\mu, &\qquad
    W_{i}^a W_{i}^a &= \frac{1}{T}\, Z^{W_i}_\mu \big[W_{4i}^a W_{4i}^a \big]_\mu.
    \label{eq:matching_fields2}
\end{alignat}
One may see that the $3$d fields are independent of the choice of the renormalization scale $\mu$ since the right-hand sides are proportional to the bare fields.
By utilizing this field redefinition, we can identify the $4$d vertices to the corresponding $3$d couplings
\begin{alignat}{3}
    g_3^2 &= g^2 (T) \,T + \mathcal{O}(g^4),& \qquad
    g_3'^2 &= g'^2 (T) \,T + \mathcal{O}(g'^4), & \qquad
    \lambda_3 &= \lambda (T) \,T + \mathcal{O}(\lambda^2), 
    \label{eq:matching_coeffs1}\\
    h _ 3 &= \frac 14 g ^ 2 (T) \, T + \mathcal{O}( g ^ 4 ), & \qquad
    h_3' &= \frac 14 g'^ 2 (T)\, T + \mathcal{O}( g'^ 4 ), & \qquad
    h_3'' &= -  \frac 12 g (T)\, g'(T)\, T +  \mathcal{O}( g^2 g'^ 2 ).
    \label{eq:matching_coeffs2}
\end{alignat}
where the $4$d couplings are evaluated at the scale $T$ so that the leading logs are taken into account.
Again we drop the higher order terms in the counting rule of Eq.~\eqref{eq:counting_rule}.
Once the matching is performed at the hard scale $T$, these couplings do not run below this scale since they are free from the UV divergences within the $3$d theory.
The Debye screening masses are obtained by computing the loop corrections of higher Kaluza--Klein modes on the two-point functions, which yields
\begin{equation}
    m_{\rm D}^2 (T) = \frac{1}{3} g^2 (T) \, T^2 \, \qty( \frac{5}{2} +  n_f ) + \mathcal{O}(g^4), \qquad
    m_{\rm D}'^2 (T) = \frac{1}{6} g'^2 (T) \, T^2 \, \qty( 1 + \frac{10}{3}  n_f ) + \mathcal{O}(g'^4).
    \label{eq:Debye_mass}
\end{equation}
Similarly, the $3$d Higgs mass is obtained as
\begin{equation}
    m_3^2 (T) = m_\Phi^2 (T) + \frac{T^2}{2} \qty[ \lambda (T) + \frac{3}{8} g^2 (T) + \frac{1}{8} g'^2 (T) + \frac{1}{2} y_t^2 (T) ] + \mathcal{O}(\lambda^2).
    \label{eq:Higgs_mass}
\end{equation}
Here the $4$d parameters on the right-hand sides are evaluated at the scale $T$.
On the contrary to the $3$d couplings, the $3$d mass parameters run below the hard scale $T$.
Hence the expressions given above in Eqs.~\eqref{eq:Debye_mass} and \eqref{eq:Higgs_mass} can be regarded as the boundary condition of the renormalization group equation at the scale $T$.
For clarity, we explicitly state the scale dependence of the $3$d masses and set it to be $T$ on the left-hand sides.

When the $4$d negative Higgs mass squared overcomes the thermal mass squared, \textit{i.e.,} $m_3^2 < 0$, it is convenient to expand the Higgs doublet around $v_3$ as
\begin{equation}
    \Phi ({\bm x}) = 
    \begin{pmatrix}
        \phi^+ ({\bm x}) \\
        \frac{v_3 + H ({\bm x}) + i \phi^0 ({\bm x})}{\sqrt{2}}
    \end{pmatrix}, \qquad
    \Phi^\dag ({\bm x}) =
    \begin{pmatrix}
        \phi^- ({\bm x}) & \frac{v_3 + H ({\bm x}) - i \phi^0 ({\bm x})}{\sqrt{2}}
    \end{pmatrix}.
    \label{eq:Higgs_dofs}
\end{equation}
We are mostly interested in a regime where $ v_3 \sim \sqrt{T}$ in the following discussion.
In this regime, the $3$d electroweak gauge fields are gapped because of the Higgs mechanism rather than the confinement owing to $g_3 v_3 > g_3^2$, and hence the $3$d EFT remains perturbative.
This allows us to perform the perturbative calculations based on Eq.~\eqref{eq:SM3} in Sec.~\ref{sec:magnetic}.

Note that, for a bit smaller case of $\sqrt{T} > v_3$ (but still $v_3 > g_3$), one may further integrate out the $3$d matter fields such as $Y_\tau$ and $W_\tau^a$ because the Debye screening mass is larger than that via the Higgs mechanism, \textit{i.e.}, $m_{\rm D}^2 > h_3 v_3^2$ and $m_{\rm D}'^2 > h_3' v_3^2$.
In this case, we may obtain the magnetostatic $3$d SM, which is solely composed of the $3$d gauge fields, $Y_i$ and $W_i^a$, and the Higgs field, $\Phi$.

\section{Phases of \texorpdfstring{\boldmath{$3$}}{3}d SM}
\label{sec:phases}

In this section, we extend the Landau paradigm to the generalized symmetry for the EWSB in the 3d SM.
Naively, the high- and low-temperature phases are distinguished by the symmetry breaking of $\mathrm{SU}(2)_{\mathrm L}\times \mathrm{U}(1)_Y$ to $\mathrm{U}(1)_{\mathrm{em}}$ induced by the VEV of the Higgs field $\Phi$.
However, the Higgs field is gauge dependent, and hence it is not a good order parameter \cite{Elitzur:1975im}.\footnote{
    In other words, we cannot characterize the phase by the gauge symmetry breaking.
    We should use the global symmetries in the Landau paradigm.
}
Moreover, there is no globally charged operator to distinguish the two phases, and it has been believed that the EWSB is the crossover.
We confirm that this is the case even in light of the generalized symmetries (\textit{i.e.,} symmetry breaking patterns are the same in both phases).
To this end, we first identify the symmetries of the $3$d SM including the generalized ones as well as the operators charged under the symmetries.
We then study the symmetry breaking pattern in high- and low-temperature limits by computing the VEV of the charged operators.
Finally, we comment on the potential implications on the cosmological magnetic fields.
Here we focus on internal symmetries [\textit{i.e.,} $CPT$ symmetry given in Eq.~\eqref{eq:cpt} and the emergent $P$ symmetry is excluded].
We distinguish the Lie algebra $\mathfrak{g}$ and Lie group $G$ in this section.

\subsection{Symmetry and anomaly}
\label{sec:symandanomaly}
The Wilson and 't Hooft operators are key ingredients of the generalized symmetry.
To identify them, we discuss the gauge group of the $3$d SM~\eqref{eq:SM3} (see also~\cite{Tong:2017oea}).
Although the Lie algebra of the 3d SM is $\mathfrak{su}(2)_{\mathrm L} \oplus \mathfrak{u}(1)_Y$, there can be several candidates of the global form of the gauge group.
This is related to the fact that the matter field listed in Tab.~\ref{tab:3d_matter} is invariant under the $\mathbb{Z}_2$ action
\begin{align}
    \mathbb{Z}_2: e^{2\pi i Y}(-1)^{2s}, \label{eq:Z2_identification}
\end{align}
where $s$ is the $\mathfrak{su}(2)_{\mathrm L}$ spin of the field.
Consequently, there are two candidates of the global form of the gauge group:
\begin{align}
    &G=\mathrm{SU}(2)_{\mathrm L} \times \mathrm{U}(1)_Y,
    &&\tilde{G}=(\mathrm{SU}(2)_{\mathrm L} \times \mathrm{U}(1)_Y)/\mathbb{Z}_2.
\end{align}
Notice that there are four variations of the global form of the gauge group in the SM~\cite{Tong:2017oea}.
Here we have found two variations since we concentrate on the electroweak sector.

\begin{figure}[ht]
    \centering
    \begin{tikzpicture}
    \draw[->] (-2,0) -- (3,0) node[right] {};
    \draw[->] (0,-2) -- (0,3) node[above] {};
    \draw[->, color=green!80!black, opacity=0.5] (-2,-2) -- (3,3) node[right] {$Q$};
    \draw[thick] (2,2.5) -- (2,3); 
    \draw[thick] (2,2.5) -- (2.5,2.5); 
    
    \foreach \x in {-2,-1,0,1,2}
        \foreach \y in {-2,-1,0,1,2}
            \fill (\x, \y) circle (1pt);
    
    \draw[->, thick, blue] (0,0) -- (1,0) node[below right] {$(1,0)$};
    \draw[->, thick, red] (0,0) -- (0,1) node[above left] {$\left(0,\frac{1}{\sqrt{2}}\right)$};

    \node at (-0.5,2.8) {$\sqrt{2}T^3$};
    \node at (3.2,-0.3) {$2Y$};
    \node at (2.3,2.8) {$G$};
    
    \begin{scope}[xshift=7cm]
    \draw[->] (-2,0) -- (3,0) node[right] {};
    \draw[->] (0,-2) -- (0,3) node[above] {};
    \draw[->, color=green!80!black, opacity=0.5] (-1,-2) -- (1.5,3) node[left] {$Q_{\mathrm M}$};
    \draw[thick] (2,2.5) -- (2,3); 
    \draw[thick] (2,2.5) -- (2.5,2.5); 

    \foreach \x in {-1,0,1}
        \foreach \y in {-2,0,2}
            \fill (\x, \y) circle (1pt);

    \draw[->, thick, red] (0,0) -- (0,2) node[below right] {$\left(0,\sqrt{2}\right)$};
    \draw[->, thick, blue] (0,0) -- (1,0) node[above left] {$\left(1,0\right)$};

    \node at (-0.5,2.8) {$\sqrt{2}T^3_{\mathrm M}$};
    \node at (3.2,-0.3) {$Y_{\mathrm M}$};
    \node at (2.3,2.8) {$G$};
    \end{scope}

    \end{tikzpicture}
    \caption{Electric (left) and magnetic (right) charge lattices of $G=\mathrm{SU}(2)_{\mathrm L}\times {\mathrm U}(1)_Y$ theory.
    The green lines correspond to the electric and magnetic charges of $\mathfrak{u}(1)_\mathrm{em}$.}
    \label{fig:charge_lattices_withoutZ2}
\end{figure}

These two theories, $G$ and $\tilde{G}$, have different charge lattices~\cite{Aharony:2013hda,Tong}.
The charge lattice is the set of the points corresponding to the allowed electric and magnetic charges of the gauge group.
The electric charge lattice of the gauge group $G$ is (see also the left panel of Fig.~\ref{fig:charge_lattices_withoutZ2})\footnote{
    Here we assume that the periodicity of $\mathrm{U(1)}_Y$ is $4\pi$ so that the charge $Y=1/2$ is allowed, but $Y=1/4$ is not allowed.
    Strictly speaking, depending on the choice of the periodicity, there are infinite possibilities of the charge lattices.
}

\begin{align}
    &\text{(Electric)}_{G}: \quad \Lambda_{\mathrm e} = \left\{ \left(n_1, \frac{n_2}{\sqrt{2}}\right)\in 2Y\times\sqrt{2}T^3  \, \middle|\, n_{1,2} \in \mathbb{Z} \right\},
\end{align}
where $n_1=2Y$ is the double of the charge of $\mathfrak{u}(1)_Y$ (we normalize $n_1$ in such a way that the Higgs field has $n_1=1$), and $T^3$ is one of the generator of $\mathfrak{su}(2)_{\mathrm L}$ Lie algebra.\footnote{
    The commutation relation is given by $[T^a,T^b]=i\epsilon^{abc}T^c$. The eigenvalue of $T^3$ is $\pm1/2$ for the fundamental representation.
}
For instance, the Higgs field and $W$ bosons correspond to $(1,\pm 1/\!\sqrt{2})$ and $(0,\pm \sqrt{2})$ in $\Lambda_{\mathrm e}$, respectively.
It is important that the points with $n_1+n_2\in 2\mathbb{Z}$ are populated by the dynamical states, while the points with $n_1+n_2\in 2\mathbb{Z}+1$ are not.
On the other hand, the electric charge lattice of the gauge group $\tilde{G}$ is (see also the left panel of Fig.~\ref{fig:charge_lattices_withZ2})
\begin{align}
    &\text{(Electric)}_{\tilde{G}}: \quad \Lambda_{\mathrm e} = \left\{ \left(n_1, \frac{n_2}{\sqrt{2}}\right)  \, \middle|\, n_{1,2} \in \mathbb{Z}, \, n_1 + n_2 \in 2\mathbb{Z} \right\}.
\end{align}
Note that we have the additional constraint $n_1 + n_2 \in 2\mathbb{Z}$ because of $\mathbb{Z}_2$ identification introduced in \eqref{eq:Z2_identification}.

On top of the electrically charged states, there are states magnetically charged under the gauge group.
The set of allowed magnetic charges again forms the lattice, and is called the magnetic charge lattice.
The magnetic charge lattice is given by the dual lattice of the electric charge lattice, where the inner product is defined as
\begin{align}
    \left(n_1, \frac{n_2}{\sqrt{2}}\right)\cdot
    \left(m_1, \frac{m_2}{\sqrt{2}}\right)
    = n_1 m_1 + \frac{1}{2}n_2 m_2.
\end{align}
Then, the magnetic charge lattices are (see also the right panels of Figs.~\ref{fig:charge_lattices_withoutZ2} and \ref{fig:charge_lattices_withZ2})
\begin{align}
    &\text{(Magnetic)}_{G}: \quad \Lambda_{\mathrm m} = \left\{ \left(m_1, \sqrt{2} m_2\right)  \, \middle|\, m_{1,2} \in \mathbb{Z} \right\},\\
    &\text{(Magnetic)}_{\tilde{G}}: \quad \Lambda_{\mathrm m} = \left\{ \left(\frac{m_1}{2}, \frac{m_2}{\sqrt{2}}\right)  \, \middle|\, m_{1,2} \in \mathbb{Z}, \, m_1 + m_2 \in 2\mathbb{Z} \right\}.
\end{align}

\begin{figure}[ht]
    \centering
    \begin{tikzpicture}
    \draw[->] (-1,0) -- (3,0) node[right] {};
    \draw[->] (0,-1) -- (0,3) node[above] {};
    \draw[->, color=green!80!black, opacity=0.5] (-1,-2) -- (1.5,3) node[right] {$Q$};
    \draw[thick] (2,2.5) -- (2,3); 
    \draw[thick] (2,2.5) -- (2.5,2.5); 

    \foreach \x in {-1,0,1}
        \foreach \y in {-1,0,1,2}
            \fill (\x + 0.5*\y, \y) circle (1pt);
    
    \draw[->, thick, blue] (0,0) -- (1,0) node[below right] {$(2,0)$};
    \draw[->, thick, red] (0,0) -- (0.5,1) node[above left] {$\left(1,\frac{1}{\sqrt{2}}\right)$};
    
    \node at (-0.5,2.8) {$\sqrt{2}T^3$};
    \node at (3.2,-0.3) {$2Y$};
    \node at (2.3,2.8) {$\tilde{G}$};    
    \begin{scope}[xshift=7.5cm]
    \draw[->] (-1,0) -- (3,0) node[right] {};
    \draw[->] (0,-1) -- (0,3) node[above] {};
    \draw[->, color=green!80!black, opacity=0.5] (-1,-1) -- (2.3,2.3) node[left] {$Q_{\mathrm M}$};
    \draw[thick] (2,2.5) -- (2,3); 
    \draw[thick] (2,2.5) -- (2.5,2.5); 

    \foreach \x in {-2,0,2}
        \foreach \y in {-1,0,1,2}
            \fill (\x + 1.0*\y, \y) circle (1pt);

    \draw[->, thick, red] (0,0) -- (0,2) node[below right] {$\left(0,\sqrt{2}\right)$};
    \draw[->, thick, blue] (0,0) -- (1,-1) node[above right] {$\left(\frac{1}{2},-\frac{1}{\sqrt{2}}\right)$};

    \node at (-0.5,2.8) {$\sqrt{2}T^3_{\mathrm M}$};
    \node at (3.2,-0.3) {$Y_{\mathrm M}$};
    \node at (2.3,2.8) {$\tilde{G}$};
    \end{scope}

    \end{tikzpicture}
    \caption{Electric (left) and magnetic (right) charge lattices of $\tilde{G}=(\mathrm{SU}(2)_{\mathrm L}\times \mathrm{U}(1)_Y)/\mathbb{Z}_2$ theory.
    The green lines correspond to the electric and magnetic charges of $\mathfrak{u}(1)_\mathrm{em}$.}
    \label{fig:charge_lattices_withZ2}
    \end{figure}

Each point in the electric charge lattice corresponds to the charge of the Wilson loop, and each point in the magnetic charge lattice corresponds to the charge of the 't Hooft operator.
In particular, the Wilson/'t Hooft operators corresponding to the points where no dynamical states are populated are not screened, and hence they can be viewed as the order parameters of the generalized symmetry.
In this way, given charge lattices, we identify the electric 1-form and magnetic 0-form symmetries acting on the Wilson loop and the 't Hooft operator, respectively.
First, we have the electric symmetry acting on the spatial and temporal Wilson loops (denoted by $\mathbb{Z}_2^{[1]}$ and $\mathbb{Z}_2^{[0]}$, respectively)
\begin{align}
    \mathbb{Z}_2^{[1]}: \quad &\left|\left(0,\frac{1}{\sqrt{2}}\right)\text{Wilson loop} \right\rangle \to -\left|\left(0,\frac{1}{\sqrt{2}}\right)\text{Wilson loop} \right\rangle,\\
    \mathbb{Z}_2^{[0]}: \quad & \left|\mathrm{U}(1)_Y/\mathrm{SU}(2)_{\mathrm L}\text{ Polyakov loops}\right\rangle \to - \left|\mathrm{U}(1)_Y/\mathrm{SU}(2)_{\mathrm L}\text{ Polyakov loops}\right\rangle,
\end{align}
where the Polyakov loops (temporal Wilson loops) are defined as the trace of the operators appeared in Eq.~\eqref{eq:Polyakov_line}.
As the $(0,1/\!\sqrt{2})$ Wilson loop is allowed only for the gauge group $G$, the $\mathbb{Z}_2^{[1]}$ and $\mathbb{Z}_2^{[0]}$ symmetries exist only for $G$ theory.
Next, we have the magnetic symmetry acting on the 't Hooft operator (denoted by $\mathrm{U}(1)_{\mathrm M}^{[0]}$)
\begin{align}
    \mathrm{U}(1)_{\mathrm M}^{[0]}\text{ for $G$}: \quad &\left|\qty(Y_{\mathrm M},T_{\mathrm M}^3)\text{ monopole op.} \right\rangle \to e^{2\pi i\theta Y_{\mathrm M}}\left|\qty(Y_{\mathrm M},T_{\mathrm M}^3)\text{ monopole op.} \right\rangle, \nonumber\\
    \mathrm{U}(1)_{\mathrm M}^{[0]}\text{ for $\tilde{G}$}: \quad &\left|\qty(Y_{\mathrm M},T_{\mathrm M}^3)\text{ monopole op.} \right\rangle \to e^{4\pi i\theta Y_{\mathrm M}}\left|\qty(Y_{\mathrm M},T_{\mathrm M}^3)\text{ monopole op.} \right\rangle. \label{eq:magnetic_symmetry}
\end{align}
where $\theta=[0,2\pi]$.
The different values in the exponent for $G$ and $\tilde{G}$ are due to the fact that $Y_{\mathrm M}$ is an integer for $G$ and can be a half integer for $\tilde{G}$.
We also have the emergent symmetry which only appears in the IR:
\begin{align}
    C:\quad\Phi\to i\sigma^2\Phi^*, \quad Y_i \to -Y_i, \quad W_{i,\tau} \to -W_{i,\tau}^T,
    \qquad
    \mathbb{Z}_2^{\tau}:\quad Y_\tau \to -Y_\tau, \quad W_\tau \to -W_\tau.
\end{align}

Putting altogether, the total symmetry (in the IR theories) is
\begin{align}
    G: \quad & \left(C\ltimes \mathrm{U}(1)_{\mathrm M}^{[0]}\right) \times \mathbb{Z}_2^{[0]}\times \mathbb{Z}_2^{[1]}\times \mathbb{Z}_2^{\tau}, \nonumber\\
    \tilde{G}: \quad & \left(C\ltimes \mathrm{U}(1)_{\mathrm M}^{[0]} \right) \times \mathbb{Z}_2^{\tau}.
\end{align}
We list the symmetry of the theory in Tab.~\ref{table:symmetry}.
\begin{table}[h]
    \centering
    \begin{tabular}{|c|c|c|}
        \hline
         & $G$ & $\tilde{G}$ \\
        \hline
        \textbf{0-form} & $\mathbb{Z}_2^{[0]}, \mathrm{U}(1)_{\mathrm M}^{[0]}, C, \mathbb{Z}_2^{\tau}$ & $\mathrm{U}(1)_{\mathrm M}^{[0]}, C, \mathbb{Z}_2^{\tau}$ \\
        \hline
        \textbf{1-form} & $\mathbb{Z}_2^{[1]}$ &  \\
        \hline
    \end{tabular}
    \caption{Symmetries of the $G=\mathrm{SU}(2)_{\mathrm L}\times \mathrm{U}(1)_Y$ and $\tilde{G}=(\mathrm{SU}(2)_{\mathrm L} \times \mathrm{U}(1)_Y)/\mathbb{Z}_2$ theories.}
    \label{table:symmetry}
\end{table}

Let us discuss the 't Hooft anomaly of the theory.
The equation \eqref{eq:magnetic_symmetry} suggests the mixed anomaly between the $\mathbb{Z}_2$ electric symmetry and the $\mathrm U(1)$ magnetic symmetry in $G$ theory.
The $\mathrm{U}(1)_{\mathrm M}^{[0]}$ symmetry can be viewed as $\mathbb{R}$ symmetry with $\mathbb{Z}$ gauging.
However, this $\mathbb{Z}$ gauge symmetry is broken if we gauge $\mathbb{Z}_2^{[1]}$ symmetry due to the presence of fractional $Y_{\mathrm M}$ charged state.
Consequently, the mixed anomaly rules out the trivially gapped phase.
An analogous anomaly is discussed in Ref.~\cite{Komargodski:2017dmc}.

\subsection{Electroweak crossover}
\label{sec:crossover}
Here we discuss the electroweak crossover and its relation with the breaking pattern of the symmetry summarized in Tab.~\ref{table:symmetry}.
We focus on the $G$ theory for simplicity, but the discussion can be applied to the $\tilde{G}$ theory as well.

\begin{itemize}
    \item High temperature ($m_3^2 \gg g_3^4$) 
    
    At high temperature corresponding to large positive Higgs mass squared, all the 3d matter fields ($\Phi, Y_\tau$ and $W_{\tau}^a$) can be integrated out perturbatively.
    Since the VEV of these fields are zero, $\mathbb{Z}_2^{[0]}$ is spontaneously broken while $C$ and $\mathbb{Z}_2^{\tau}$ are unbroken.
    The IR theory becomes the pure gauge theory in 3d.
    The gauge algebra $\mathfrak{su}(2)_{\mathrm L}$ is confining (\textit{i.e.,} gauge bosons get the magnetic mass) while $\mathfrak{u}(1)_Y$ is in the Coulomb phase.
    We consider the VEV of $\mathbb{Z}_2^{[1]}$ Wilson loop, which can be viewed as the worldline of the probe particles with charge $(0,\pm1/\!\sqrt{2})$.
    In the presence of the dynamical Higgs field $\Phi$, the interaction between the probes is mediated by $\mathfrak{u}(1)_Y$ gauge boson.
    As the force between the probes is attractive and scales as $1/r$ with respect to the distance $r$, we observe (see \textit{e.g.,} \cite{Komargodski})
    \begin{align}
        (\text{energy}) \sim g_3'^2 \log (r g_3'^2).
    \end{align}
    The overall sign is positive, and hence the probes are weakly confined.
    Let us consider the Wilson loop with the rectangular shape and the length of vertical and horizontal directions $L_{T}$ and $L_X$, respectively.
    By viewing the vertical direction as the Euclidean time direction, the expectation value of the Wilson loop is
    \begin{align}
        \left\langle\left(0,\frac{1}{\sqrt{2}}\right)\text{Wilson loop}\right\rangle
        \sim\exp\left(-L_T (\text{energy})\right)
        \sim \exp\left(-L_T g_3'^2 \log (L_X g_3'^2)\right)
        =\left(\frac{1}{L_X g_3'^2}\right)^{L_T g_3'^2}.
    \end{align}
    We obtain the power law behavior of the Wilson loop VEV.
    We can view this as a Berezinski--Kostelitz--Thouless (BKT) type breaking of the $\mathbb{Z}_2^{[1]}$ symmetry.

    We also observe that the magnetic symmetry is spontaneously broken as $\mathfrak{u}(1)_Y$ is in the Coulomb phase (see \textit{e.g.,} \cite{Tong}).
    The charged object is the monopole operator with charge $(1,0)$.
    The massless $\mathfrak{u}(1)_Y$ gauge boson is interpreted as the Nambu--Goldstone mode of the spontaneously broken $\mathrm{U}(1)_{\mathrm M}^{[0]}$.
    \item Low temperature ($g_3 v_3 \gg g_3^2$) 
    
    At the low temperature corresponding to large negative Higgs mass squared, the Higgs field develops the VEV.
    The gauge symmetry $\mathfrak{su}(2)_{\mathrm L}\oplus\mathfrak{u}(1)_Y$ is broken down to $\mathfrak{u}(1)_\mathrm{em}$, and all the 3d fields other than the photon are gapped.
    Consequently, $\mathbb{Z}_2^{[0]}$ symmetry is spontaneously broken while $C$ and $\mathbb{Z}_2^{\tau}$ are unbroken.\footnote{
        Even though $\Phi$ develops the VEV, the gauge invariant operator $\Phi^\dagger\Phi$ is not charged under $C$.
    }
    Regarding the electric line operators, the $\mathbb{Z}_2^{[1]}$ Wilson loop becomes Wilson loop of $\mathfrak{u}(1)_\mathrm{em}$.
    This exhibits the power law behavior of the VEV.

    For the 't Hooft operator with charge $(Y_{\mathrm M},T_{\mathrm M}^3)$, it becomes 't Hooft operator of $\mathfrak{u}(1)_\mathrm{em}$ with the charge
    \begin{align}
        \left(1,\frac{1}{\sqrt{2}}\right)\cdot\qty(Y_{\mathrm M}, \sqrt{2}T_{\mathrm M}^3)=Y_{\mathrm M}+T_{\mathrm M}^3=:Q_{\mathrm M}. \label{eq:electric_charge}
    \end{align}
    Based on the picture of the superconductivity, the Higgsing in the electric sector corresponds to the confinement in the magnetic sector.
    The confinement charge is
    \begin{align}
        \left(1,-\frac{1}{\sqrt{2}}\right)\cdot\qty(Y_{\mathrm M}, \sqrt{2}T_{\mathrm M}^3)=Y_{\mathrm M}-T_{\mathrm M}^3. \label{eq:confinement_charge}
    \end{align}

    Now, let us argue that $\mathrm{U}(1)_{\mathrm M}^{[0]}$ is spontaneously broken.
    We consider the two-point correlation function of $(\pm1,0)$ monopole operators.
    At first sight, $(\pm1,0)$ operator has the confinement charge $\pm1$, but the dynamical $(0,\pm\sqrt{2})$ magnetically charged object can screen it.
    Consequently, the behavior of the two-point function is the same as that of $\pm(1,\sqrt{2})$ monopole operators.
    This is nothing but the correlation function of the 't Hooft operator of $\mathfrak{u}(1)_\mathrm{em}$ in the Coulomb phase.
    Therefore, we see that $\mathrm{U}(1)_{\mathrm M}^{[0]}$ is spontaneously broken.    
\end{itemize}

The result of the symmetry breaking pattern at the high and low temperatures are summarized in Tab.~\ref{table:phases}.
\begin{table}[h]
    \centering
    \begin{tabular}{|c|c|c|c|c|c|}
        \hline
        $G$ & $\mathbb{Z}_2^{[0]}$ & $\mathbb{Z}_2^{[1]}$ & $\mathrm{U}(1)_{\mathrm M}^{[0]}$ & $\mathbb{Z}_2^{\tau}$ & $C$ \\
        \hline
        \textbf{High T} & SSB & BKT & SSB & Unbroken & Unbroken \\
        \hline
        \textbf{Low T} & SSB & BKT & SSB & Unbroken & Unbroken \\
        \hline
    \end{tabular}
    \caption{High and low-temperature phases of the $G=\mathrm{SU}(2)_{\mathrm L}\times \mathrm{U}(1)_Y$ theory.}
    \label{table:phases}
\end{table}
We see that the symmetry breaking pattern of the high- and low-temperature phases are the same, and these two regions could be smoothly connected.
This is what is observed in Monte Carlo simulation~\cite{Kajantie:1996qd,DOnofrio:2015gop}.

Even though this is likely the case, we comment on a possibility consistent with the analysis of the mixed anomaly.
Suppose that the $\mathrm{SU}(2)_{\mathrm L}$ is confining and the Higgs field forms a bound state $S=\epsilon_{\alpha\beta}\phi^\alpha \phi^\beta$ which is an $\mathfrak{su}(2)_{\mathrm L}$ singlet but has charge $1$ under $\mathfrak{u}(1)_Y$.
If the bound state $S$ develops the VEV, then the $\mathbb{Z}_2^{[1]}$ is spontaneously broken but $\mathrm{U}(1)_{\mathrm M}^{[0]}$ remains unbroken.
This is the scenario consistent with the mixed anomaly, and the intermediate-temperature regime is described by a topological theory without massless degrees of freedom.
The massless excitation (or equivalently the primordial magnetic field) generated before the electroweak crossover just decays without leaving the ordinary magnetic field in the low-temperature regime.
If this could be the case, the discussion in the following sections does not apply.

\begin{figure}[ht]
    \centering
    \begin{tikzpicture}
        \draw[->, thick] (-5,-1) -- (-5,5) node[anchor=south] {$T$ or $m_3^2$};
        
        \node[text=red] at (-3,2.5) {$\mathbb{Z}_2^{[1]}$: BKT};
        \node[text=red] at (-3,1.5) {$\mathrm{U}(1)_{\mathrm M}^{[0]}$: SSB};

        \begin{scope}[xshift=7.5cm]

        \draw[->, thick] (-5,-1) -- (-5,5) node[anchor=south] {$T$ or $m_3^2$};
        \node[text=red] at (-3,4.5) {$\mathbb{Z}_2^{[1]}$: BKT};
        \node[text=red] at (-3,3.5) {$\mathrm{U}(1)_{\mathrm M}^{[0]}$: SSB};
        \node[text=blue] at (-3,2.5) {$\mathbb{Z}_2^{[1]}$: SSB};
        \node[text=blue] at (-3,1.5) {$\mathrm{U}(1)_{\mathrm M}^{[0]}$: unbroken};
        \node[text=red] at (-3,0.5) {$\mathbb{Z}_2^{[1]}$: BKT};
        \node[text=red] at (-3,-0.5) {$\mathrm{U}(1)_{\mathrm M}^{[0]}$: SSB};
        \draw[thick] (-5,3) -- (-1,3); 
        \draw[thick] (-5,1) -- (-1,1); 
    \end{scope}
    \end{tikzpicture}
    \caption{Possible phase diagrams consistent with the mixed anomaly.
    Left: Electroweak crossover. This scenario is supported by Monte Carlo simulation.
    Right: An intermediate phase with mass gap.}
    \label{fig:phase_diagram}
\end{figure}

\subsection{Implications on the cosmological magnetic field}
\label{sec:implications_magnetic}
We are interested in the cosmological magnetic field.
The $\mathrm{U}(1)_Y$ magnetic field is generated at high temperature, and, after the completion of the electroweak crossover, it becomes $\mathrm{U}(1)_\mathrm{em}$ magnetic field.

Here we make some comments on the cosmological magnetic field based on the analysis in this section.
First, the existence of the massless mode is guaranteed by the SSB of $\mathrm{U}(1)_{\mathrm M}^{[0]}$ at all temperatures (assuming the left scenario in Fig.~\ref{fig:phase_diagram}).
It would be interesting to directly check the nonvanishing expectation value of the 't Hooft operator in the lattice simulation.

The second comment concerns the stability of the magnetic field. 
At high temperature, the large scale magnetic field at the high temperature survives without suffering from (the magnetic version of) the Schwinger effect thanks to the Bianchi identity of $\mathrm{U}(1)_Y$. 
On the other hand, at low temperature, the stability of $\mathrm{U}(1)_\mathrm{em}$ is not clear since the Bianchi identity of $\mathrm{U}(1)_\mathrm{em}$, namely $A_{jk}$ in Eq.~\eqref{eq:calW_def}, does not hold.
Is there Schwinger effect for the $\mathrm{U}(1)_\mathrm{em}$ magnetic field?
Interestingly, it turns out that, although the object magnetically charged under $\mathrm{U}(1)_\mathrm{em}$ exists, it is confined and does not screen the magnetic field.
To argue that this is the case, we recall that there are two basis vectors in magnetic charge lattice (right panel of Fig.~\ref{fig:charge_lattices_withoutZ2}). 
One has $(1,0)$ magnetic charge, and the other has $(0,\sqrt{2})$ magnetic charge.
The monopole with $(1,0)$ magnetic charge is not a dynamical object while the monopole with $(0,\sqrt{2})$ magnetic charge is a dynamical object in the theory.
In fact, in the low-temperature phase, the pair of $(0,\pm\sqrt{2})$ monopoles is described as the Nambu monopole~\cite{Nambu:1977ag} (see Fig.~\ref{fig:Nambu_monopole}).
\begin{figure}[ht]
    \centering
    \begin{tikzpicture}
        \node[draw, circle, minimum size=1cm] (M) at (0,0) {\textcolor{red}{$M$}};
        \node[above left=0.0cm of M] {$(0, \sqrt{2})$ monopole};
    
        \node[draw, circle, minimum size=1cm] (Mbar) at (4,0) {\textcolor{red}{$\overline{M}$}};
        \node[above right=0.0cm of Mbar] {$(0, -\sqrt{2})$ anti-monopole};
    
        \draw (M) -- (Mbar) node[midway, below, blue] {flux tube};
    \end{tikzpicture}
    \caption{The Nambu monopole in the SM. The monopole has $(0, \sqrt{2})$ magnetic charge, which is the minimal unit in the magnetic charge lattice (right panel of Fig.~\ref{fig:charge_lattices_withoutZ2}).
    The monopole has both electromagnetic and confinement charges from \eqref{eq:electric_charge} and \eqref{eq:confinement_charge}.
    Consequently, the monopole is confined and is connected by the flux tube.}
    \label{fig:Nambu_monopole}
\end{figure}
The Nambu monopole has both electromagnetic and confinement charges.
As a result, even though it has $\mathrm{U}(1)_\mathrm{em}$ magnetic charge, it is confined and does not screen the magnetic field.

The final remark is the speculation about the stable configuration of the magnetic field.
Before the electroweak crossover, the constant $\mathrm{U}(1)_Y$ magnetic field is a stable configuration (left panel of Fig.~\ref{fig:magnetic_field}).
The Bianchi identity implies that the $\mathrm U(1)_Y$ magnetic flux is conserved ever, namely $g'\int \dd\vec S\cdot\vec B_Y=\mathrm{const.}$
After the electroweak crossover, the $\mathrm{U}(1)_Y$ magnetic field is no longer stable configuration due to the energy cost.
The stable configuration involves the creation of the Nambu monopole-antimonopole pairs, and the bulk magnetic field becomes the $\mathrm{U}(1)_\mathrm{em}$ magnetic field (right panel of Fig.~\ref{fig:magnetic_field}).
Here, by assuming that the massive $Z$ flux vanishes to minimize the energy cost, we have $e\int \dd\vec S\cdot\vec B_{\rm em}=g'\int \dd\vec S\cdot\vec B_Y$, namely the number of the electromagnetic fluxes is the same as the original $\mathrm U(1)_Y$ ones and thus is conserved, as is clear in the illustration in Fig.~\ref{fig:magnetic_field}.
\begin{figure}[ht]
    \centering
    \begin{tikzpicture}
        \draw[thick] (-2, 0) ellipse (0.6 and 1.2); 
        \draw[thick] (-2, -1.2) -- (2, -1.2);        
        \draw[thick] (-2, 1.2) -- (2, 1.2);          
        \draw[thick] (2, 0) ellipse (0.6 and 1.2);

        \node[thick, red, scale=1.4] at (-2.0, 0) {+};
        \node[thick, red, scale=1.4] at (2.1, 0) {--};
        \node[red] at (-2, 1.5) {\textcolor{red}{$(N,0)$}};
        \node[red] at (2, 1.5) {\textcolor{red}{$(-N,0)$}};
        \node[below=0.1cm] at (0, -1.1) {\textcolor{blue}{$\mathrm{U}(1)_Y$ magnetic field}};

        \draw[->, blue, thick] (-1.55, 0.7) -- (1.5, 0.7);
        \draw[->, blue, thick] (-1.4, 0) -- (1.4, 0);
        \draw[->, blue, thick] (-1.55, -0.7) -- (1.5, -0.7);

        \node[scale=1.4] at (-2, 2.2) {High T};

        \begin{scope}[shift={(8,0)}]
        \draw[thick] (-2, 0) ellipse (0.6 and 1.2); 
        \draw[thick] (-2, -1.2) -- (2, -1.2);        
        \draw[thick] (-2, 1.2) -- (2, 1.2);           
        \draw[thick] (2, 0) ellipse (0.6 and 1.2);

        \node[thick, purple, scale=1.4] at (-2.0, 0) {+};
        \node[thick, purple, scale=1.4] at (2.1, 0) {--};
        \node[purple] at (-2, 1.5) {\textcolor{purple}{$(N, N\sqrt{2})$}};
        \node[purple] at (2, 1.5) {\textcolor{purple}{$(-N, -N\sqrt{2})$}};
        \node[below=0.1cm] at (0, -1.1) {\textcolor{blue}{U(1)$_{\text{em}}$ magnetic field}};

        \draw[->, blue, thick] (-1.55, 0.7) -- (1.5, 0.7);
        \draw[->, blue, thick] (-1.4, 0) -- (1.4, 0);
        \draw[->, blue, thick] (-1.55, -0.7) -- (1.5, -0.7);

        \node[draw, circle, purple, minimum size=0.5cm] at (-2, 0.7) {\textcolor{purple}{$M$}};
        \node[draw, circle, purple, minimum size=0.5cm] at (-2, -0.7) {\textcolor{purple}{$M$}};
        \node[draw, circle, purple, minimum size=0.5cm] at (2, 0.7) {\textcolor{purple}{$\overline{M}$}};
        \node[draw, circle, purple, minimum size=0.5cm] at (2, -0.7) {\textcolor{purple}{$\overline{M}$}};

        \node[scale=1.4] at (-2, 2.2) {Low T};
        \end{scope}
    \end{tikzpicture}
    \caption{The behavior of the magnetic field from high to low temperature.
    Left: At the high temperature before the electroweak crossover, the large scale $\mathrm{U}(1)_Y$ magnetic field is not screened thanks to the Bianchi identity.
    Right: At low temperature after the electroweak crossover, the constant $\mathrm{U}(1)_Y$ magnetic field is no longer stable profile as it carries the confined charge~\eqref{eq:confinement_charge}.
    To remove the flux tube, the Nambu monopole-antimonopole pairs are created. In this way, the $\mathrm{U}(1)_Y$ magnetic field becomes the $\mathrm{U}(1)_\mathrm{em}$ magnetic field.}
    \label{fig:magnetic_field}
\end{figure}

\section{Nambu--Goldstone mode and unconfined magnetic flux}
\label{sec:magnetic}

As discussed in the previous section, the magnetic $\mathrm{U}(1)_{\mathrm M}^{[0]}$ symmetry originated from the original $1$$+$$3$d $1$-form symmetry of $\mathrm{U}(1)_Y$ is always in the SSB phase, given that the {\it Left} phase diagram in Fig.~\ref{fig:phase_diagram} is correct.
This implies that the associated Nambu--Goldstone boson contributes to the macroscopic behavior of the SM, which is nothing but an unconfined magnetic flux, \textit{i.e.,} the ${\rm U}(1)_Y$ magnetic flux in the high-temperature limit and the ${\rm U}(1)_{\rm em}$ one in the low-temperature limit.
Moreover, since the EWSB is believed to be a crossover transition, we should have the Nambu--Goldstone boson associated with the SSB of $\mathrm{U}(1)_{\mathrm M}^{[0]}$ at any temperatures.
In other words, there must exist an unconfined magnetic field $\vec B_{\rm c\mkern-7.5mu/}$, which interpolates those high- and low-temperature limits.
What can we learn about $\vec B_{\rm c\mkern-7.5mu/}$?
In this section, we address the fate of the cosmological magnetic field during the electroweak crossover.

\subsection{Evolution of magnetic flux}\label{sec:unconfined}
Here we perturbatively calculate the transition of the ${\rm U}(1)_Y$ magnetic flux into the ${\rm U}(1)_{\rm em}$ magnetic flux.
In the low-temperature phase, $g_3 v_3 \gg g_3^2$, the unconfined magnetic flux is the ${\rm U}(1)_{\rm em}$ one.
Indeed, one may diagonalize the mass matrix in the Lagrangian density \eqref{eq:SM3} and see that the only massless physical degrees of freedom is the ${\rm U}(1)_{\rm em}$ magnetic field\footnote{Note that the direction of the rotation in the operator space, {\it i.e.}, the sign in front of $\sin\theta_{\rm w}$ in Eq.~\eqref{eq:photon_def}, is opposite to the one in some literature depending on their diverse conventions \cite{Romao:2012pq}.}
\begin{align}
    B_i
        \coloneq\epsilon^{ijk}\partial_jA_k,\qquad
    A_i
        \coloneq\cos\theta_{\rm w}Y_i-\sin\theta_{\rm w}W^3_i,
    \label{eq:photon_def}
\end{align}
where we define the weak mixing angle $\theta_{\rm w}$ so that $\sin\theta_{\rm w}=g_3'/\sqrt{g_3^2+g_3'^2}$ and $\cos\theta_{\rm w}=g_3/\sqrt{g_3^2+g_3'^2}$.
As is explicit in Eq.~\eqref{eq:photon_def}, $Y_i$ and $W^3_i$ mix by an angle $\theta_{\rm w}$ to yield the massless field $A_i$ and the massive $Z$ boson $Z_i\coloneq\cos\theta_{\rm w}W^3_i+\sin\theta_{\rm w}Y_i$ in the perpendicular direction.

Then, a naive expectation is that the massless field during the transition is specified by a time-dependent angle $\theta_{\rm eff}(T)$, which interpolates $\theta_{\rm eff}(T\gg T_{\rm EW})=0$ in the symmetric and $\theta_{\rm eff}(T\ll T_{\rm EW})=\theta_{\rm w}$ in the broken phases \cite{DOnofrio:2015gop,Kamada:2016cnb}, where $T_{\rm EW}\sim\mathcal O(100)\,{\rm GeV}$ is the electroweak energy scale.
On the other hand, a possible complication would be that the $T^3_{\rm M}$ direction is not as distinctive as it is in the broken phase when the symmetry is not completely broken down to ${\rm U}(1)_{\rm em}$.
The Nambu--Goldstone bosons (or equivalently the longitudinal modes of the massive gauge bosons) are no longer safely integrated out during the transition, implying that their excitation may introduce a contamination of $W^{1,2}_i$ into the massless mode. 

To reconcile with the complication mentioned above, let us remark that
\begin{align}
    Y_{ij}\quad\text{and}\quad
    \mathcal W_{ij}
        \coloneq-\dfrac{\Phi^\dagger\sigma^a\Phi}{\Phi^\dagger\Phi}W_{ij}^a
        =W_{ij}^3+\mathcal O(g_3/m_W)
    \label{eq:calW_def}
\end{align}
are ${\rm SU}(2)_{\rm L}\times{\rm U}(1)_Y$ gauge-independent.
Note that this operation can be regarded as a projection of the gauge field $W^a_i$ onto $n^a \equiv - \Phi^\dag \sigma^a \Phi/ \Phi^\dag \Phi$ direction.
Since the $\mathcal O(g_3/m_W)$ correction vanishes in the $m_W\to\infty$ limit, we can rewrite Eq.~\eqref{eq:photon_def} in a gauge-independent way,
\begin{align}
    B_i
        =\dfrac{1}{2}\epsilon^{ijk}A_{jk},\qquad
    A_{jk}
        =\cos\theta_{\rm w}Y_{ij}-\sin\theta_{\rm w}\mathcal W_{ij}\qquad\qquad
    \text{in the low-temperature limit.}
\end{align}

On top of this, we suppose that the massless mode always lies in a certain direction in the operator space spanned by $Y$ and $\mathcal W$.
This direction, $\cos\theta_{\rm eff}Y-\sin\theta_{\rm eff}\mathcal W$, defines the effective mixing angle $\theta_{\rm eff}(T)$.
For convenience in the perturbative calculation of the evolution of $\theta_{\rm eff}(T)$, we introduce ${\rm SU}(2)_{\rm L}\times{\rm U}(1)_Y$-invariant magnetic fields
\begin{align}
    B_{Yi} \coloneq \dfrac{1}{2}\epsilon^{ijk}Y_{jk}, \qquad
    B_{\mathcal W i} \coloneq \dfrac{1}{2}\epsilon^{ijk}\mathcal W_{jk}.
    \label{eq:magneticfield_def}
\end{align}
At the one-loop level, one may straightforwardly compute the correlation functions of the magnetic fields of $B_{Yi}$ and $B_{\mathcal W i}$
\begin{align}
    \label{eq:BYBY}
    \left\langle 
        B_{Yi}(\vec p)B_{Yj}(- \vec p)
    \right\rangle'
    &\simeq
    \qty[
        \qty(
            \cos \theta_\text{w}^2 R_{\mathcal{AA}} + \sin 2\theta_\text{w} R_{\mathcal{AZ}} 
            )
        P_{ij}(\hat p)],\\
    \label{eq:BYBW}
    \left\langle 
        B_{Yi}(\vec p)B_{\mathcal W j}(-\vec p)
    \right\rangle'
    &\simeq
    \qty[
        \qty(
            - \frac{\sin 2\theta_\text{w} }{2}R_{\mathcal{AA}} + \cos 2 \theta_\text{w} R_{\mathcal{AZ}} 
            ) 
        P_{ij}(\hat p)
        - \qty(
            \cot \theta_\text{w} S_{\mathcal{AA}} + S_{\mathcal{AZ}} 
            )
        \delta_{ij}],\\
    \label{eq:BWBW}
    \left\langle
        B_{{\mathcal W}i}(\vec p)B_{{\mathcal W}j}(-\vec p)
    \right\rangle'
    &\simeq
    \qty[
        \qty(
            \sin \theta_\text{w}^2 R_{\mathcal{AA}} - \sin 2\theta_\text{w} R_{\mathcal{AZ}} )
        P_{ij}(\hat p)
        - \qty(
            \frac{\cos 2 \theta_\text{w}}{\sin^2 \theta_\text{w}} S_{\mathcal{AA}} + 2 \cot \theta_\text{w} S_{\mathcal{AZ}}
            )
        \delta_{ij}
        ],
\end{align}
where the similarity implies the negligence of $\mathcal O(\vert \vec p\vert)$ terms, and the projection operator to the massless mode is defined as $P_{ij}(\hat p)\coloneq\delta_{ij}-\hat p_i\hat p_j$ with $\hat p_i=p_i/\vert \vec p\vert$.
For notational brevity, we extract the delta function associated with the momentum conservation as $\langle X(\vec p) Y (\vec q) \rangle \eqqcolon (2 \pi)^3 \delta^3 (\vec p + \vec q) \langle X(\vec p) Y (- \vec p) \rangle'$.
The expressions of $R_{\mathcal{AA}},\,R_{\mathcal{AZ}},\,S_{\mathcal{AA}},$ and $S_{\mathcal{AZ}}$ are shown in App.~\ref{appendix:1loop_g-indep}.
One may readily see that the massless modes in proportion to $P_{ij}$ are contained all the correlators.
We would like to define the unconfined magnetic field $B_{{\rm c\mkern-7.5mu/}i}$ so that only $B_{{\rm c\mkern-7.5mu/}i}$ has the divergenceless and hence unconfined mode.
The other confined magnetic field $B_{{\rm c}i}$ is taken so that it is orthogonal to $B_{{\rm c\mkern-7.5mu/}i}$.
Specifically, the correlators are expressed as
\begin{align}
    \label{eq:Bslashedc}
    \left\langle
        B_{{\rm c\mkern-7.5mu/}i}(\vec p)B_{{\rm c\mkern-7.5mu/}j}(- \vec p)
    \right\rangle'
    &=
    Z^{\rm c\mkern-7.5mu/} \qty(
        P_{ij}(\hat p)+S_{{\rm c\mkern-7.5mu/}{\rm c\mkern-7.5mu/}}\delta_{ij}+\mathcal O(\vert \vec p\vert, g_3^4)
        ),\\
    \label{eq:Bmix}
    \left\langle 
        B_{{\rm c\mkern-7.5mu/}i}(\vec p)B_{{\rm c}j}(- \vec p)
    \right\rangle'
    &=
    {Z^{\rm c\mkern-7.5mu/}}^{\frac{1}{2}}
    \qty(
        S_{{\rm c\mkern-7.5mu/}{\rm c}}\delta_{ij}+\mathcal O(\vert \vec p\vert, g_3^4)
        ),\\
    \label{eq:Bc}
    \left\langle
        B_{{\rm c}i}(\vec p)B_{{\rm c}j}(- \vec p)
    \right\rangle'
    &=
        S_{{\rm cc}}\delta_{ij}+\mathcal O(\vert \vec p\vert,g_3^4)
        ,
\end{align}
where the wave function renormalization of the unconfined magnetic field is denoted by $Z^{\rm c\mkern-7.5mu/}$.
The unconfined magnetic field $B_{{\rm c\mkern-7.5mu/}i}$ not only has the divergenceless mode but also involves the term invalidating the Bianchi identity, \textit{i.e.}, the term proportional to $S_{\rm c\mkern-7.5mu/ \rm c\mkern-7.5mu/}$.
This result is consistent with the analysis in the previous Sec.~\ref{sec:implications_magnetic}, where a magnetically charged object exists in the theory but is confined.

Now we are ready to define the effective mixing angle, which is given by the relation between $(B_{{\rm c\mkern-7.5mu/}i}, B_{{\rm c}i})$ and $(B_{Yi}, B_{\mathcal{W}i})$ as follows:
\begin{equation}
    \label{eq:def_(un)confdB}
    B_{Y i} \eqqcolon \cos \theta_\text{eff} B_{{\rm c\mkern-7.5mu/}i} + \sin \theta_\text{eff} B_{{\rm c}i}, \qquad
    B_{\mathcal{W}i} \eqqcolon  - \sin \theta_\text{eff} B_{{\rm c\mkern-7.5mu/}i} + \cos \theta_\text{eff}  B_{{\rm c}i}.
\end{equation}
By means of the definitions of the magnetic fields given in Eqs.~\eqref{eq:Bslashedc}, \eqref{eq:Bmix}, and \eqref{eq:Bc}, the correlation functions of the magnetic fields $B_{Yi}$ and $B_{\mathcal{W}i}$ can be expressed as
\begin{align}
    \left\langle B_{Y i}(\vec p) B_{Y j}(-\vec p)\right\rangle'
        &=
        \qty[
            \cos^2 \theta_\text{eff} Z^{\rm c\mkern-7.5mu/} \qty( P_{ij} (\hat p) + S_{\rm c\mkern-7.5mu/ \rm c\mkern-7.5mu/} \delta_{ij}) + \qty(\sin 2 \theta_\text{eff} {Z^{\rm c\mkern-7.5mu/}}^{\frac{1}{2}} S_{\rm c\mkern-7.5mu/ c}
            + \sin^2 \theta_\text{eff} S_{\rm cc}) \delta_{ij}
        ], \\
    \left\langle B_{Yi} ({\vec p}) B_{\mathcal{W}j} (- {\vec p})
    \right\rangle'
    &=
    \qty[
        - \frac{1}{2}\sin 2 \theta_\text{eff} Z^{\rm c\mkern-7.5mu/} \qty( P_{ij} (\hat p) + S_{\rm c\mkern-7.5mu/ \rm c\mkern-7.5mu/} \delta_{ij}) 
        + \qty( \cos 2 \theta_\text{eff} {Z^{\rm c\mkern-7.5mu/}}^{\frac{1}{2}} S_{\rm c\mkern-7.5mu/ c}
        + \frac{1}{2} \sin 2 \theta_\text{eff} S_{\rm cc}
        ) \delta_{ij}
    ], \\
    \left\langle B_{\mathcal{W}i} ({\vec p}) B_{\mathcal{W}j} (- {\vec p})
    \right\rangle'
    &=
    \qty[
        \sin^2 \theta_\text{eff} Z^{\rm c\mkern-7.5mu/}\qty( P_{ij} (\hat p) + S_{\rm c\mkern-7.5mu/ \rm c\mkern-7.5mu/} \delta_{ij}) 
        - \qty(
            \sin 2 \theta_\text{eff} {Z^{\rm c\mkern-7.5mu/}}^{\frac{1}{2}} S_{\rm c\mkern-7.5mu/ c} - \cos^2 \theta_\text{eff} S_{\rm cc}
        ) \delta_{ij}
    ].
\end{align}
These results are compared to Eqs.~\eqref{eq:BYBY}, \eqref{eq:BYBW}, and \eqref{eq:BWBW}, which implies the following relations
\begin{align}
    \label{eq:Zcs}
    Z^{\rm c\mkern-7.5mu/} &= R_{\mathcal{AA}}, \quad
    S_{\rm c\mkern-7.5mu/\rm c\mkern-7.5mu/} = \frac{S_{\mathcal{AA}}}{R_{\mathcal{AA}}} \simeq S_{\mathcal{AA}}, \quad
    S_{c\mkern-7.5mu/\rm c} = \frac{S_{\mathcal{AZ}}}{R_{\mathcal{AA}}^{1/2}} \simeq S_{\mathcal{AZ}}, \quad
    S_{\rm cc} = S_{\mathcal{ZZ}},\\
    \cos^2 \theta_\text{eff} &= \cos^2 \theta_\text{w} + \sin 2 \theta_\text{w} \frac{R_{\mathcal{AZ}}}{R_{\mathcal{AA}}}
    \simeq  \cos^2 \theta_\text{w} + \sin 2 \theta_\text{w} R_{\mathcal{AZ}}, \label{eq:ceff}\\
    \sin^2 \theta_\text{eff} &= \sin^2 \theta_\text{w} - \sin 2 \theta_\text{w} \frac{R_{\mathcal{AZ}}}{R_{\mathcal{{AA}}}} \simeq \sin^2 \theta_\text{w} - \sin 2 \theta_\text{w} R_{\mathcal{AZ}}, \label{eq:seff} \\
    \sin 2 \theta_\text{eff} &= \sin 2\theta_\text{w} - 2 \cos 2 \theta_\text{w} \frac{R_{\mathcal {AZ}}}{R_{\mathcal{AA}}}
    \simeq \sin 2\theta_\text{w} - 2 \cos 2 \theta_\text{w} R_{\mathcal{AZ}}, \label{eq:s2eff} \\
    0 &= \cos^2 \theta_\text{eff} Z^{\rm c\mkern-7.5mu/} S_{\rm c\mkern-7.5mu/\rm c\mkern-7.5mu/} + \sin 2 \theta_\text{eff} {Z^{\rm c\mkern-7.5mu/}}^{\frac{1}{2}} S_{\rm c\mkern-7.5mu/ c}
    + \sin^2 \theta_\text{eff} S_{\rm cc} \nonumber \\
    &\simeq \cos^2 \theta_\text{w} S_{\rm c\mkern-7.5mu/\rm c\mkern-7.5mu/} + \sin 2 \theta_\text{w} S_{\rm c\mkern-7.5mu/\rm c} + \sin^2 \theta_\text{w} S_{\rm cc}.
    \label{eq:Zcs2}
\end{align}
As our formulae are valid up to the one-loop level, we neglect the $\mathcal O(g_3^4)$ terms in the similarities of the above equations.
Note here that $R_{\mathcal{AA}}\sim 1+\mathcal O(g_3^2)$ and $R_{\mathcal{AZ}}\sim\mathcal O(g_3^2)$, $S_{\mathcal{AA}}\sim S_{\mathcal{AZ}} \sim S_{\mathcal{ZZ}}\sim \mathcal O(g_3^2)$.
The last two equations can be regarded as consistency conditions for the effective mixing angle $\theta_{\rm eff}$ up to the one-loop level.
Indeed, by multiplying Eqs.~\eqref{eq:ceff} and \eqref{eq:seff}, one may readily show $\sin \theta_\text{eff} \cos \theta_\text{eff} \simeq (\cos \theta_\text{w} + \sin \theta_\text{w} R_{\mathcal{AZ}}) (\sin \theta_\text{w} - \cos \theta_\text{w} R_{\mathcal{AZ}}) \simeq \sin \theta_\text{w} \cos \theta_\text{w} - R_{\mathcal{AZ}} \cos 2 \theta_\text{eff}$, which is consistent with Eq.~\eqref{eq:s2eff}, up to $\mathcal{O}(g_3^2)$.
Similarly, by using Eq.~\eqref{eq:Zcs} [supplemented by \eqref{eq:ceff}, \eqref{eq:seff} and \eqref{eq:s2eff}], one may confirm that Eq.~\eqref{eq:Zcs2} is satisfied up to $\mathcal{O}(g_3^2)$, owing to the non-trivial relation among $S_{\mathcal{AA}}$, $S_{\mathcal{AZ}}$, and $S_{\mathcal{ZZ}}$, originated from the magnetic symmetry of ${\mathrm U}(1)_Y$.

To sum up, the effective mixing angle $\cos \theta_\text{eff}$, which characterizes a massless mode associated with the SSB of $\mathrm{U}(1)_{\mathrm M}^{[0]}$ in the operator basis of $(B_{Yi}, B_{\mathcal{W}i})$, is given by a function of $T$ as follows:
\begin{align}
    \cos\theta_{\rm eff}(T)
        &=\cos\theta_{\rm w} + R_{\mathcal{AZ}}\sin\theta_{\rm w}+\mathcal O(g_3^4)\notag\\
        &=\qty(1+\dfrac{g_3^2\sin^2\theta_{\rm w}}{4\pi m_W})\cos\theta_{\rm w}+\mathcal O(g_3^4),
    \label{eq:coseff}
\end{align}
where the $\mathcal O(g_3^2)$ correction is solely determined by $R_{\mathcal{AZ}}$, while the $\mathcal O(g_3^2)$ correction in $R_{\mathcal{AA}}$ determines the wave function renormalization of the unconfined magnetic field.
As discussed in Sec.~\ref{sec:comparison}, this interpretation is in a clear contrast with the ``effective mixing'' in the literature.

\begin{figure}[htpb]\centering
    \includegraphics[keepaspectratio, width=.8\textwidth]{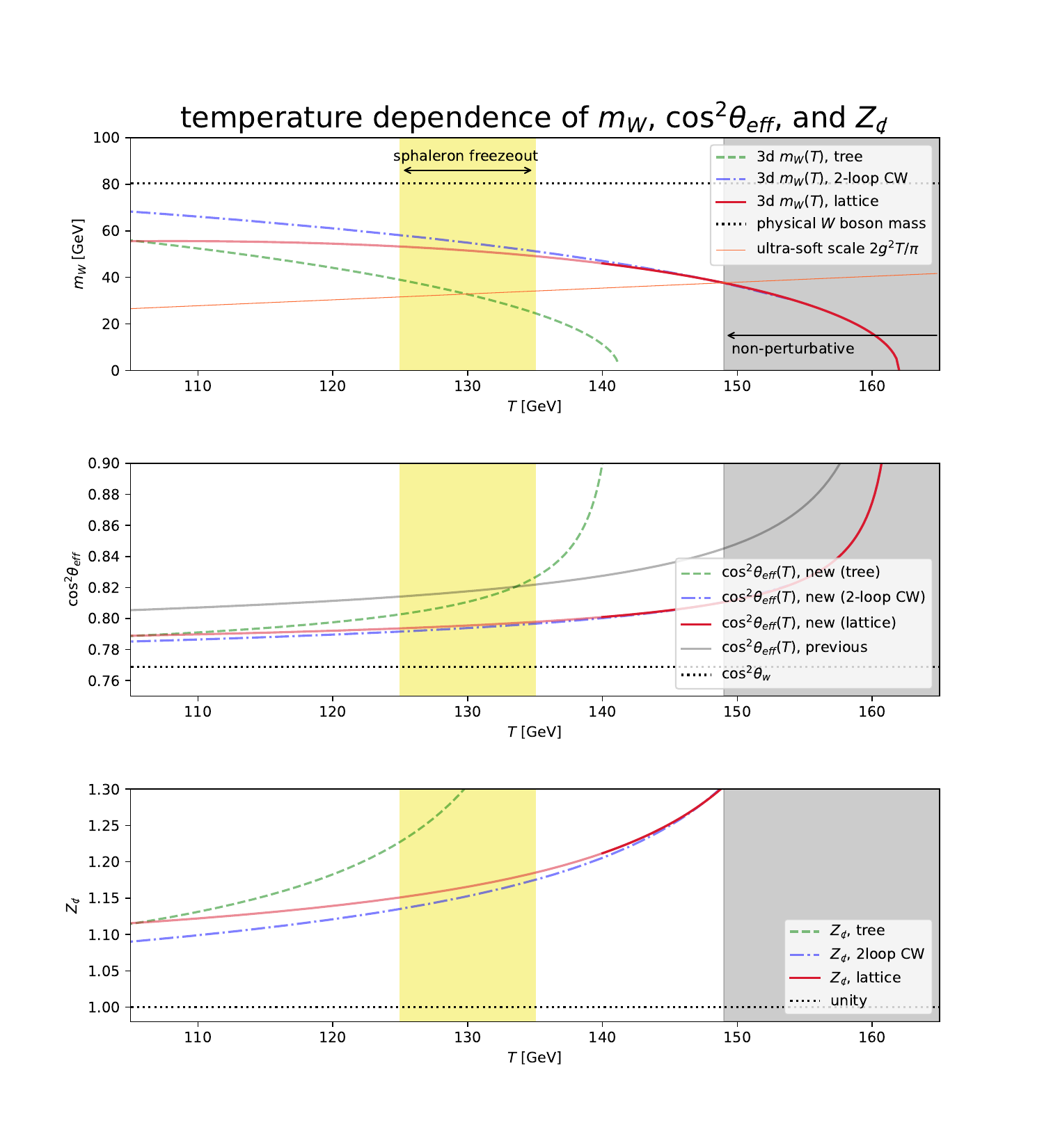}
    \caption{Temperature dependence of the 3d $W$ boson mass ({\it Top}), the effective weak mixing angle ({\it Middle}), which are to be compared with the previous definition (Light-gray line), and the wavefunction renormalization of the unconfined gauge field ({\it Bottom}). Gray-shaded temperature range corresponds to a non-perturbative regime, where the 3d $W$ boson mass is below the ultra-soft scale (pink line in {\it Top}). Yellow-shaded temperature range roughly indicates the sphaleron freezeout temperature $\sim 130\,{\rm GeV}$ \cite{DOnofrio:2014rug}. In each panel, green-dashed lines employ the tree-level relation between $m_W$ and $m_3^2$, where $m_3^2$ is given by Eq.~\eqref{eq:Higgs_mass}. Blue-dashdot lines are drawn using the two-loop Coleman--Weinberg method \cite{Kajantie:1995dw}, where we choose renormalization scales to reproduce the corresponding line drawn in Fig.~3 in Ref.~\cite{DOnofrio:2015gop}. Red-solid lines employ the lattice result \cite{DOnofrio:2015gop} (Dark part of the line) and its extrapolating formula \cite{Kamada:2016cnb} (Light part of the line). Black-dotted line in each panel is the physical $W$ boson mass, the weak mixing angle at $T=0$, and unity, respectively, for reference.}
    \label{fig:theta_eff}
\end{figure}

In Fig.~\ref{fig:theta_eff}, we plot $\cos\theta^2_{\rm eff}(T)$ as a function of the temperature $T$ ({\it Middle} panel).
Following the spirit in the literature, we regard the Higgs VEV, $v_3(T)$, as an input to determine the temperature dependence of the $W$ boson mass $m_W$ (see the last equation in Eq.~\eqref{eq:masses_def} for the relation between $v_3$ and $W_3(v_3)$).
As the input, we employ the tree-level relation between $m_3$ and $v_3$ (the first equation in Eq.~\eqref{eq:masses_def}), the two-loop Coleman--Weinberg-type computation \cite{Kajantie:1995dw}, and an empirical formula \cite{Kamada:2016cnb}, which refers to the lattice result \cite{DOnofrio:2015gop}.
Note that our perturbative formula, Eq.~\eqref{eq:coseff}, is invalid when the $W$ boson is lighter than the ultra-soft scale $\sim g^2T$ (specifically, we show $2g^2T/\pi$ in the figure).

We comment on the reason why we need to go beyond Eq.~\eqref{eq:masses_def} when we substitute $m_W(T)$ in Eq.~\eqref{eq:coseff}.
In Sec.~\ref{sec:baryon}, we will discuss the co-evolution of the cosmological magnetic field with the baryon number during the electroweak crossover.
Since the electroweak sphaleron plays a crucial role there, the fate of them much below the electroweak scale is determined mostly around the sphaleron freezeout epoch at $T_{\rm fo}\sim 130\,{\rm GeV}$ \cite{DOnofrio:2014rug}.
However, the tree-level $m_W(T)$ stays below the ultra-soft scale at the sphaleron freezeout epoch, and we would not be sure if we may trust our perturbative results unless we confirm that $m_W(T)$ is well within the perturbative regime, consulting the Coleman--Weinberg method and lattice results.
Indeed, {\it Top} panel of Fig.~\ref{fig:theta_eff} clearly shows that $m_W(T)$ is safely larger than the ultra-soft scale around the sphaleron freezeout epoch, and hence we believe that Eq.~\eqref{eq:coseff} is a reasonable formula to describe the effective mixing in the relevant epochs.

\subsection{Comparison to existence literatures}
\label{sec:comparison}

The ``effective'' weak mixing angle during the electroweak crossover has already been discussed in the literature \cite{DOnofrio:2015gop,Kamada:2016cnb}. 
In this section, we clarify a subtlety in their definition and argue that it is appropriate to interpret our $\theta_{\rm eff}(T)$ as the effective weak mixing angle.

The lattice studies \cite{Kajantie:1996qd, DOnofrio:2015gop} have investigated the long-range behavior of the correlation function of the hyper-magnetic field, $B_{Yi}\coloneq\epsilon^{ijk}\partial_jY_k$ with a given transverse momentum $\vec p_\perp=(p_1,p_2)$,
\begin{align}
    G(w)
        \coloneq\left\langle B_{Y3}(\vec p_\perp,z)B_{Y3}(-\vec p_\perp,z+w)\right\rangle,\qquad
    B_{Y3}(\vec p_\perp,z)
        \coloneq\int\!\dfrac{\dd p_3}{2\pi}\,e^{-ip_3z}B_{Y3}(\vec p_\perp,p_3).
\end{align}
In the symmetric phase, $\langle B_{Yi}(\vec p)B_{Yj}(\vec q)\rangle=(2\pi)^3\delta^3(\vec p+\vec q)P_{ij}(\hat p)$ implies
\begin{align}
    G(w)
        =(2\pi)^2\delta^2(0)\dfrac{\vert\vec p_\perp\vert}{2}\exp\qty(-\vert\vec p_\perp\vert\vert w\vert)\qquad
    \text{in the symmetric phase,}
    \label{eq:G_sym}
\end{align}
where $(2\pi)^2\delta^2(0)$ is the two-dimensional spatial volume in the transverse directions.
In the broken phase, on the other hand, $\langle B_i(\vec p)B_j(\vec q)\rangle=(2\pi)^3\delta^3(\vec p+\vec q)P_{ij}(\hat p)$ is the only contribution to the long-range behavior, implying that $G(\vert w\vert\to\infty)$ is Eq.~\eqref{eq:G_sym} multiplied by a factor $\cos^2\theta_{\rm w}$.
Throughout the intermediate regime, the lattice calculations have found that $G(\vert w\vert\to\infty)$ is parametrized as
\begin{align}
    G(w)
        =(2\pi)^2\delta^2(0)\dfrac{A_\gamma(T)\vert\vec p_\perp\vert}{2}\exp\qty(-\vert\vec p_\perp\vert\vert w\vert),
\end{align}
where $A_\gamma$ approaches unity in high temperatures and $\cos\theta^2_{\rm w}$ in low temperatures \cite{Kajantie:1996qd, DOnofrio:2015gop}.

References \cite{DOnofrio:2015gop,Kamada:2016cnb} identified $A_\gamma(T)$ as the cosine squared of the ``effective'' mixing angle.
This interpretation may sound reasonable but has a subtlety.
If we just rotate the bases schematically as
\begin{align}
    \begin{pmatrix}
        Y\\W^3
    \end{pmatrix}
    \xlongrightarrow{\tiny{\theta_{\rm eff}}}
    \begin{pmatrix}
        \rm c\mkern-7.5mu/\\\rm c
    \end{pmatrix}
    \xlongrightarrow{\tiny{\theta_{\rm w}-\theta_{\rm eff}}}
    \begin{pmatrix}
        A\\Z
    \end{pmatrix}\qquad ?,
    \label{eq:rot_only}
\end{align}
we could obtain $A_\gamma(T)=\hspace{-2.5mm}\raisebox{1.3ex}[0ex][1ex]{\tiny{?}}\hspace{1.5mm}\cos^2\theta_{\rm eff}(T)$ just as in the broken phase.
However, there is no guarantee that just the rotation results in properly normalized magnetic fields.
Indeed, the necessity of the renormalization of the wavefunction turns out clear in the perturbative calculation.
Reference \cite{Kajantie:1996qd} carried out\footnote{They integrate out the heavy scale (, one may equivalently take a limit $m_{\rm D}\to\infty$,) and hence do not include the third term within the parenthesis (see the discussion in the last paragraph in Sec.~\ref{sec:3dSM}).} the one-loop calculations of the correlation function of the hyper-magnetic field to obtain
\begin{align}
    A_{\gamma}
        =\begin{cases}
            1-\dfrac{g_3'^2}{48\pi m_3}& \text{at high $T$}\\[1.em]
            \cos^2\theta_{\rm w}\qty(
        1+\dfrac{11g_3^2\sin^2\theta_{\rm w}}{12\pi m_W}-\dfrac{g_3^2\sin^2\theta_{\rm w}}{24\pi\sqrt{m_{\rm D}^2+m_W^2}}
    )&\text{at low $T$}.
        \end{cases}
    \label{eq:Kajantie_1loop}
\end{align}
We reproduce this result in App.~\ref{sec:one-loop}.
The main point here is that we cannot directly promote Eq.~\eqref{eq:Kajantie_1loop} as the effective mixing angle $\theta_{\rm eff}$ for the following reasons.
First, if we take Eq.~\eqref{eq:Kajantie_1loop} as the effective mixing angle, the one-loop $Y$$-$$W^3$ correlation function at high temperatures should not vanish because the $Y$$-$$W^3$ correlation function includes the long-range magnetic flux proportional to $\sin\theta_{\rm eff}\cos\theta_{\rm eff}$, given Eq.~\eqref{eq:rot_only}.
Our one-loops calculations, however, show that the $Y$$-$$W^3$ correlation function vanishes identically at high temperatures.
Second, the $W^3$$-$$W^3$ correlation function contains the long-range magnetic flux proportional to $\sin^2\theta_{\rm eff}$, given Eq.~\eqref{eq:rot_only}.
Hence, if this were the mixing angle, the summation of massless modes in $Y$$-$$Y$ and $W^3$$-$$W^3$ correlation functions should be unity.
However, this is not fulfilled and it rather gives a factor greater than unity, which is nothing but the wavefunction renormalization of the unconfined magnetic field.
For these reasons, we should not interpret the one-loop correction in Eq.~\eqref{eq:Kajantie_1loop} as the mixing angle of the fields, but it also involves a correction of the residue at the massless-pole of the unconfined magnetic field.
Instead of Eq.~\eqref{eq:rot_only}, it should be better to consider
\begin{align}
    \begin{pmatrix}
        Y\\W^3
    \end{pmatrix}
    \xlongrightarrow[{\tiny{Z^{\rm c\mkern-7.5mu/}}}]{\tiny{\theta_{\rm eff}}}
    \begin{pmatrix}
        \rm c\mkern-7.5mu/\\\rm c
    \end{pmatrix}
    \xlongrightarrow[{\tiny{(Z^{\rm c\mkern-7.5mu/})^{-1}}}]{\tiny{\theta_{\rm w}-\theta_{\rm eff}}}
    \begin{pmatrix}
        A\\Z
    \end{pmatrix}\qquad ?,
    \label{eq:rot_norm}
\end{align}
where $Z^{\rm c\mkern-7.5mu/}$ indicates the wavefunction renormalization of the unconfined magnetic field.

At low temperatures, the correction in Eq.~\eqref{eq:Kajantie_1loop} includes both contributions from mixing and a factor to be renormalized.
We calculated the correlation functions of $\vec B_A$ and $\vec B_Z$ in App.~\ref{sec:1loop_broken} to find
\begin{align}
    \left\langle B_{i}(\vec p)B_{j}(-\vec p)\right\rangle'
        &=
        \qty(R_{AA}P_{ij}(\hat p)+S_{AA}\delta_{ij}+\mathcal O(\vert \vec p\vert, g_3^4)),\\
    \left\langle B_{i}(\vec p)B_{Zj}(-\vec p)\right\rangle'
        &=
        \qty(R_{AZ}P_{ij}(\hat p)+S_{AZ}\delta_{ij}+\mathcal O(\vert \vec p\vert, g_3^4)),\\
    \left\langle B_{Zi}(\vec p)B_{Zj}(-\vec p)\right\rangle'
        &=
        \qty(S_{ZZ}\delta_{ij}+\mathcal O(\vert \vec p\vert, g_3^4)),
\end{align}
where $B_{Zi}\coloneq\epsilon^{ijk}\partial_jZ_k$, and the expressions of $R_{AA}$ and $R_{AZ}$ are explicit in App.~\ref{sec:1loop_broken}.
According to Eq.~\eqref{eq:rot_norm}, $R_{AZ}$ determines the rotation angle $\theta_{\rm w}-\theta_{\rm eff}$, while $R_{AA}$ determines the renormalization $Z^{\rm c\mkern-7.5mu/}$, at the $\mathcal O(g_3^2)$ order.
However, $R_{AZ}$ and $R_{AA}$ are gauge-dependent, although the combined result at low temperatures in Eq.~\eqref{eq:Kajantie_1loop} is gauge-independent.
For example, $R_{AZ}$ even vanishes at one-loop in the unitary gauge.
The results in the $R_\xi$-gauge are shown in App.~\ref{sec:1loop_broken}.

\begin{figure}[t]\centering
    \includegraphics[keepaspectratio, width=0.5\textwidth]{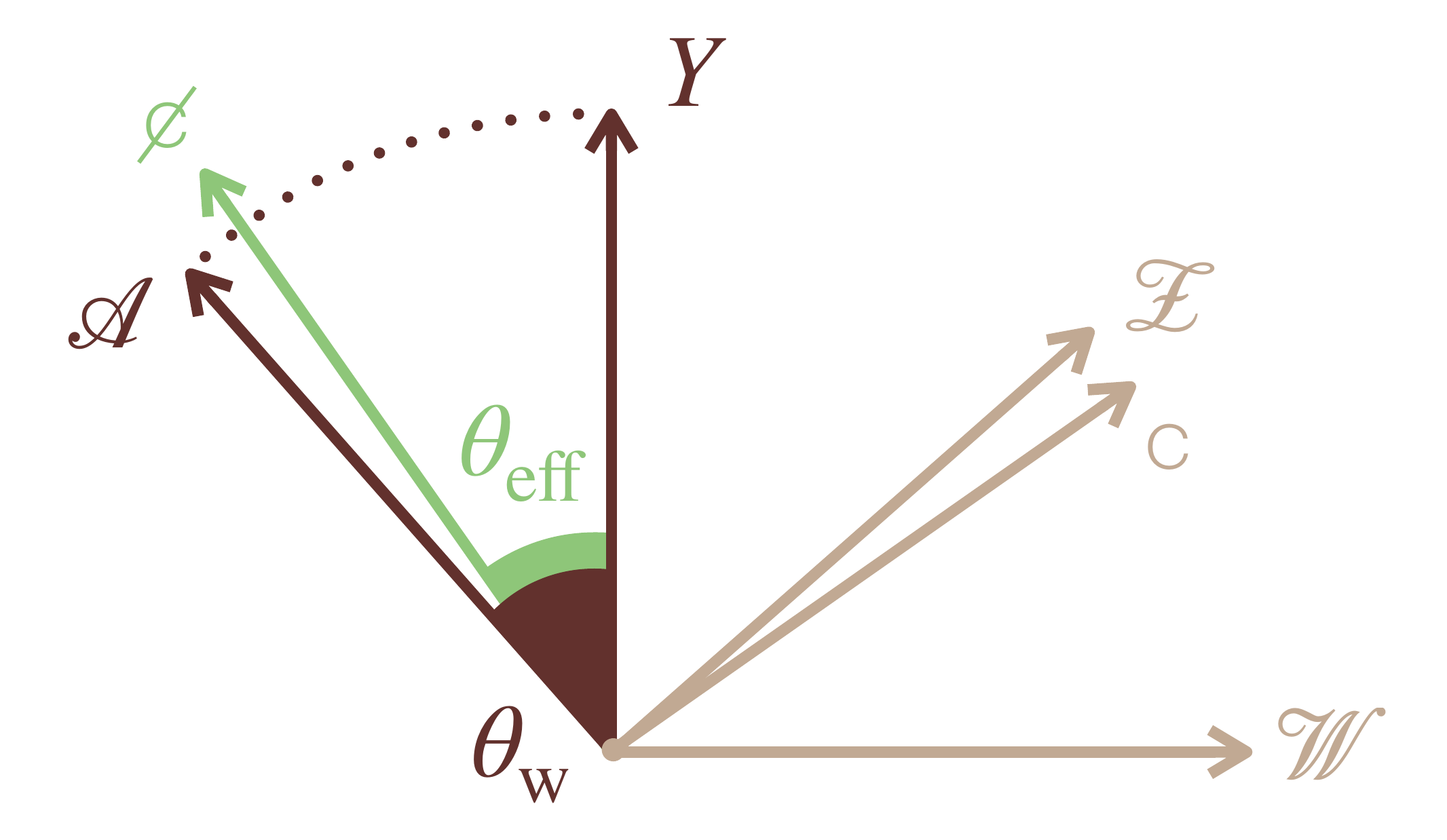}
    \caption{\label{fig:mixing_sche}A schematic visualization of how the unconfined magnetic field ${\rm c\mkern-7.5mu/}$ is realized out of the fundamental fields in the Standard Model. The bases are the ${\rm U}(1)_Y$ field $Y$ and the gauge-independent generalization of the $W^3$ field, denoted as $\mathcal W$. The gauge-independent generalization of the electromagnetic field $\mathcal A$ is the ${\rm U}(1)_Y$ field just tilted in the direction of $-\mathcal W$ by the weak mixing angle $\theta_{\rm w}$. However, the unconfined magnetic field ${\rm c\mkern-7.5mu/}$ is the ${\rm U}(1)_Y$ field not only tilted in the direction of $-\mathcal W$ by the effective mixing angle $\theta_{\rm eff}$ but also renormalized properly.}
\end{figure}

To summarize the two-fold subtlety with the ``effective'' mixing angle in the literature \cite{DOnofrio:2015gop,Kamada:2016cnb},
\begin{itemize}
    \item[i)] $A_\gamma$ includes not only the contribution from mixing but also the renormalization factor $Z^{\rm c\mkern-7.5mu/}$.
    \item[ii)] However, if we try to separate the mixing and the renormalization contributions in the $A-Z$ basis, the mixing contribution looks gauge-dependent.
\end{itemize}
To overcome the gauge-dependence to define the effective mixing angle, we propose to consider a situation [instead of Eqs.~\eqref{eq:rot_only} and \eqref{eq:rot_norm}]
\begin{align}
    \begin{pmatrix}
        Y\\\mathcal W
    \end{pmatrix}
    \xlongrightarrow[\tiny{Z^{\rm c\mkern-7.5mu/}}]{\tiny{\theta_{\rm eff}}}
    \begin{pmatrix}
        \rm c\mkern-7.5mu/\\\rm c
    \end{pmatrix}
    \xlongrightarrow[\tiny{(Z^{\rm c\mkern-7.5mu/})^{-1}}]{\tiny{\theta_{\rm w}-\theta_{\rm eff}}}
    \begin{pmatrix}
        \mathcal A\\\mathcal Z
    \end{pmatrix},
    \label{eq:rot_norm_g-inv}
\end{align}
where $\mathcal W,\,\mathcal A,$ and $\mathcal Z$ are gauge-independent fields, which reduce to $W^3$ in the symmetric phase and $A$ and $Z$ in the broken phase.
The situation is visualized in Fig.~\ref{fig:mixing_sche}.
In Sec.~\ref{sec:unconfined}, we have proposed an explicit construction of $\mathcal W,\,\mathcal A,$ and $\mathcal Z$ and discussed the effective weak mixing angle determined in this way.

\section{Revisiting the baryon asymmetry of the Universe from the helicity decay}
\label{sec:baryon}
Based on the understanding developed in this paper, we revisit the baryon asymmetry of the Universe from the helicity decay of the magnetic field.
We will argue that the generation of the baryon asymmetry could be significantly modified compared to the previous estimates in the literature.

\subsection{Chiral anomaly in the SSB phase of \texorpdfstring{$\mathrm{U}(1)_{\mathrm M}^{[0]}$}{U(1)M}}
\label{sec:baryon_anomaly}

Here we rephrase the baryogenesis by the helicity decay from our perspective.
Let us start with the $B$\,$+$\,$L$ ABJ anomaly equation, whose integrated form over the spacetime volume yields\footnote{
    Here, we assume that the spatial manifold is $\mathbb{R}^3 + \{\infty\} \simeq \mathbb{S}^3_\infty$.
}
\begin{align}
    \Delta Q_{B+L} = 2 \cdot 3\, \qty( \Delta N_{\rm CS}^{\mathrm{SU}(2)_{\mathrm L}} -  \Delta H_{ Y} ),
    \quad
    \Delta H_{Y} \coloneq - \frac{g'^2}{8 \pi^2} \int \dd t \dd^3 x\, \left\langle \vec{E}_{Y} \cdot \vec{B}_{Y} \right\rangle,
    \label{eq:QBL}
\end{align}
where $\Delta Q_{B+L}$ is the change in the $B+L$ charge, $\Delta N_{\rm CS}^{\mathrm{SU}(2)_{\mathrm L}}$ is the change in the $\mathrm{SU}(2)_{\mathrm L}$ Chern--Simons number and $\Delta H_Y$ is the change in the hypercharge helicity.\footnote{Here the normalization of $\Delta H_Y$ is unusual compared with the literature.}
In the usual discussion of the baryon transport, we assume that the change of the hypermagnetic helicity $\Delta H_Y$ is zero since we do not have the hypermagnetic field in the global thermal equilibrium.
At temperatures higher than $\sim 130\,$GeV, the thermal fluctuations allow the $\mathrm{SU}(2)_{\mathrm L}$ Chern--Simons number to change, $\Delta N_{\rm CS}^{\mathrm{SU}(2)_{\mathrm L}} \neq 0$, by overcoming the potential barrier via the sphaleron process.

However, in a more general situation (slightly) away from the global thermal equilibrium, we have to take into account the contribution from the hypermagnetic helicity.
As discussed in the previous sections, the 3d SM has the magnetic ${\mathrm U}(1)_{\mathrm M}^{[0]}$ symmetry, which is spontaneously broken at finite temperatures.
It is natural to expect that there exists a corresponding Nambu--Goldstone mode, \textit{i.e.,} the unconfined magnetic flux $\vec{B}_{\rm c\mkern-7.5mu/}$, which is responsible for the transport phenomena associated with the conservation law.
For instance, the magnetohydrodynamics (MHD) can be understood as the EFT of quantum electrodynamics in the SSB phase of the magnetic $1$-form symmetry~\cite{Grozdanov:2016tdf,Liu:2018kfw}.
In a similar way, we expect that the unconfined magnetic flux $\vec{B}_{\rm c\mkern-7.5mu/}$ can play the role of transport phenomena in the electroweak plasma.

Suppose that the unconfined magnetic field $\vec{B}_{\rm c\mkern-7.5mu/}$ at cosmological scales is generated in the early Universe.
To estimate its effect on the baryon asymmetry, we need to evaluate the r.h.s of Eq.~\eqref{eq:QBL} by taking the expectation value under the corresponding coherent state of the massless mode in $\vec{B}_{\rm c\mkern-7.5mu/}$.
The exact calculation of this expectation value is, however, beyond the scope of this paper.
Instead, we will adopt a ``semi-classical'' approximation, where we replace the unconfined magnetic field $\vec{B}_{\rm c\mkern-7.5mu/}$ with the corresponding classical field configuration and other operators are taken to be zero.
The hypercharge helicity $\Delta H_Y$ can be evaluated in this approximation as follows:
\begin{align}
    \left.\Delta H_Y \right|_{\vec{B}_{\rm c\mkern-7.5mu/}}
    &= \frac{g'^2}{16 \pi^2} \int \dd t\, \frac{\dd}{\dd t}\, \int\dd^3 x\,\left\langle \vec{A}_Y \cdot \vec{B}_Y \right\rangle_{\vec{B}_{\rm c\mkern-7.5mu/}}
    \simeq
    \frac{g'^2}{16 \pi^2} \int \dd t\, \frac{\dd}{\dd t}\, \cos^2 \theta_{\rm eff}\, \int\dd^3 x\, \vec{A}_{\rm c\mkern-7.5mu/} \cdot \vec{B}_{\rm c\mkern-7.5mu/} \\
    & = \int \dd t\, \frac{\dd}{\dd t}\, H_{\rm c\mkern-7.5mu/},
    \label{eq:DeltaH_Y}
\end{align}
where the magnetic helicity of the unconfined magnetic flux is defined as
\begin{equation}
    H_{\rm c\mkern-7.5mu/} \coloneq \frac{g'^2 \cos^2 \theta_{\rm eff}}{16 \pi^2} \int \dd^3 x\, \vec{A}_{\rm c\mkern-7.5mu/} \cdot \vec{B}_{\rm c\mkern-7.5mu/}.
    \label{eq:def_unconfined_helicity}
\end{equation}
The normalization of the magnetic helicity in Eq.~\eqref{eq:def_unconfined_helicity} is motivated by the result of Sec.~\ref{sec:implications_magnetic} based on the magnetic $0$-form symmetry, where the number of the $\mathrm{U}(1)_Y$ magnetic flux is the same as the number of the $\mathrm{U}(1)_{\rm em}$ magnetic flux in the low temperature phase, \textit{i.e.,} $g' \int \dd \vec{S} \cdot \vec{B}_Y \simeq g' \cos \theta_\text{w} \int \dd \vec{S} \cdot \vec{B}_{\rm em}$.
This suggests that the number of the unconfined magnetic flux remains the same during the electroweak crossover, namely $g' \int \dd \vec{S} \cdot \vec{B}_Y \simeq g' \cos \theta_{\rm eff} \int \dd \vec{S} \cdot \vec{B}_{\rm c\mkern-7.5mu/}$.
Therefore, the magnetic helicity in this definition $H_{\rm c\mkern-7.5mu/}$ does not evolve by the change of the effective mixing angle $\theta_{\rm eff}$.

The only effect that can change the hypercharge helicity is the magnetic diffusion, which dissipates the unconfined magnetic flux $\vec{B}_{\rm c\mkern-7.5mu/}$ by the interactions with the thermal plasma.
In the hydrodynamic regime, the unconfined electric flux $\vec{E}_{\rm c\mkern-7.5mu/}$ obeys the constraint equation
\begin{equation}
    \label{eq:generalized_ohms_law}
    \vec{E}_{\rm c\mkern-7.5mu/} =  \frac{1}{\sigma_{\rm c\mkern-7.5mu/}} \vec{\nabla} \times \vec{B}_{\rm c\mkern-7.5mu/} - \frac{1}{\sigma_{\rm c\mkern-7.5mu/}} \vec{J}_\text{CME}
    - \vec{v} \times \vec{B}_{\rm c\mkern-7.5mu/} ,
\end{equation}
where the electric conductivity is given by $\sigma_{\rm c\mkern-7.5mu/}$, the chiral magnetic effect (CME) current is $\vec{J}_\text{CME}$~\cite{Fukushima:2008xe}, and the velocity of the fluid is $\vec{v}$.
Utilizing Eq.~\eqref{eq:generalized_ohms_law}, we can rewrite the hypercharge helicity as
\begin{equation}
    \left.\Delta H_Y \right|_{\vec{B}_{\rm c\mkern-7.5mu/}} \simeq - \int \dd t \dd^3 x\,  \frac{g'^2 \cos^2 \theta_{\rm eff}}{8 \pi^2} \left( \frac{1}{\sigma_{\rm c\mkern-7.5mu/}} \vec{\nabla} \times \vec{B}_{\rm c\mkern-7.5mu/} - \frac{1}{\sigma_{\rm c\mkern-7.5mu/}} \vec{J}_\text{CME} \right) \cdot \vec{B}_{\rm c\mkern-7.5mu/}.
\end{equation}

As explained in the end of Sec.~\ref{sec:implications_magnetic}, the conservation of the number of $\vec{B}_{\rm c\mkern-7.5mu/}$ also implies the absence of the confined magnetic flux, $\vec{B}_{\rm c} = 0$, during the electroweak crossover because of the energy cost.
Neglecting the non-perterbative quantum contributions from the creation of Nambu monopoles (that will be discussed in the next Sec.~\ref{sec:non-perturbative}), we can still realize this absence classically by developing the $\vec{B}_\mathcal{W}$ condensate as $\vec{B}_\mathcal{W} = - \sin \theta_{\rm eff} \vec{B}_{\rm c\mkern-7.5mu/}$ [see also Eq.~\eqref{eq:def_(un)confdB} and discussion on the $\mathrm{SU}(2)_{\mathrm L}$ gauge dressing in the next Sec.~\ref{sec:non-perturbative}].
For this reason, we expect the time evolution of the $\mathrm{SU}(2)_{\mathrm L}$ Chern--Simons number $N_{\rm CS}^{\mathrm{SU}(2)_{\mathrm L}}$, contrary to $H_Y$, which is evaluated as
\begin{align}
    \left. \Delta N_{\rm CS}^{\mathrm{SU}(2)_{\mathrm L}} \right|_{\vec{B}_{\rm c\mkern-7.5mu/}} &= - \frac{g^2}{8 \pi^2} \int \dd t \dd^3 x\, \left\langle \vec{E}_{W}^a \cdot \vec{B}_{W}^a \right\rangle
    \simeq - \frac{g^2}{8 \pi^2} \int \dd t \dd^3 x\, \left\langle \vec{E}_{\mathcal{W}} \right\rangle \cdot \left\langle \vec{B}_{\mathcal{W}}  \right\rangle
    \\
    & \simeq \frac{g^2}{16 \pi^2} \int \dd t \frac{\dd}{\dd t}\,  \sin^2 \theta_{\rm eff}\, \int \dd^3 x\, \vec{A}_{\rm c\mkern-7.5mu/} \cdot \vec{B}_{\rm c\mkern-7.5mu/}
    \simeq \cot^2 \theta_\text{w} \int \dd t\, \frac{\dd }{\dd t} \tan^2 \theta_\text{eff}\, H_{\rm c\mkern-7.5mu/} \\
    & \simeq
    - \int \dd t \dd^3 x\,  \frac{g^2 \sin^2 \theta_{\rm eff}}{8 \pi^2} \left( \frac{1}{\sigma_{\rm c\mkern-7.5mu/}} \vec{\nabla} \times \vec{B}_{\rm c\mkern-7.5mu/} - \frac{1}{\sigma_{\rm c\mkern-7.5mu/}} \vec{J}_\text{CME} \right) \cdot \vec{B}_{\rm c\mkern-7.5mu/}
    + \cot^2 \theta_\text{w} \int \dd t\, \frac{\dd \tan^2 \theta_\text{eff}}{\dd t}\, H_{\rm c\mkern-7.5mu/},
\end{align}
In the second similarity, we have inserted the identity $\delta^{ab} = n^a n^b + P^{ab}_{\mathrm c}$ with $(P_{\mathrm c} n)^a = 0$ and $(P_{\mathrm c}^2)^{ab} = P_{\mathrm c}^{ab}$, and assumed that only the projection onto $n^a n^b$ involves the unconfined magnetic flux and $P^{ab}_{\mathrm c}$ does not contribute to the long-range correlation.

To sum up, $\vec{B}_{\rm c\mkern-7.5mu/}$ at cosmological scales contributes to the r.h.s of Eq.~\eqref{eq:QBL} as
\begin{align}
    \label{eq:B+L}
    \Delta Q_{B+L} &= 3 \cdot 2\,  \qty(\Delta N_{\rm CS}^{\mathrm{SU}(2)_{\mathrm L}}  - \Delta H_Y )_{\vec{B}_{\rm c\mkern-7.5mu/}} + 3 \cdot 2\, \Delta N_{\rm CS}^{\mathrm{SU}(2)_{\mathrm L}} \Big|_\text{NP} \\
    & \simeq
    3 \cdot 2\,  \cot^2 \theta_\text{w}\, \Delta \qty[ \qty( \tan^2 \theta_\text{eff} - \tan^2 \theta_\text{w} )  H_{\rm c\mkern-7.5mu/}]
    + 3 \cdot 2\, \Delta N_{\rm CS}^{\mathrm{SU}(2)_{\mathrm L}} \Big|_\text{NP},
\end{align}
where $\Delta N_{\rm CS}^{\mathrm{SU}(2)_{\mathrm L}} \big|_\text{NP}$ involves non-perturbative contributions such as the sphaleron process.
The first term in the r.h.s acts as a source term for the baryon asymmetry, while the second term is responsible for the washout of the baryon asymmetry.
The first term has two contributions: one from the decay of the unconfined magnetic flux via the magnetic diffusion process at small scales $\Delta H_{\rm c\mkern-7.5mu/}$, and the other from the change of the effective mixing angle $\Delta \theta_\text{eff}$.
The first effect has been discussed in the literature \cite{Fujita:2016igl,Kamada:2016eeb}, and the second effect has been later discussed in \cite{Kamada:2016cnb}.
Our new understanding of the effective mixing angle, at least, yields an order one change of the resultant baryon asymmetry because our gauge independent effective mixing angle $\theta_\text{eff}$ is different from that in the literature (see Eq.~\eqref{eq:coseff} and Fig.~\ref{fig:theta_eff}).

Note here that, although our estimation is based on a similar spirit to the previous literature \cite{Fujita:2016igl,Kamada:2016eeb,Kamada:2016cnb}, there are crucial differences not only in the definition of the effective mixing angle but also in the treatment of the magnetic helicity.
In the previous literature \cite{Kamada:2016cnb}, it is assumed that the amplitude of the unconfined magnetic flux $\vec{B}_{\rm c\mkern-7.5mu/}$ remains the same during the electroweak crossover except for the magnetic diffusion, which means $B_Y \simeq B_{\rm c\mkern-7.5mu/} \simeq B_\text{em}$ contrary to our assumption $g' B_Y \simeq g' \cos \theta_\text{eff}B_{\rm c\mkern-7.5mu/} \simeq e B_\text{em}$.\footnote{The difference arises because the conservation of the energy density of the unconfined magnetic field is assumed in Ref.~\cite{Kamada:2016cnb}.
Here we do not make this assumption since the energy density of the universe is dominated by the other components. There is no particular reason to impose the energy conservation only in the magnetic field sector.
Instead, our ansatz respects the conservation of the number of the unconfined magnetic flux.
}
This indicates that the approximately conserved magnetic helicity is $H_{\rm c\mkern-7.5mu/}^\text{KL} = \int_{\bm x}\vec{A}_{\rm c\mkern-7.5mu/} \cdot \vec{B}_{\rm c\mkern-7.5mu/}$, which gives a slightly different result from ours, \textit{i.e.,}
\begin{equation}
    \label{eq:KL}
    \Delta Q_{B+L} =
    - \frac{3}{16 \pi^2} \qty(g^2 + g'^2)\, \Delta \qty[ \qty( \cos 2 \theta_\text{eff} - \cos 2 \theta_\text{w} ) H_{\rm c\mkern-7.5mu/}^\text{KL}]
    + 3 \cdot 2\, \Delta N_{\rm CS}^{\mathrm{SU}(2)_{\mathrm L}} \Big|_\text{NP},
\end{equation}
See also discussion at the end of Sec.~\ref{sec:non-perturbative}.

This is not the end of the story.
In the previous literature, it has been implicitly assumed that the local change of the $\mathrm{SU}(2)_{\mathrm L}$ Chern--Simons number is only due to the sphaleron process.
For example, after the sphaleron decoupling, $\Delta N_{\rm CS}^{\mathrm{SU}(2)_{\mathrm L}} \big|_{\text{NP}} = 0$ is assumed, where $N_{\rm CS}^{\mathrm{SU}(2)_{\mathrm L}}$ can be evaluated on the trivial topological sector.\footnote{
    By this, we mean that we choose the gauge field configuration corresponding to the trivial element of $\pi_3(\mathrm{SU}(2)_{\mathrm L})=\mathbb{Z}$.
}
Hence, all the change of the first term in the r.h.s. is converted to the baryon asymmetry after the sphaleron decoupling.
If this is the case, we can compute the $B+L$ charge today by equilibrating the source rate $(\dd/\dd t)(\tan^2 \theta_\text{eff} - \tan^2 \theta_\text{w}) H_{\rm c\mkern-7.5mu/}$ and the sphaleron washing-out rate
at the time of the sphaleron decoupling~\cite{Kamada:2016cnb}.
However, it is not clear whether we can simply assume that all the magnetic helicity decays into the baryon asymmetry, since an unknown $\mathrm{SU}(2)_{\mathrm L}$ non-perturbative effect could induce the transition between the topological sectors.

In the next section, we will provide the proof of the existence for the novel non-perturbative effect other than the sphaleron, which yields $\Delta N_{\rm CS}^{\mathrm{SU}(2)_{\mathrm L}} \big|_{\text{NP}} \neq 0$ even after the sphaleron decoupling and thereby could change the calculation of the baryon asymmetry.

\subsection{Novel non-perturbative effect in the presence of magnetic flux}
\label{sec:non-perturbative}

The goal of this section is to demonstrate the existence of the novel non-perturbative effect, and to point out the importance of keeping track of the final fate of the magnetic helicity for understanding the baryon asymmetry.
In particular, we will show that the change of the effective mixing angle $\Delta \theta_\text{eff}$ can be accompanied by the novel non-perturbative effect that yields $\Delta N_{\rm CS}^{\mathrm{SU}(2)_{\mathrm L}} \big|_{\text{NP}} \neq 0$.
To this end, we consider the magnetic field profile at the time of the sphaleron decoupling, and discuss the change of the effective mixing angle.

The massless magnetic field $\vec{B}_{\rm c\mkern-7.5mu/}$ computed in Sec.~\ref{sec:magnetic} forms a profile as schematically shown in a left panel of Fig.~\ref{fig:crossover_magnetic_fields}.
As the crossover continues, the massless direction varies according to Eq.~\eqref{eq:def_(un)confdB}.
Consequently, the original profile would become the mixture of the massless and massive modes, as shown in the right panel of Fig.~\ref{fig:crossover_magnetic_fields}.
Suppose that the magnetic helicity carried by the unconfined magnetic flux decreases by $\Delta( N_{\rm CS}^{\mathrm{SU}(2)_{\mathrm L}} - H_Y )_{\vec{B}{\rm c\mkern-7.5mu/}}$ from the left to the right panel of Fig.~\ref{fig:crossover_magnetic_fields}.
At this moment, the magnetic helicity $\Delta( N_{\rm CS}^{\mathrm{SU}(2)_{\mathrm L}} - H_Y )_{\vec{B}{\rm c\mkern-7.5mu/}}$ is carried mainly by the network of $B_{\rm c}$ loops.
The question is how the decay of the $B_{\rm c}$ network contributes to the baryon asymmetry.

\begin{figure}[ht]
    \centering
    \begin{tikzpicture}[scale=1.2, line width=1pt,
        ->/.style={-{Latex[length=6pt, width=4pt]}},
        circ/.style={draw, thick},
        redc/.style={draw=red, thick},
        bluec/.style={draw=blue, thick},
        arrow inside/.style={postaction={decorate},decoration={markings,mark=at position #1 with {\arrow{>}}}},
    ]

    \def\r{1}
    \def\shift{1.2}
    \def\right{4.8}

    \draw[sagegreen, arrow inside=0.5] (180:\r) arc[start angle=180, end angle=540, radius=\r]
    node[above, midway, xshift=-13pt, yshift=5.5pt] {\textcolor{sagegreen}{$B_{\rm c\mkern-7.5mu/}$}};

    \draw[->] (\shift + 1.2, 0) -- +(1.2, 0);
    \draw[->] (\shift + 1.2, 0) -- +(1.2, 0);

    \begin{scope}[shift={(\right,0)}]
        \draw[sagegreen, arrow inside=0.5] (180:\r) arc[start angle=180, end angle=540, radius=\r]
        node[above, midway, xshift=-13pt, yshift=5.5pt] {\textcolor{sagegreen}{$B_{\rm c\mkern-7.5mu/}$}};
        \draw[sepia, arrow inside=0] (0.5,0) circle(\r)
        node[above, midway, yshift=6pt, scale=0.8] {\textcolor{sepia}{$B_{\rm c}$}};
    \end{scope}

    \end{tikzpicture}
    \caption{A model of the time evolution of the magnetic field at the electroweak crossover.
    Left: the magnetic field consists of the deconfined one $B_{\rm c\mkern-7.5mu/}$.
    Right: as the crossover continues, the magnetic field becomes the mixture of deconfined $B_{\rm c\mkern-7.5mu/}$ and the confined magnetic field $B_{\rm c}$.}
    \label{fig:crossover_magnetic_fields}
\end{figure}

Now we argue that the $B_{\rm c}$ network does not necessarily decay into baryon asymmetry.
To illustrate this idea, we consider the following toy model: the $\mathrm{SU}(2)$ gauge group, the $\mathrm{SU}(2)$ adjoint and fundamental Higgs fields, and two fundamental Weyl fermions. 
We consider gauge symmetry breaking in two steps.
First, the $\mathrm{SU}(2)$ gauge symmetry is broken into $\mathrm{U}(1)$ by the VEV of the adjoint Higgs field.
We further break the $\mathrm{U}(1)$ gauge symmetry by turning on the VEV of the fundamental Higgs field.
The first step leads to the 't Hooft--Polyakov monopole~\cite{tHooft:1974kcl,Polyakov:1974ek}, and the second step leads to the $\mathrm{U}(1)$ gauge string on which the $\mathrm{U}(1)$ magnetic flux is confined.
In this model, the integrated anomaly equation is written as
\begin{align}
    \Delta Q_{\mathrm{chi}}= 2 \Delta N_\text{CS},
    \label{eq:SU(2)_conservation}
\end{align}
where $\Delta Q_{\mathrm{chi}}$ is the change of chiral charge, and $\Delta N_\text{CS}$ is the change of the $\mathrm{SU}(2)$ Chern--Simons number.

Let us consider the profile of the $\mathrm{U}(1)$ gauge string as shown in the left panel of Fig.~\ref{fig:Hopf_link}. 
The Gauss' linking number of the magnetic flux lines is $1$ corresponding to $N_\text{CS}=2$ (see App.~\ref{sec:linking} for the computation).\footnote{
    The magnetic flux is quantized in a unit of $(1/2) \times g B / (2 \pi)$, where $g$ is the $\mathrm{SU}(2)$ gauge coupling and $1/2$ is the unit charge of fundamental fermions.
    The quantized helicity in this case is $N_{\rm M} = (1/2)^2 \times g^2 \int \dd^3 x \,\vec{A} \cdot \vec{B} / (4 \pi^2)$, which implies $\Delta N_{\rm M} =  g^2\int \dd^4 x \,F_{\mu\nu} \tilde F^{\mu\nu}/(32 \pi^2)$. Hence, after the SSB of $\mathrm{SU}(2) \to \mathrm{U}(1)$, we have $N_\text{CS}\big|_{\mathrm{SU}(2) \to \mathrm{U}(1)} = N_{\rm M}$.
    When the Gauss' linking number of the magnetic flux line is $1$, we have $N_{\rm M} = 2$, and hence $N_\text{CS} = 2$.
    Here we fix the large gauge transformation so that $N_{\rm M} = 0$ corresponds to the trivial topological sector.
    See also App.~\ref{sec:linking}.
}
We study the possible way to untie the link to the profile with the Gauss' linking number zero shown in the right panel of Fig.~\ref{fig:Hopf_link}.
There are two ways to untie the link.
The first option is to continuously deform the red loop so that the final configuration is unlinked (middle top panel in Fig.~\ref{fig:Hopf_link}).
As shown as a black dot in the top-middle panel, this process should include a moment when two loops intersect at a point.
The chiral charge is generated at the intersection in accordance with the conservation law.

The second option is to split the string into the monopole and the antimonopole (middle bottom panel in Fig.~\ref{fig:Hopf_link}).
This is possible because the fundamental group of the vacuum manifold is trivial once $\mathrm{U}(1)$ is embedded in $\mathrm{SU}(2)$.
In other words, the magnetic monopole can be at the end of the flux tube.
By moving the monopole pairs and combining them, we can untie the link without having the intersection.
In this process, the chiral charge may not be generated (for instance, we can do this process adiabatically).

\begin{figure}[ht]
    \centering
    \begin{tikzpicture}[scale=1.2, line width=1pt,
        ->/.style={-{Latex[length=6pt, width=4pt]}},
        circ/.style={draw, thick},
        redc/.style={draw=red, thick},
        bluec/.style={draw=blue, thick},
        arrow inside/.style={postaction={decorate},decoration={markings,mark=at position #1 with {\arrow{>}}}},
    ]

    \def\r{1}
    \def\shift{1.2}
    \def\center{4.8}
    \def\right{10.2}

    \begin{scope}[shift={(0,0)}]
    \begin{scope}[shift={(\shift,0)}]
        \draw[bluec] (0:\r) arc[start angle=0, end angle=210, radius=\r];
        \draw[bluec, arrow inside=0.5] (210:\r) arc[start angle=210, end angle=230, radius=\r];
        \draw[bluec] (250:\r) arc[start angle=250, end angle=360, radius=\r];
    \end{scope}
    \draw[redc, arrow inside=0.5] (180:\r) arc[start angle=180, end angle=90, radius=\r];
    \draw[redc] (90:\r) arc[start angle=90, end angle=60, radius=\r];
    \draw[redc] (45:\r) arc[start angle=45, end angle=-180, radius=\r];
    \end{scope}

    \draw[->] (\shift + 1.2, 0.3) -- +(1.2, 0.5);
    \draw[->] (\shift + 1.2, -0.3) -- +(1.2, -0.5);
    \draw[->] (\center + 3.3, 0.7) -- +(1.2, -0.3);
    \draw[->] (\center + 3.3, -0.7) -- +(1.2, 0.3);

    \begin{scope}[shift={(\center,1.3)}]
    \draw[redc, arrow inside=0.5] (0:\r) arc[start angle=0, end angle=-360, radius=\r];
    \draw[bluec, arrow inside=0] (2*\r,0) circle(\r);
    \fill[black] (\r,0) circle(2.5pt); 
    \end{scope}

    \begin{scope}[shift={(\center+0.5,-1.3)}]
        \begin{scope}[shift={(\shift,0)}]
        \draw[bluec] (0:\r) arc[start angle=0, end angle=210, radius=\r];
        \draw[bluec, arrow inside=0.5] (210:\r) arc[start angle=210, end angle=230, radius=\r];
        \draw[bluec] (250:\r) arc[start angle=250, end angle=360, radius=\r];
        \end{scope}
        \draw[redc, arrow inside=0.5] (180:\r) arc[start angle=180, end angle=90, radius=\r];
        \draw[redc] (90:\r) arc[start angle=90, end angle=60, radius=\r];
        \draw[redc] (45:\r) arc[start angle=45, end angle=15, radius=\r];
        \draw[redc] (-15:\r) arc[start angle=-15, end angle=-180, radius=\r];
        \fill[green!60!black] (15:\r) circle(2pt) node[above right, scale=0.9] {$M$};
        \fill[green!60!black] (-15:\r) circle(2pt) node[below right, scale=0.9] {$\overline{M}$};
    \end{scope}

    \begin{scope}[shift={(\right,0)}]
        \draw[redc, arrow inside=0.5] (0:0.8*\r) arc[start angle=0, end angle=-360, radius=0.8*\r];
        \draw[bluec, arrow inside=0] (1.8,0) circle(0.8*\r);
    \end{scope}

    \end{tikzpicture}
    \caption{Two processes to untie the link. Left: The Hopf link (Gauss' linking number $1$) of the magnetic field, where each loop represents ${\mathrm U}(1)$ gauge string.
    Middle top: When the red loop is continuously moved to untie the link, we encounter the intersection point (black dot), where the chiral fermions are produced.
    Middle bottom: The red loop is split into the monopole $M$ and the anti-monopole $\overline{M}$. By moving the monopole pairs and combining them, we can untie the link without having the intersection.
    Right: The profile of the magnetic field with the Gauss' linking number $0$.}
    \label{fig:Hopf_link}
\end{figure}

How is the latter option compatible with Eq.~\eqref{eq:SU(2)_conservation}?
The key is the $S^1$ moduli of the magnetic monopole.
The magnetic monopole solution is parameterized by the moduli whose value takes at $S^1$ (see \textit{e.g.}, \cite{Harvey:1996ur} for a review).
The $\mathrm{SU}(2)$ Chern--Simons number changes continuously as we vary the value of the moduli.
In this way, the monopole profile can accommodate the $N_\text{CS}$ such that $\Delta N_\text{CS}=0$.
Namely, starting from the Hopf link of $N_\text{CS} = 2$, we may end up with the trivial linking of the magnetic flux while the vacuum state lies in a different topological sector with $N_\text{CS} = 2$, which in total gives $\Delta N_\text{CS} = 0$.\footnote{
    Similar (virtual) processes have been discussed in the context of the axion mass in quantum electrodynamics~\cite{Fan:2021ntg}.
}
We note that, even though we also have an antimonopole, as long as the distance between the monopole pair is large enough compared to the electroweak length, focusing on the $S^1$ moduli of the single monopole should be a good approximation.
This process is analogous to the sphaleron/instanton process.
The non-vanishing chiral charge is absorbed in the superposition of the different values of the Chern--Simons number (\textit{i.e.}, the $\theta$ vacua).
We emphasize that the existence of these processes is naturally expected from the symmetry point of view.
In general, both the chiral charge and the magnetic helicity can be erased by the $\mathrm{SU}(2)$ non-perturbative effect.
The former is done by the sphaleron/instanton process, and the latter is done by the produced monopole pairs.

The above argument provides an explanation of why we should not take the conversion of magnetic helicity to baryon asymmetry for granted.
Having understood the toy model of $\mathrm{SU}(2) \to \mathrm{U}(1)$, we turn to the case of EWSB.
The loops in question are the flux tubes of the confined magnetic field $B_{\rm c}$.
As in the case of the $\mathrm{SU}(2)$ toy model, the fundamental group of the vacuum manifold is trivial, and the decay of the $B_{\rm c}$ flux tube network can be non-trivial.
However, the dynamics of electroweak theory is more complicated than the $\mathrm{SU}(2)$ toy model.
Nevertheless, one may describe the process in a gauge invariant way. As $N_\text{CS}$ is not gauge invariant, we instead consider the gauge invariant quantity $\delta\coloneq N_\text{CS} - H_Y - N_H$, where $N_H$ is the winding number of the Higgs field corresponding to $\pi_3(\mathrm{SU}(2)_{\mathrm L}\times\mathrm U(1)_Y/\mathrm U(1)_\text{em})=\mathbb{Z}$~\cite{Krauss:1999ng}. Notice that $N_H$ is well-defined when there are no zeros of the Higgs field. The state without $B_{\rm c}$ flux is characterized by $\delta = 0$.

To simplify the situation, let us suppose that the unconfined magnetic flux is immediately converted from $\mathrm U(1)_Y$ to the $\mathrm U(1)_\text{em}$, and the Higgs field develops the VEV at this moment.
In this case, all the hypercharge magnetic helicity is first carried by the $B_{\rm c}$ flux right after the transition, which is characterized by $\delta = - H_Y^0$ assuming that $N_H=0$,\footnote{
    It would be interesting to study the initial condition of the Higgs winding number.
    The Z-string may decay into the skyrmion of $CP^1$ model~\cite{Hindmarsh:1991jq,Gibbons:1992gt,Preskill:1992bf,Hindmarsh:1992yy} and $N_H$ becomes the skyrmion number.
    As this is also interpreted as the linking number of the preimage of the map $S^3\to S^2$~\cite{Manton:2004tk}, the Higgs winding number may be non-zero.
} with $H_Y^0$ being the initial hypercharge magnetic helicity.
As this configuration is unstable, its helicity decays from $\delta = - H_Y^0$ to $\delta = 0$.
We can consider two cases: (i) $\Delta  N_H = - H_Y^0$ and (ii) $\Delta ( N_{\rm CS}^{\mathrm{SU}(2)_{\mathrm L}} - H_Y ) = H_Y^0$.
In the first case (i), the baryon asymmetry is not generated while, in the second case (ii), the baryon asymmetry is generated.
See Fig.~\ref{fig:baryon-transition} for an illustration.
In a realistic situation, the transition should be gradual owing to the crossover nature.
Yet, one may model the crossover transition as a collection of many sudden transitions, where the above qualitative understanding may still be valid.
Below, we remark three possibilities.
\begin{itemize}
    \item The $B_{\rm c}$ flux tube decays into the pair of Nambu monopoles, which converts the $B_{\rm c}$ flux to the $B_{\rm c\mkern-7.5mu/}$ flux.
    In this case, the baryon asymmetry is not generated similarly to the case of the $\mathrm{SU}(2)$ toy model.
    Now one may immediately see that this is the example where the change of helicity is compensated by the non-perturbative process, namely the r.h.s. of Eq.~\eqref{eq:B+L} becomes zero.
    \item The $B_{\rm c}$ flux tube is similar to the $Z$-string, which is known to be unstable~\cite{Vachaspati:1991dz,Achucarro:1999it}.
    Consequently, the $B_{\rm c}$ flux tube is not concentrated in the small region, although this corresponds to a confined magnetic field.
    Rather, the $B_{\rm c}$ flux is spread over the large region, and would become the texture/skyrmion~\cite{Preskill:1992bf,Hindmarsh:1992yy}.
    This situation is similar to the cold baryogenesis~\cite{Krauss:1999ng,Garcia-Bellido:1999xos,Konstandin:2011ds} (see also Refs.~\cite{Turok:1990in,Turok:1990zg} for numerical simulations for the texture dynamics).
    If the transition from $\delta = - H_Y^0$ to $\delta = 0$ occurs via the $\mathrm{SU}(2)_{\mathrm L}$ gauge dressing, the baryon asymmetry is generated as $\Delta ( N_{\rm CS}^{\mathrm{SU}(2)_{\mathrm L}} - H_Y ) = H_Y^0$.
    On the other hand, if this occurs via the change of the Higgs winding, the baryon asymmetry is not generated as $\Delta  N_H = - H_Y^0$.
    \item If the $B_{\rm c}$ flux tube is neither cut by the Nambu monopoles, accompanied by the change of the Higgs winding, nor dressed by the $\mathrm{SU}(2)_{\mathrm L}$ gauge field, the $B_{\rm c}$ flux tube just shrinks to a point.
    During the course of this process, the $B_{\rm c}$ flux tubes disentangle their linking and reduce the magnetic helicity carried by them.
    This case may correspond to Eq.~\eqref{eq:KL} studied in Ref.~\cite{Kamada:2016cnb}.
\end{itemize}
We hope to study the details of the above scenarios in the future.

\begin{figure}[htbp]
    \centering
    \begin{tikzpicture}[scale=2, thick]
    
      \draw[->] (0,-1.2) -- (0,1.4) node[above] {$N_{\text{CS}}^{\mathrm{SU}(2)_{\mathrm L}}$};
      \draw[->] (-1.4,0) -- (1.2,0) node[right] {$N_H$};
    
      \draw[red, thick] (-1.1,-1.1) -- (0.9,0.9);
      \node[red] at (1.35,1.05) {{\scriptsize $N_{\text{CS}}^{\mathrm{SU}(2)_{\mathrm L}} - N_H = 0\,(\delta = - H_Y^0)$}};
    
      \draw[blue, dashed] (-1.3,-0.3) -- (0.3,1.3);
      \node[blue] at (1.20,1.4) {{\scriptsize$N_{\text{CS}}^{\mathrm{SU}(2)_{\mathrm L}} - N_H = H_Y^0\,(\delta = 0)$}};
      \filldraw[fill=teal] (0,0) circle (1.2pt);
      \filldraw[fill=teal] (-1,0) circle (1.2pt); 
      \filldraw[fill=teal] (0,1) circle (1.2pt); 
      \node[align=left, anchor=north west, text=teal] at (0,0) {\scriptsize initial\\\scriptsize $(N_H = N_{\text{CS}}^{\mathrm{SU}(2)_{\mathrm L}} = 0,\, H_Y = H_Y^0)$};
    
      \draw[->, thick, teal] (0,0) to[bend left=15] (-0.95,-0.05);
      \node[below, text=teal] at (-0.6,-0.06) {\scriptsize$\Delta Q_{B+L} = 0$};
    
      \draw[->, thick, teal] (0,0) to[bend right=15] (0.05,0.95);
      \node[left, text=teal] at (0.03,0.27) {\scriptsize$\Delta Q_{B+L} \neq 0$};
    
    \end{tikzpicture}
    \caption{A schematic illustration of the transition from the initial state with $B_{\rm c}\neq0$ to the final state with $B_{\rm c}=0$ in the $N_{\text{CS}}^{\mathrm{SU}(2)_{\mathrm L}} - N_H$ plane.
    As an example, the initial state is taken to be $N_{\text{CS}}^{\mathrm{SU}(2)_{\mathrm L}} = N_H = 0$ and $H_Y = H_Y^0$.
    The decay of the $B_{\rm c}$ happens by either (i) $\Delta N_H = - H_Y^0$  or (ii) $\Delta N_{\text{CS}}^{\mathrm{SU}(2)_{\mathrm L}} = H_Y^0$, or mixture of them.
    These correspond to $\Delta Q_{B+L} = 0$ and $\Delta Q_{B+L} \neq 0$, respectively.
    To be more precise, $\Delta H_Y=-H_Y^0$ is also allowed, in which case $\Delta Q_{B+L} \neq 0$ and the helicity of $B_{\rm c\mkern-7.5mu/}$ disappears.
    }
    \label{fig:baryon-transition}
\end{figure}

\section{Summary and discussion}
\label{sec:disc}

In this paper, we revisit the electroweak crossover from the perspective of generalized symmetries.
After reviewing the $3$d EFT of the SM (Sec.~\ref{sec:hotSM}), we identify the generalized symmetries of the $3$d SM (see Tab.~\ref{table:symmetry} in Sec.~\ref{sec:symandanomaly}).
We have confirmed that the order parameters of the generalized symmetries do not exhibit discontinuity at EWSB and therefore are consistent with the crossover picture (see Tab.~\ref{table:phases} in Sec.~\ref{sec:crossover}).
In particular, the magnetic $1$-form symmetry associated with ${\rm U}(1)_Y$ leads to the magnetic $0$-form symmetry after the dimensional reduction, which is always spontaneously broken in the high-/low-temperature phases.
This implies the existence of long-range magnetic fields as the corresponding Nambu--Goldstone boson in both phases.
We provide a physical interpretation of this result by putting a pair of non-dynamical ${\rm U}(1)_Y$ monopoles as a probe, which implies that the ${\rm U}(1)_Y$ magnetic flux is smoothly converted to the ${\rm U}(1)_\text{em}$ magnetic flux by the dressing of dynamical monopoles, \textit{i.e.,} Nambu monopoles (see Fig.~\ref{fig:magnetic_field} in Sec.~\ref{sec:implications_magnetic}).
Motivated by this formal understanding, we identify the long-range magnetic field, and define the effective mixing angle between the ${\rm U}(1)_Y$ and ${\rm SU}(2)_{\mathrm L}$ magnetic fields in a gauge invariant way (see Fig.~\ref{fig:theta_eff} in Sec.~\ref{sec:unconfined}).
Importantly, we have demonstrated that the effective mixing angle defined in the previous literature contains an overall factor of the wavefunction renormalization, which cannot be interpreted as the mixing angle (see Sec.~\ref{sec:comparison}).

Finally, we discuss the implications of our understanding on the baryogenesis via decaying magnetic helicity (Sec.~\ref{sec:baryon}).
The effects of our findings are threefold: (i) the refined definition of the effective mixing angle (Sec.~\ref{sec:baryon_anomaly}), (ii) the number of the unconfined magnetic flux is approximately conserved, and (iii) the novel non-perturbative effect, which unties the knot of confined magnetic flux without changing the Chern--Simons number of $\mathrm{SU}(2)_{\mathrm L}$ (Sec.~\ref{sec:non-perturbative}).
The first/second effects change the resultant baryon asymmetry by one order of magnitude, as the source term from the time evolution would be modified by just a couple of factors compared the previous definition (see Fig.~\ref{fig:theta_eff}).
The third effect would significantly reduce the resultant baryon asymmetry, as the magnetic helicity carried by the unconfined magnetic flux initially is not necessarily converted to the baryon asymmetry.
Rather, a novel non-perturbative effect may directly convert it to a non-trivial topological sector of the groundstate, namely the first term of the r.h.s.\ in Eq.~\eqref{eq:B+L} is compensated by the second term of the r.h.s., yielding $\Delta Q_{B+L} = 0$.
However, as emphasized at the end of Sec.~\ref{sec:non-perturbative}, the final outcome is not yet clear because the details of physical processes are involved mainly due to the crossover nature of EWSB, which is worthwhile investigating in the future.
At this stage, we may at least warn that the scenarios based on this baryogenesis mechanism may have not only the quantitative uncertainties (the former two points) but also the qualitative uncertainty (the third one),
and we should be careful in interpreting the results of the previous works.

Similar cautions also apply to the magnetogenesis before the EWSB.
As mentioned in the introduction, the recent works~\cite{Fujita:2016igl,Kamada:2016eeb,Kamada:2020bmb} have claimed that any magnetogenesis scenario before the EWSB suffers from either the baryon overproduction or the baryon isocurvature perturbations.
However, these analysis are also based on the previous understanding of the magnetic flux and its dynamics, which does not capture the threefold effects discussed above.
We will come back to this issue in a separate publication~\cite{Hamada:2025ooy}.

\paragraph{Acknowledgements}

We thank Kohei Kamada for discussions at the early stage of this project. We also thank Hajime Fukuda, Yoshimasa Hidaka, and Ryo Yokokura for discussions.
This work is supported by MEXT Leading Initiative for Excellent Young Researchers Grant No.~JPMXS0320210099 [YH] and JSPS KAKENHI Grant Nos.~JP24H00976 [YH], JP24K07035 [YH], JP24KF0167 [YH], JP22K14044 [KM], and JP23KJ0642 [FU].
This work is also supported by World Premier International Research Center Initiative (WPI Initiative), MEXT, Japan.

\appendix

\section{Feynman rules}
\label{sec:feynman}
In this appendix, we specify the conventions for calculating the one-loop corrections in the broken phase and list the Feynman rules with our convention.

The relevant terms in the three-dimensional Lagrangian density for the finite-temperature EFT of the Standard Model $\mathcal L_{{\rm ESM}_3}$, defined in Eq.~\eqref{eq:SM3}, is
\begin{align}
    {\mathcal L}_{{\rm ESM}_3}
        =&\;\dfrac{1}{4}W_{ij}^{a}W_{ij}^{a}+\dfrac{1}{4}Y_{ij}Y_{ij}
        +\left(D_i\Phi\right)^{\dagger}D_i\Phi+m_3^2\Phi^{\dagger}\Phi+\lambda_3\left(\Phi^{\dagger}\Phi\right)^2\notag\\[.5em]
        &+\frac{1}{2} \qty(\partial _ i Y_{\tau})^{2}
        + \frac{1}{2} m_{\rm D}'^{2}Y_{\tau}^{2}
        + \frac{1}{2} \qty( D _ i W^ a _{\tau})^{2}
        + \frac{1}{2}m_{\rm D}^{2}W_\tau ^ a W_\tau^a\notag\\[.5em]
        &
        + h_{3}' \Phi^{\dagger}\Phi Y_{\tau}^{2}
        + h_{3}'' Y_\tau \Phi ^  \dagger W_\tau ^ a \sigma^a  \Phi
        + h_{3}  \Phi^{\dagger}\Phi W^a_{\tau} W^a_\tau,
    \label{eq:SM3_rel}
\end{align}
where the $\mathcal O(g^4)$ contribution, the Casimir energy $\Lambda_T$, and the ${\rm SU}(3)_{\mathrm c}$ terms are implicitly omitted.
The field strengths of the ${\rm SU}(2)_{\rm L}$ gauge fields $W^{a}_i$ and the ${\rm U}(1)_Y$ gauge field $Y_i$ are
\begin{align}
    W_{ij}^{a}
        \coloneq\partial_iW_j^{a}-\partial_jW_i^{a}+g_3\epsilon^{abc}W_i^{b}W_j^{c},\qquad
    Y_{ij}
        \coloneq\partial_iY_j-\partial_jY_i,
\end{align}
where $\epsilon^{abc}$ is the totally antisymmetric tensor with the sign convention $\epsilon^{123}=+1$, and the three-dimensional covariant derivative is, depending on the representation of the fields,
\begin{align}
    D_i\Phi
        \coloneq\biggl(\partial_i-\dfrac{ig_3\sigma^a}{2}W_i^{a}+\dfrac{ig'_3}{2}Y_i\biggr)\Phi,\qquad
    D_iW^a_\tau
        \coloneq\biggl(\delta^{ac}\partial_i+g_3\epsilon^{abc}W_i^{b}\biggr)W^c_\tau,
    \label{eq:cov_deriv}
\end{align}
where $\sigma^a$ is the $a$-th Pauli matrix.

Note that the mass dimensions of the fields and the parameters are different from the original four-dimentional theory, {\it i.e.}, $[\Phi]=[Y_\mu]=[W_\mu^{a}]=[g_3]=[g'_3]=1/2$, and $[m_3]=[\lambda_3]=[m_{\rm D}]=[m'_{\rm D}]=[h_3]=[h'_3]=[h''_3]=1$, as is explicit in the matching conditions \eqref{eq:matching_fields1}, ..., \eqref{eq:Higgs_mass}.

\subsection{High-temperature phase}
In the high-temperature phase, where $m_3^2>0$, we take the $R_\xi$ gauge, by imposing  $\partial_iY_i=\partial_iW^a_i=0$, and introduce the associated ghost fields \cite{Peskin:1995ev}.
\begin{align}
    \mathcal L_{\rm sym}
        \coloneq\mathcal L_{{\rm ESM}_3}+\dfrac{1}{2\xi}(\partial_iY_i)^2+\dfrac{1}{2\xi}(\partial_iW^a_i)(\partial_jW^a_j)
        +\partial_i\bar c_Y\partial_i c_Y
        +\partial_i\bar c^a_W\partial_i c^a_W
        -g_3\epsilon^{abc}\bar c^a_W\partial_i(W^b_ic^c_W)
    \label{eq:full_lagrangian_sym}
\end{align}
is the Lagrangian density that we deal with in the high-temperature phase.

From this Lagrangian density, \eqref{eq:full_lagrangian_sym}, one can readily read off the Feynman rules, once we specify the assignment of a $-1$ factor to each diagram.
For a generic field $X$, we may add a source term $+JX$ to the Lagrangian density of the free theory, and then the partition function of the full theory is $Z_\beta[J]=\exp\bigl(-S_{\rm E}^{\rm int}[-\delta_J]\bigr)Z_{\beta,0}[J]$, where $\delta_J$ is the functional derivative with respect to $J$, $S_{\rm E}^{\rm int}[X]$ is the interaction part of the Euclidean action, and $Z_{\beta,0}[J]$ is the partition function corresponding to the free theory.
Then, a diagram that has $V_n$ $n$-point vertices, $n=1,2,\cdots$, should include a factor $(-1)^{\sum_n(1+n)V_n}=(-1)^{E+\sum_nV_n}$, where $E$ is the number of external lines.
To account for this, we assign a factor $-1$ to every interaction vertex and every external source, in addition to the $-1$ factor for each ghost loop.
In evaluating each vertex, the spatial derivative $\partial_i$ in the real space is translated as $+ip_i$ in terms of the {\it incoming} momentum $\vec p$.

\begin{figure}[H]\centering
    \includegraphics[keepaspectratio, width=1.\textwidth]{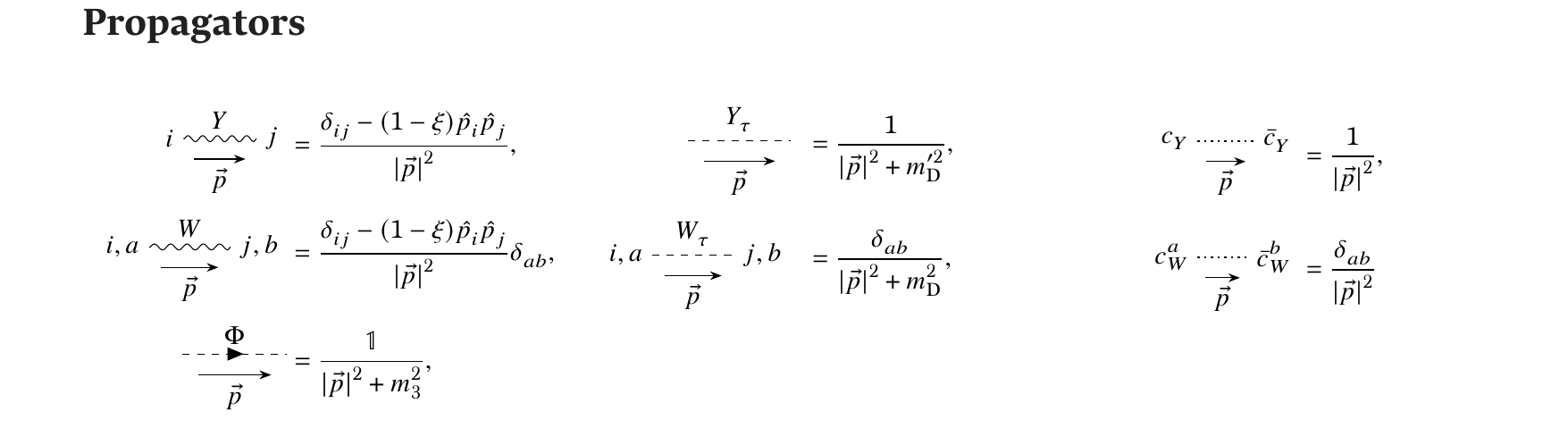}
    \caption{Propagators of the fields in the {\it high-temperature} phase. The filled arrows specify the sign of hypercharges. $\mathbb 1$ is the identity matrix in the ${\rm SU}(2)_{\rm L}$ fundamental representation, where its indices are implicit. Note that the non-abelian gauge field, $W_i^a$ has non-perturbative masses in the $\vec p\to0$ limit \cite{Aguilar:2008xm}.}
\end{figure}
\begin{figure}[H]\centering
    \includegraphics[keepaspectratio, width=1.\textwidth]{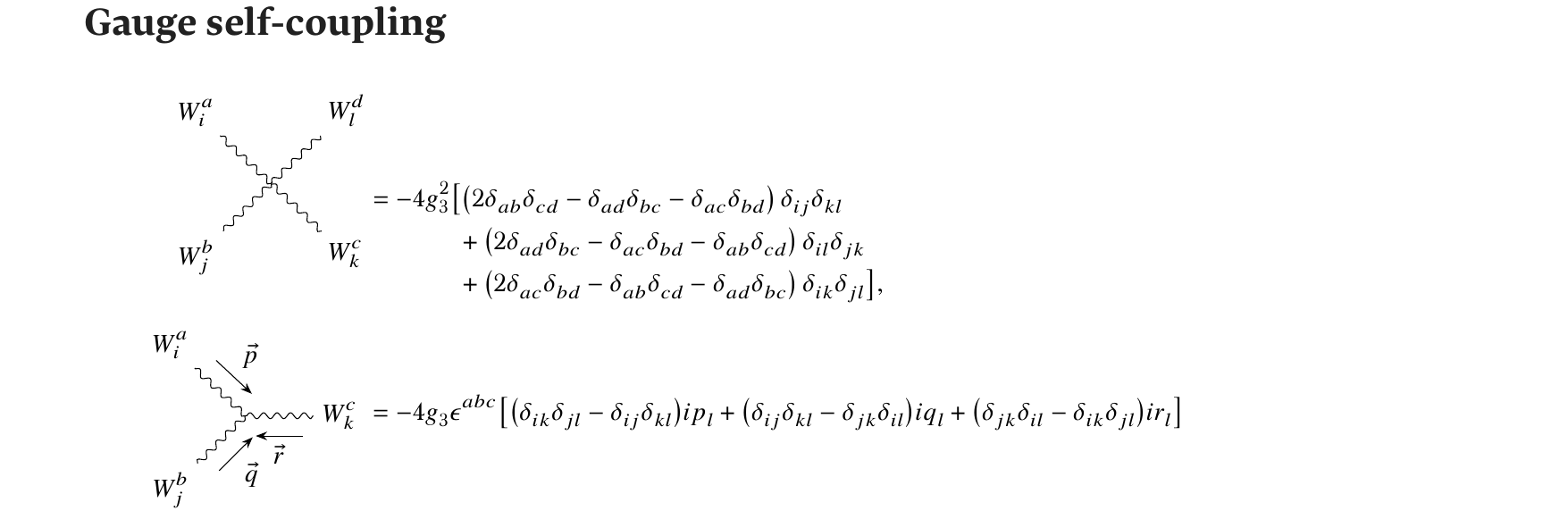}
    \caption{Gauge self-couplings in the {\it high-temperature} phase.}
\end{figure}
\begin{figure}[H]\centering
    \includegraphics[keepaspectratio, width=1.\textwidth]{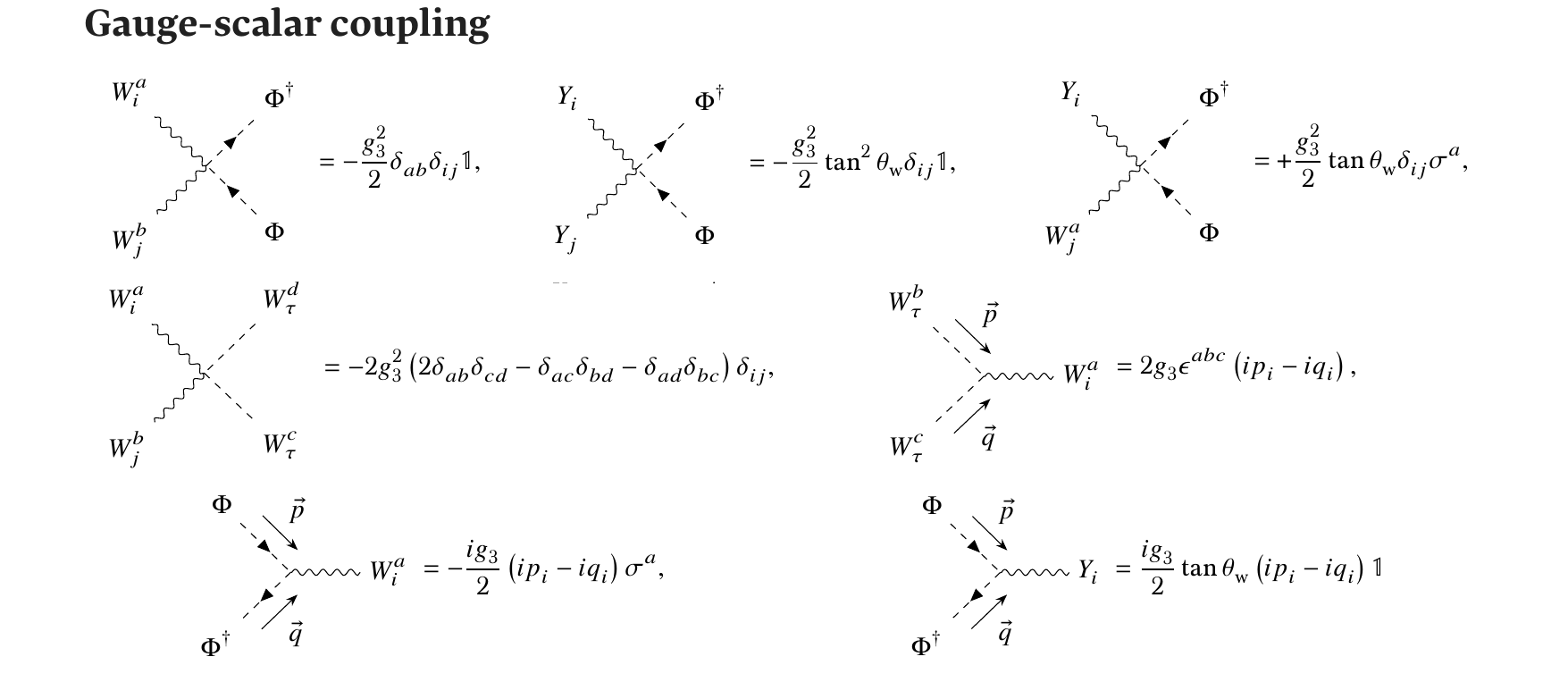}
    \caption{Gauge-scalar interactions in the {\it high-temperature} phase. $\mathbb 1$ and $\sigma^a$ are the identity and the Pauli matrices in the ${\rm SU}(2)_{\rm L}$ fundamental representation, where their indices are implicit.}
\end{figure}
\begin{figure}[H]\centering
    \includegraphics[keepaspectratio, width=1.\textwidth]{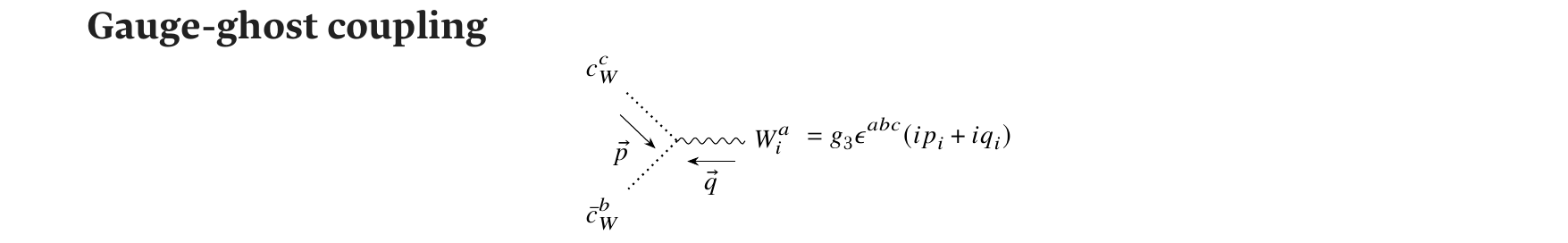}
    \caption{Gauge-ghost interactions in the {\it high-temperature} phase.}
\end{figure}
\begin{figure}[H]\centering
    \includegraphics[keepaspectratio, width=1.\textwidth]{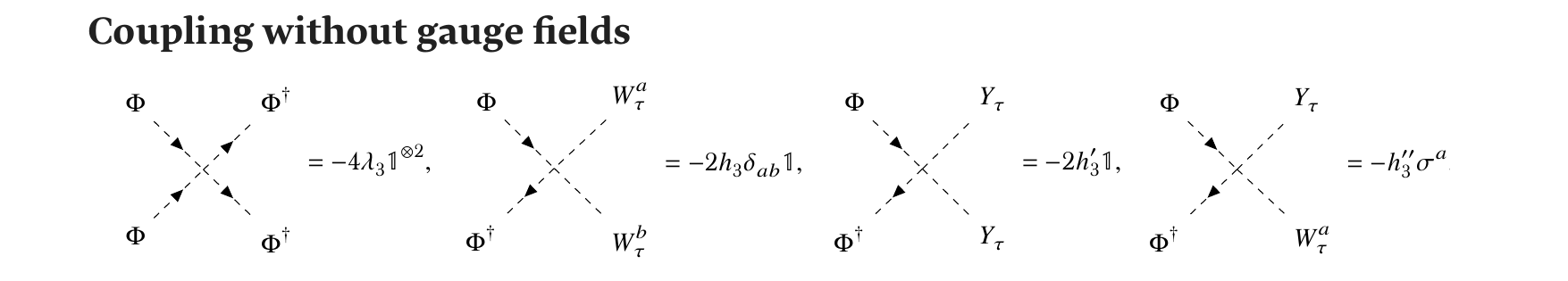}
    \caption{Interactions without gauge fields in the {\it high-temperature} phase. $\mathbb 1$ and $\sigma^a$ are the identity and the Pauli matrices in the ${\rm SU}(2)_{\rm L}$ fundamental representation, where their indices are implicit.}
\end{figure}

\subsection{Low-temperature phase}
In the low-temperature phase, where $m_3^2<0$, it is useful to introduce these parameters,
\begin{align}
    v_3
        \coloneq\sqrt{\dfrac{-m_3^2}{\lambda_3}},\qquad
    m_{H}^2
        \coloneq-2m_3^2,\qquad
    m_Z
        \coloneq\dfrac{\sqrt{g_3^2+g_3'^2}}{2}v_3,\qquad
    m_W
        \coloneq\dfrac{g_3}{2}v_3.
    \label{eq:masses_def}
\end{align}
The vacuum in this phase is specified by $\Phi^{\dagger}\Phi=v_3^2/2$, and we introduce the Higgs $H$ and the Nambu--Goldstone $\phi^{0,\pm}$ degrees of freedom as perturbations around the vacuum as in Eq.~\eqref{eq:Higgs_dofs}.
As for the electric components of the gauge fields, we introduce $W_\tau^\pm=(W_\tau^1\mp iW_\tau^2)/\sqrt{2}$ to let the $Q=Y+T^3$ charge explicit.
By diagonalizing the mass matrix of the neutral components, $Y_\tau$ and $W_\tau^3$, we introduce a weak mixing angle for the electric components, $\theta_{\tau}$ {\it s.t.}
\begin{align}
    \tan 2\theta_{\tau}
        =\dfrac{2\tan\theta_{\rm w}}{1-\tan^2\theta_{\rm w}+\frac{m_{\rm D}^2-m'^2_{\rm D}}{m_W^2}}
        +\mathcal O(g^2).
\end{align}
Correspondingly, we introduce the mass basis
\begin{align}
    \tilde Y_\tau
        \coloneq\cos\theta_\tau Y_\tau-\sin\theta_\tau W^3_\tau,\qquad
    \tilde W^3_\tau
        \coloneq\cos\theta_\tau W^3_\tau+\sin\theta_\tau Y_\tau.
    \label{eq:ele_massbasis}
\end{align}
We will discuss Feynman rules for the interaction vertex in terms of $Y_\tau$ and $W_\tau^3$ to avoid complication, since one can readily rewrite them into the ones in terms of $\tilde Y_\tau$ and $\tilde W_\tau^3$ by substituting the relation \eqref{eq:ele_massbasis}.
As for the three-dimensional gauge fields, $Y_i$ and $W_i^a$, we diagonalize the mass matrix, by introducing the photon $A_i$ and the massive gauge bosons $Z_i$ and $W^{\pm}_i$,
\begin{align}
    A_i
        \coloneq\cos\theta_{\rm w}Y_i-\sin\theta_{\rm w}W_i^{3},\qquad
    Z_i
        \coloneq\cos\theta_{\rm w}W_i^{3}+\sin\theta_{\rm w}Y_i,\qquad
    W_i^{\pm}
        =\dfrac{1}{\sqrt{2}}\qty(W_i^{1}\mp iW_i^{2}).
    \label{eq:AZWpm_def}
\end{align}
We also impose the $R_\xi$ gauge-fixing conditions \cite{Fujikawa:1972fe} on the three-dimensional gauge fields $Y_i$ and $W_i^a$
\begin{align}
    G_X=0,\qquad
    \qty(G_A,G_Z,G_\pm)
        \coloneq\qty(\partial_iA_i,\,
        \partial_iZ_i+\xi m_Z\phi^{0},\,
        \partial_iW_i^{\pm}+i\xi m_W\phi^{\pm})
    \label{eq:gauge_fixing}
\end{align}
and introduce the ghost fields, $\bar c_X$ and $c_X$ for each of the gauge fixing conditions, where $X=A,Z,\pm$.
Accordingly, by adding gauge-fixing terms and the ghost Lagrangian, we obtain the Lagrangian density that we are going to handle (See, {\it e.g.}, Refs.~\cite{Peskin:1995ev,Romao:2012pq}):
\begin{align}
    \mathcal L_{\rm broken}
        \coloneq&\;\mathcal L_{{\rm ESM}_3}+\dfrac{1}{2\xi}\sum_{X}G_{\bar X}G_X+\sum_{X,Y}\bar c_{\bar X}\dfrac{\delta G_X}{\delta\alpha_Y}c_Y,
    \label{eq:full_lagrangian_broken}
\end{align}
where $X$ and $Y$ run $A,\,Z,\,\pm$, we define $\bar A\coloneq A,\,\bar Z\coloneq Z,\,\bar\pm\coloneq\mp$, and $\alpha_X$ parametrizes an infinitesimal gauge transformation, namely $\alpha_A\coloneq\cos\theta_{\rm w}\alpha_Y-\sin\theta_{\rm w}\alpha_3$, $\alpha_Z\coloneq\sin\theta_{\rm w}\alpha_Y+\cos\theta_{\rm w}\alpha_3$, and $\alpha_\pm\coloneq(\alpha_1\mp i\alpha_2)/\sqrt{2}$, for $U=\exp{-ig_3\sigma^a\alpha_a/2+ig_3'n_Y\alpha_Y}\in{\rm SU}(2)_{\rm L}\times{\rm U}(1)_Y$.

\begin{figure}[H]\centering
    \includegraphics[keepaspectratio, width=1.\textwidth]{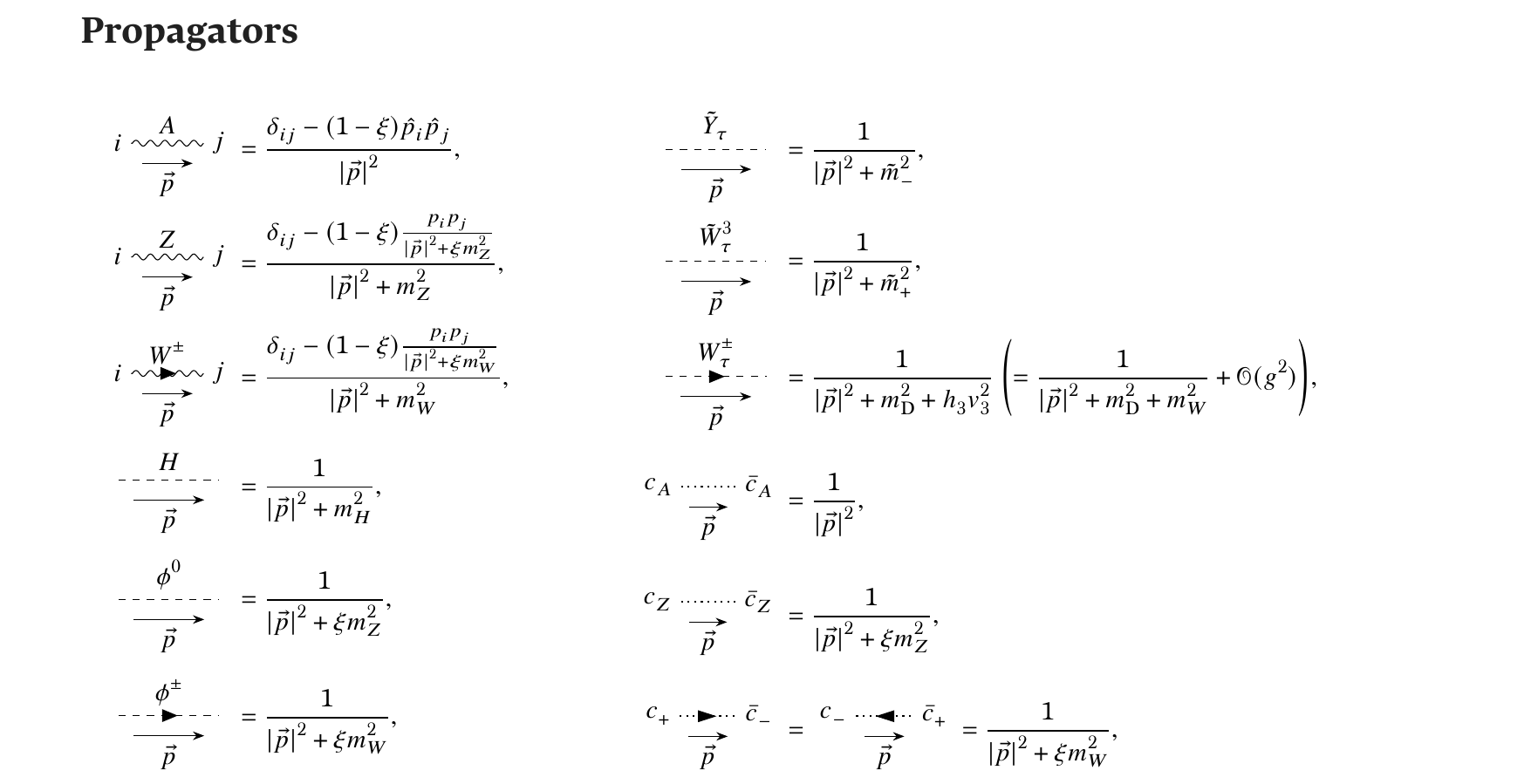}
    \caption{\label{fig:prop_broken}%
    Propagators and mixing of the fields in the {\it low-temperature} phase. The filled arrows specify the sign of $Q=Y+T^3$ charges.
    }
\end{figure}

\begin{figure}[H]\centering
    \includegraphics[keepaspectratio, width=1.\textwidth]{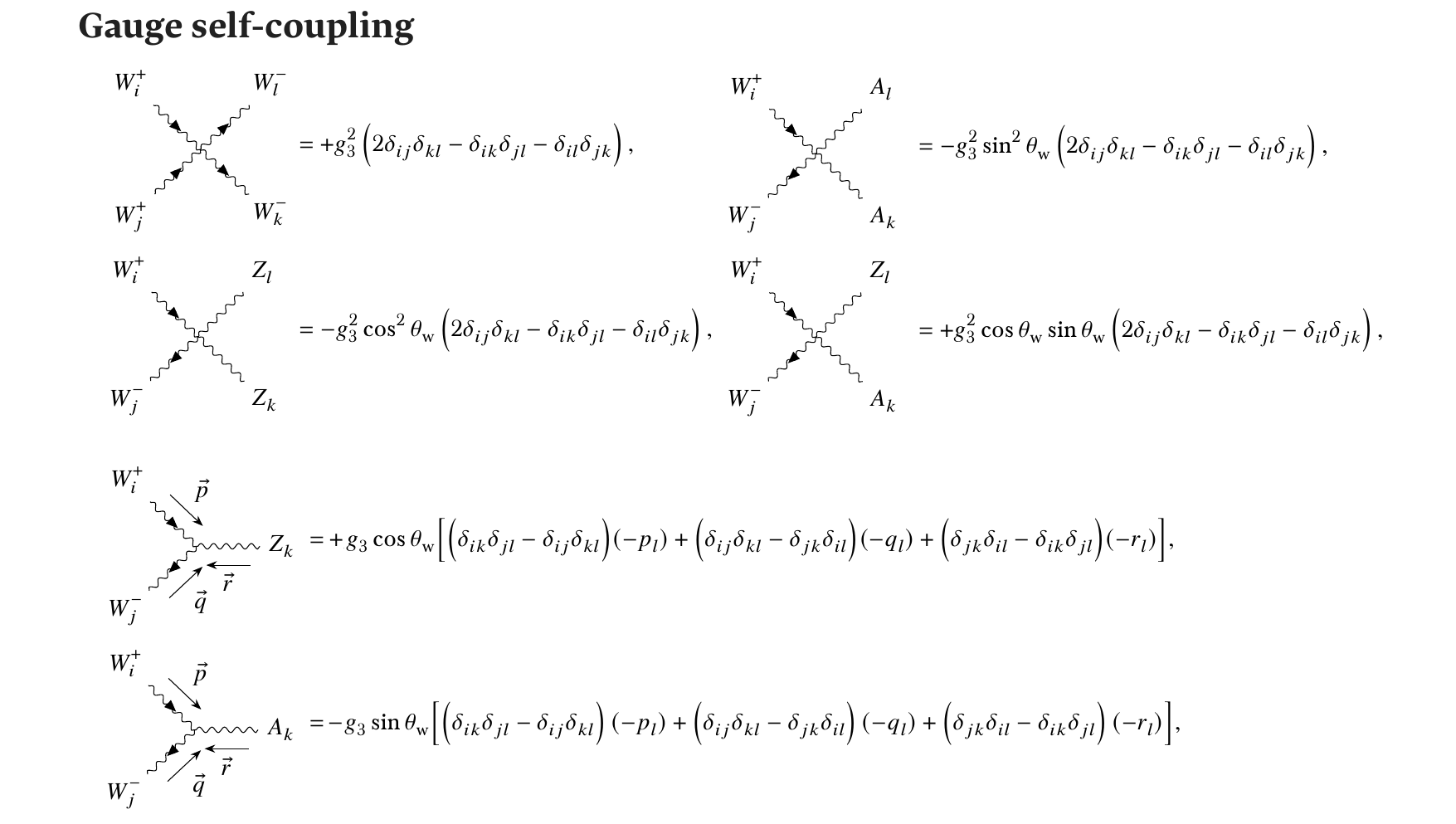}
    \caption{\label{fig:gself_broken}%
    Gauge self-couplings in the {\it low-temperature} phase.}
\end{figure}

\begin{figure}[H]\centering
    \includegraphics[keepaspectratio, width=1.\textwidth]{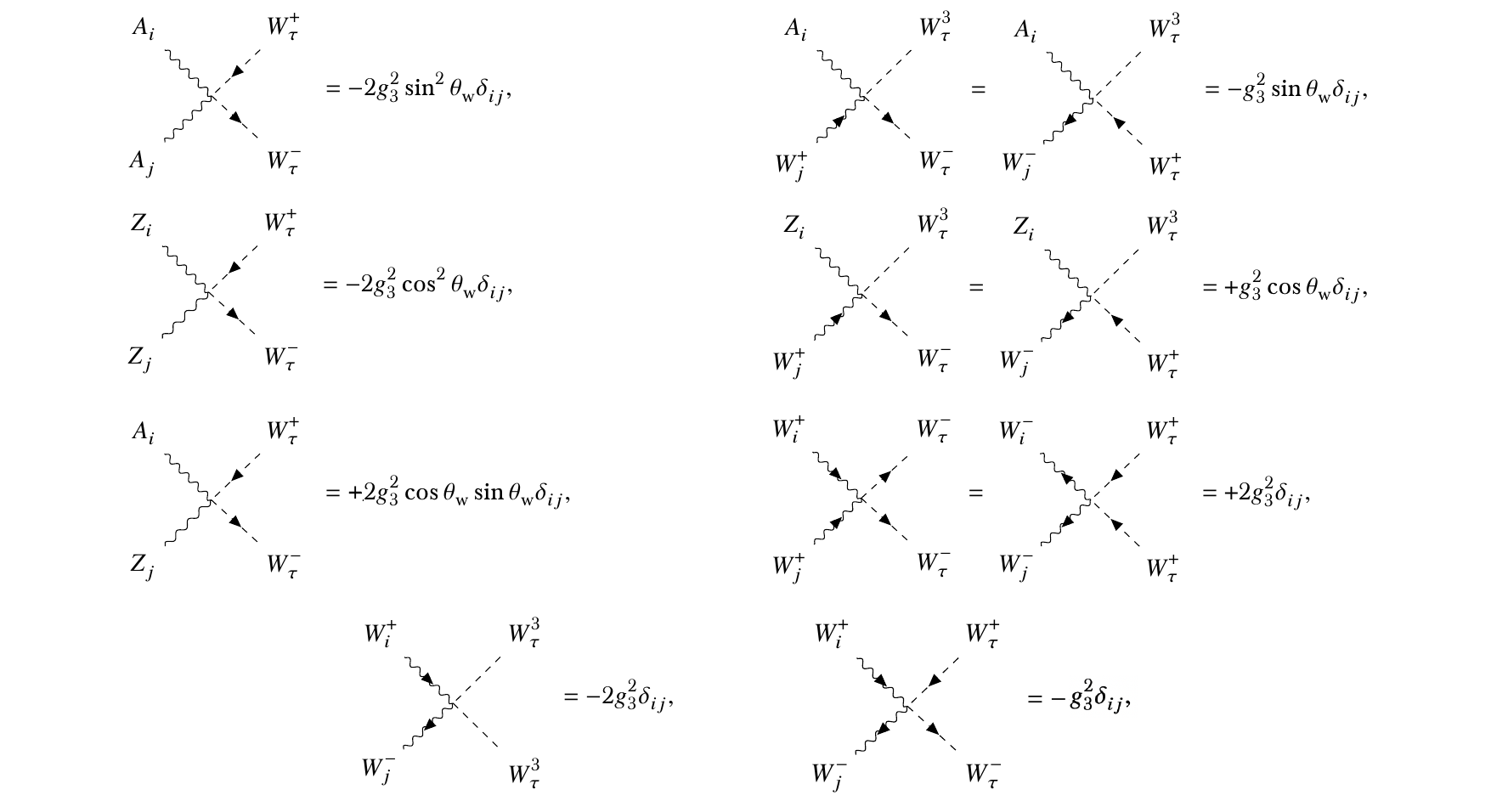}\vspace{1mm}
    \includegraphics[keepaspectratio, width=1.\textwidth]{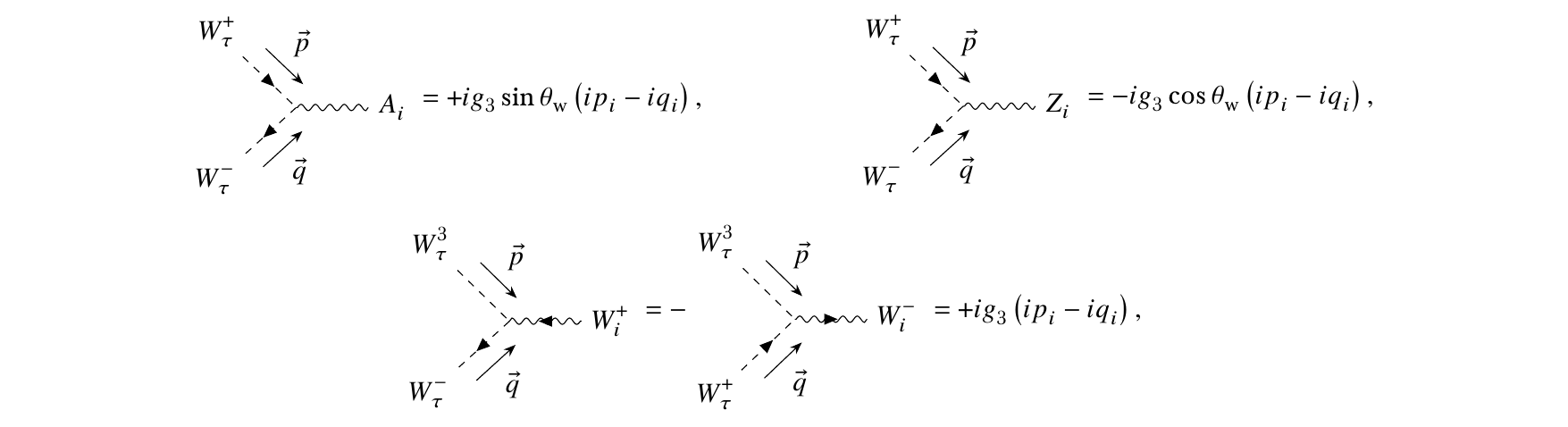}
    \caption{\label{fig:g-2_broken}%
    Gauge-scalar interactions in the {\it low-temperature} phase.}
\end{figure}

\begin{figure}[H]\centering
    \includegraphics[keepaspectratio, width=1.\textwidth]{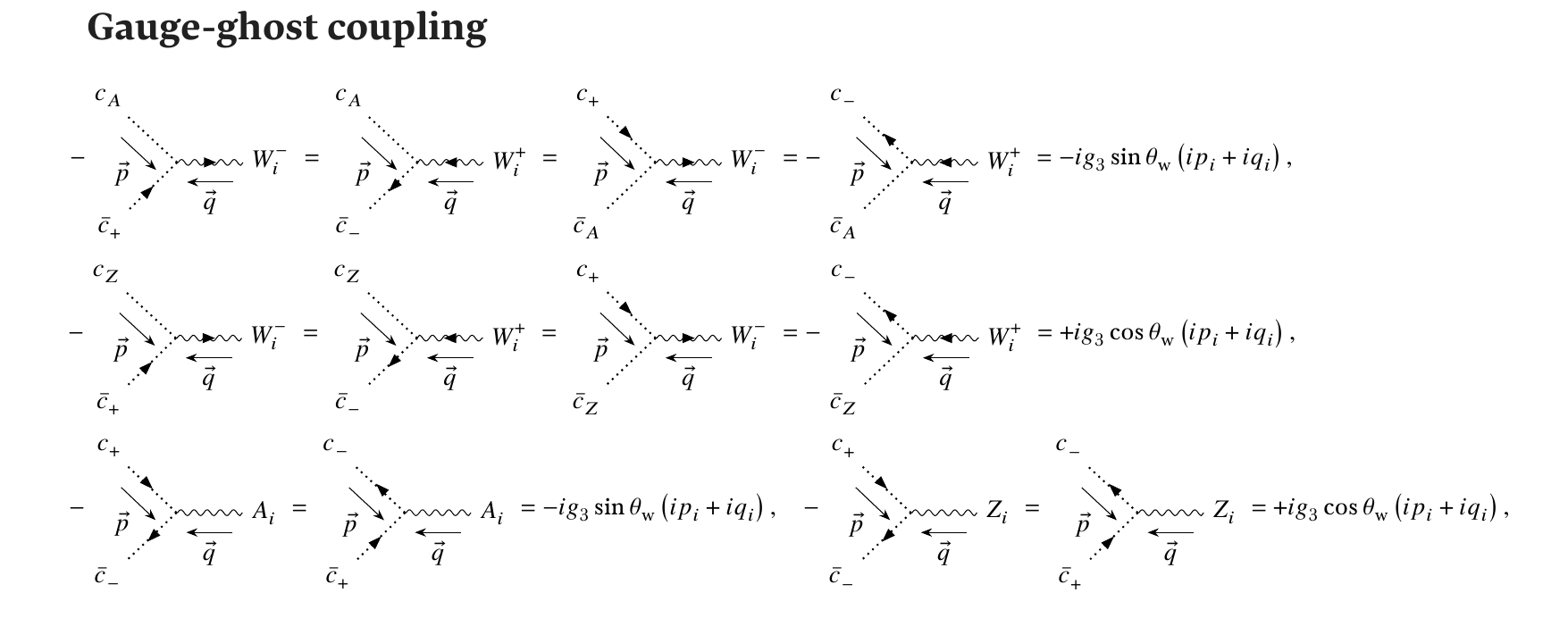}
    \caption{\label{fig:g-gh_broken}%
    Gauge-ghost interactions in the {\it low-temperature} phase.}
\end{figure}

\begin{figure}[H]\centering
    \includegraphics[keepaspectratio, width=.9\textwidth]{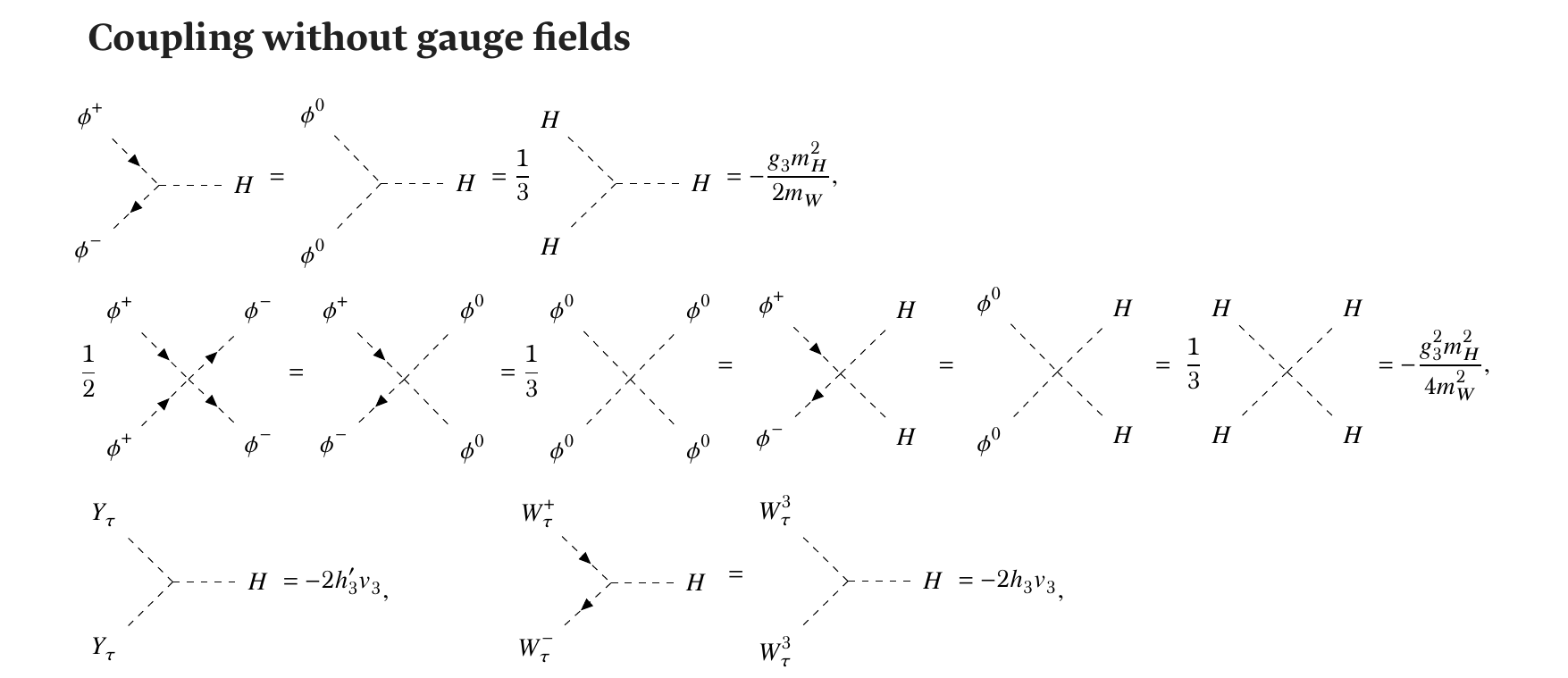}
    \includegraphics[keepaspectratio, width=.9\textwidth]{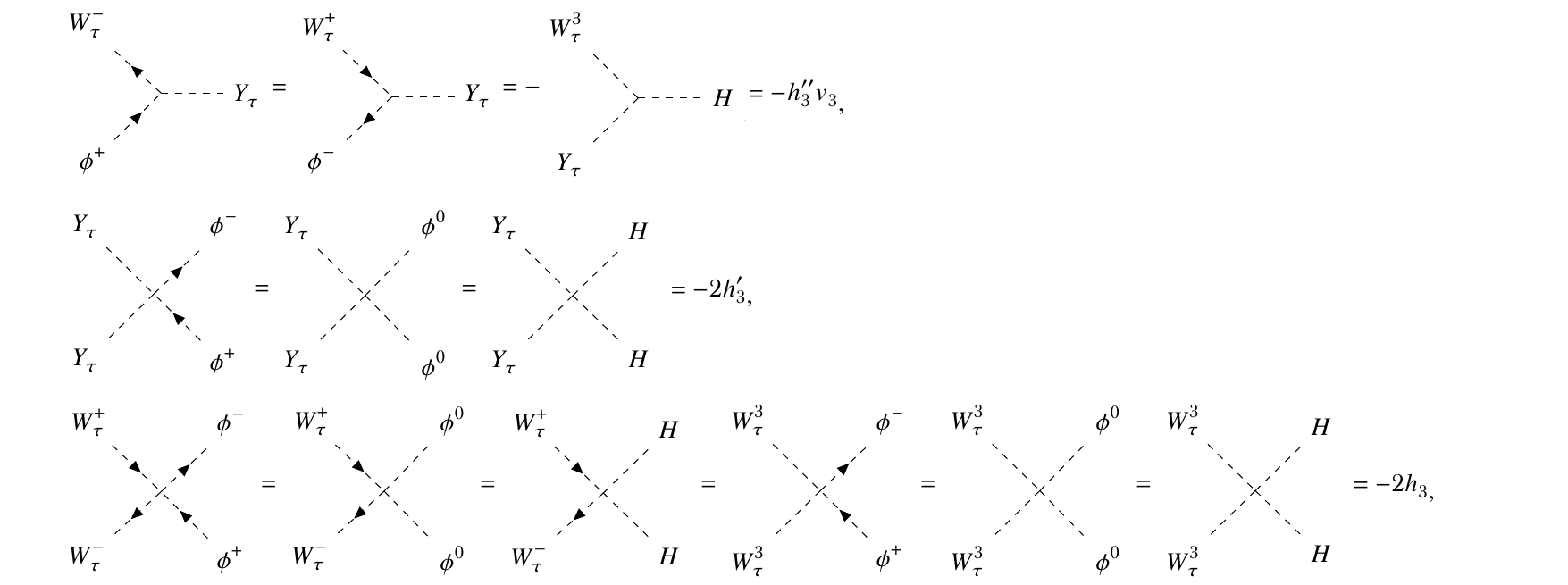}
    \includegraphics[keepaspectratio, width=.9\textwidth]{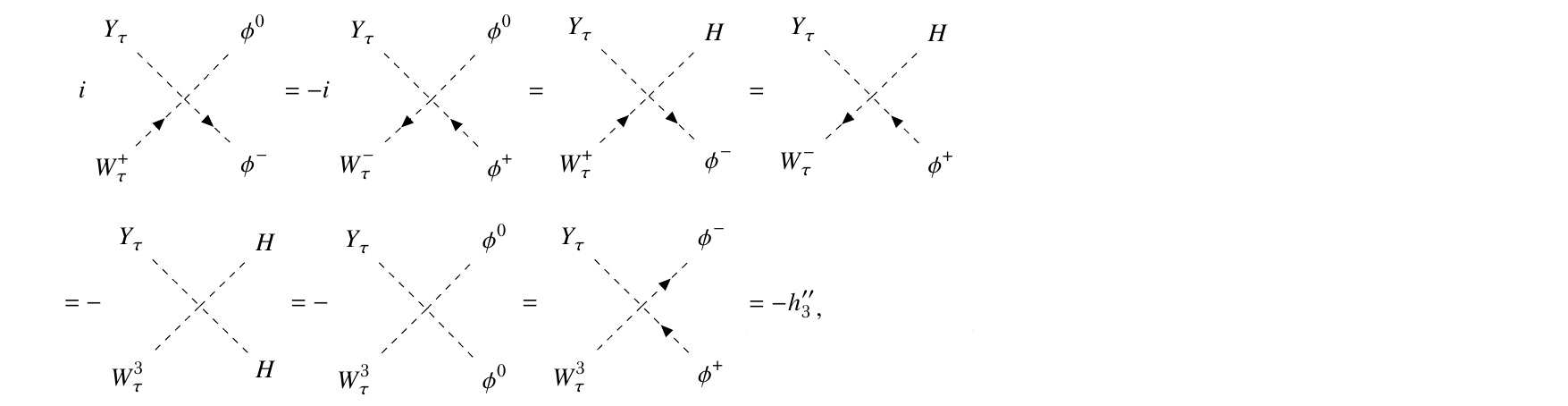}
    \includegraphics[keepaspectratio, width=.9\textwidth]{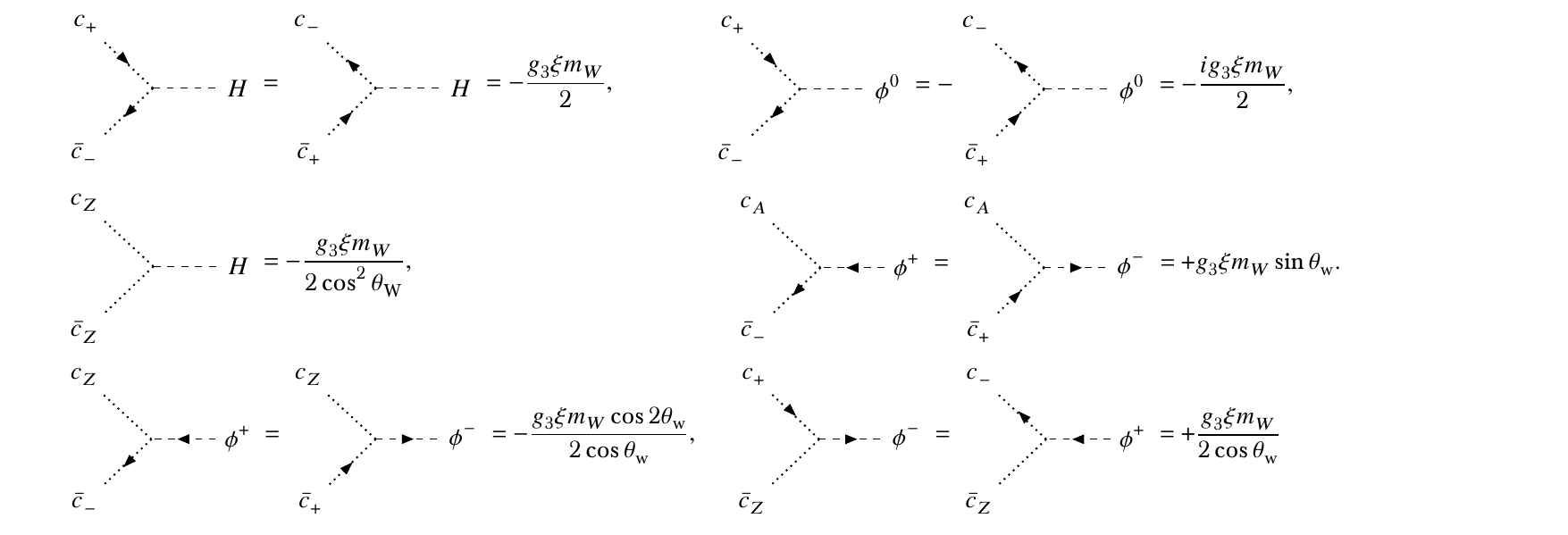}
    \caption{\label{fig:wog_broken}%
    Interactions without gauge fields in the {\it low-temperature} phase.}
\end{figure}

Our convention of assigning a $-1$ factor to each diagram is the same as in the high-temperature phase.
With these conventions, we read off the Feynman rules in the low-temperature phase from the Lagrangian density, \eqref{eq:full_lagrangian_broken}, as in Figs.~\ref{fig:prop_broken}, \ref{fig:gself_broken}, \ref{fig:g-2_broken}, \ref{fig:g-gh_broken}, and \ref{fig:wog_broken}, where we define
\begin{align}
    \tilde m^2_{\pm}
        \coloneq\dfrac{1}{2}\,\qty(\dfrac{m_W^2}{\cos^2\theta_{\rm w}}+m^2_{\rm D}+m'^2_{\rm D}\pm\sqrt{\frac{m_W^4}{\cos^4\theta_{\rm w}}+2(1-\tan^2\theta_{\rm w})\qty(m^2_{\rm D}-m'^2_{\rm D})m_W^2+\qty(m^2_{\rm D}-m'^2_{\rm D})^2}\,).
\end{align}

\begin{figure}[H]\centering
    \includegraphics[keepaspectratio, width=1.\textwidth]{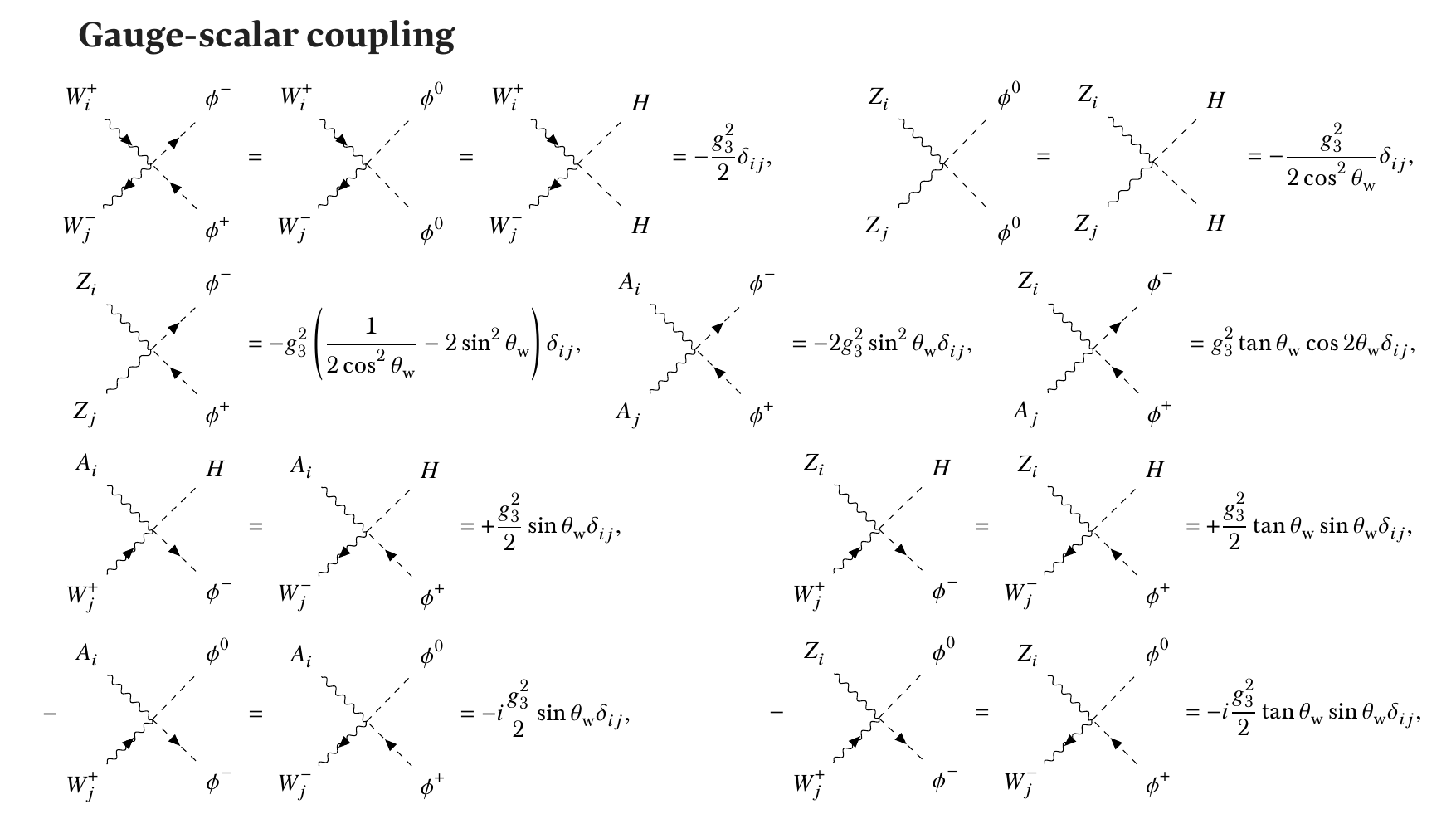}\vspace{1mm}
    \includegraphics[keepaspectratio, width=1.\textwidth]{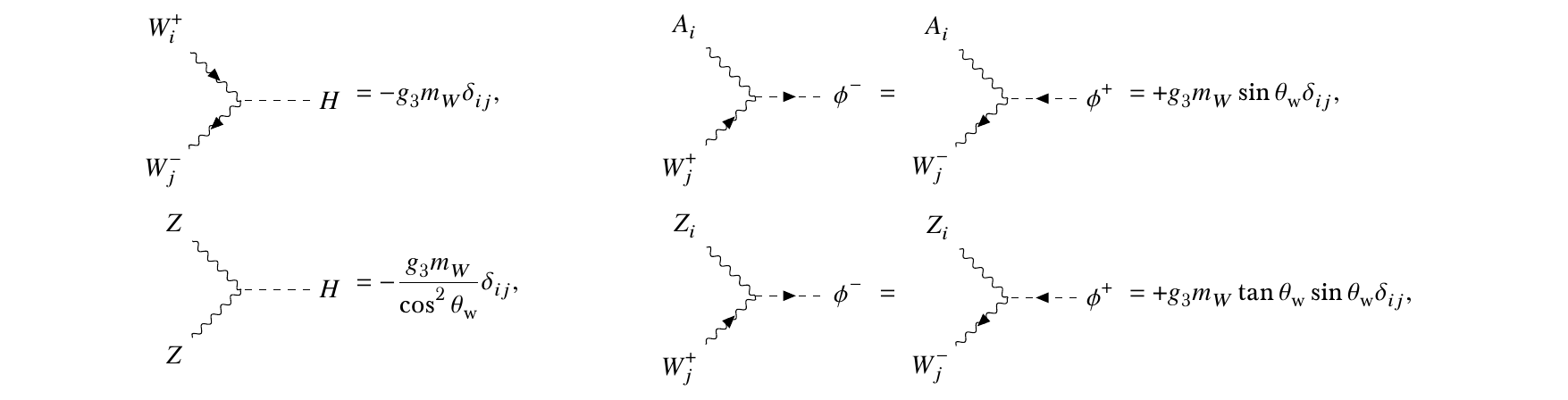}\vspace{1mm}
    \includegraphics[keepaspectratio, width=1.\textwidth]{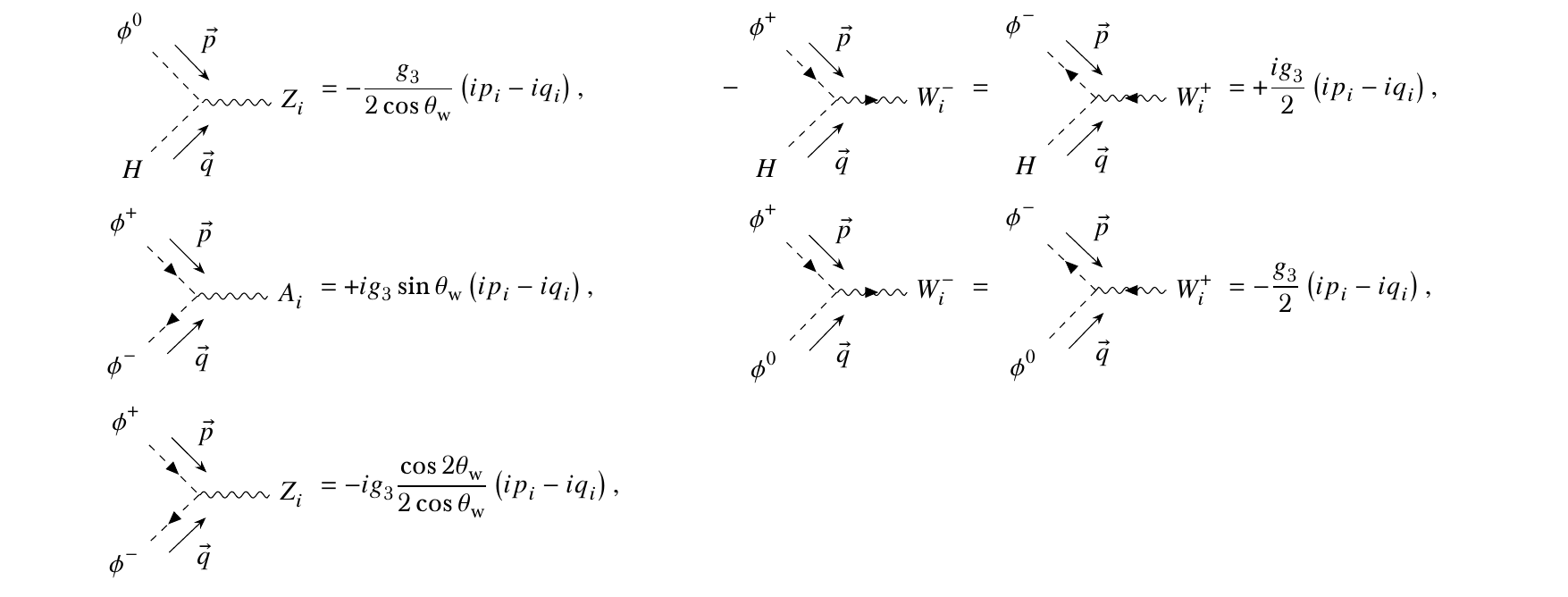}
    \captionsetup{labelformat=empty}
\end{figure}

\section{One-loop corrections in the \texorpdfstring{$R_\xi$}{Rxi} gauge}
\label{sec:one-loop}
In this appendix, we calculate one-loop corrections to identify the unconfined magnetic field for a finite $m_3^2$ by using the Feynman rules listed in App.~\ref{sec:feynman}.

\subsection{High-temperature phase}
\label{sec:1loop_sym}
In the $m_3^2\gg g^4T^2$ limit, $B_{Yi}\coloneq\epsilon^{ijk}\partial_jY_k$ is the unconfined hyper-magnetic field, while the non-abelian magnetic fields are confined because of their self-interactions.
We here consider the one-loop corrections in the high-temperature regime where $0<m_3^2<\infty$.

\begin{figure}[H]\centering
    \includegraphics[keepaspectratio, width=1.\textwidth]{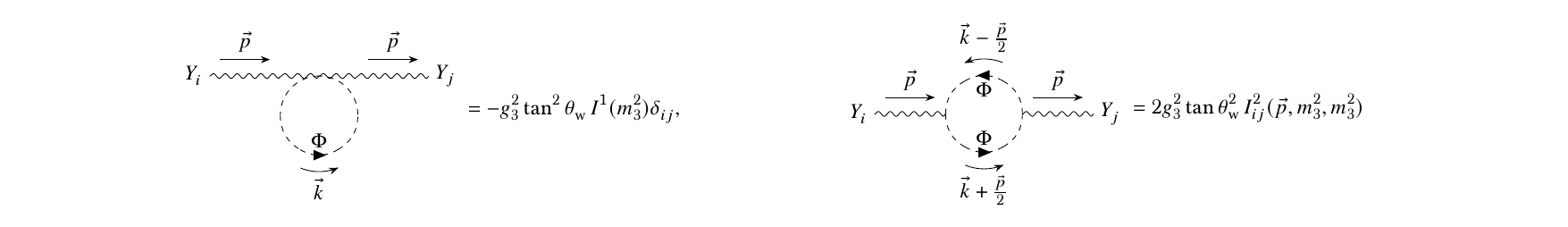}
    \caption{One-loop corrections to the $Y$ propagator in the {\it high-temperature} phase.\label{fig:diagrams_sym}}
\end{figure}

We first calculate the one-loop correction of the $\vec B_{Y}$ correlation function.
The $Y$ propagator obtains correction from $\Phi$ loops.
By using the short-hand notation of loop integrals defined in App.~\ref{sec:formulae}, we evaluate the relevant diagrams, which add up to
\begin{align}
    \sum\text{diagrams in Fig.~\ref{fig:diagrams_sym}}
    =\Pi_{YY}(\vert\vec p\vert)P_{ij}(\hat p),\qquad
    \Pi_{YY}(\vert\vec p\vert)
    \coloneq-\dfrac{g_3^2\tan^2\theta_{\rm w}}{48\pi m_3}\vert\vec p\vert^2+\mathcal O(\vert\vec p\vert^4),
\end{align}
where $P_{ij}(\hat p)\coloneq\delta_{ij}-\hat p_i\hat p_j$ is the projection tensor.
The $\vec B_{Y}$ correlation function at the one-loop order in the high-temperature phase is
\begin{align}
    \langle B_{Yi}(\vec p)B_{Yj}(-\vec p)\rangle'
    &=\epsilon^{ikl}\epsilon^{jmn}ip_ki(-p_m)
    \qty(\dfrac{P_{ln}(\hat p)}{\vert\vec p\vert^2-\Pi_{YY}(\vert\vec p\vert)}+\xi\dfrac{\hat p_l\hat p_n}{\vert\vec p\vert^2})
    \notag\\
    &=\qty(1-\dfrac{g_3^2\tan^2\theta_{\rm w}}{48\pi m_3})P_{ij}(\hat p)+\mathcal O(\vert\vec p\vert^2),
\end{align}
which reproduces the result in Ref.~\cite{Kajantie:1996qd}.

Next, we discuss the one-loop correction of the $Y$$-$$W$ mixing vertices, which are absent at the tree level.
The relevant diagrams are the ones in Fig.~\ref{fig:diagrams_sym}, except that one external line is replaced by $W^a$.
However, these diagrams vanish because $\mathrm{Tr}\,\sigma^a=0$, and therefore we obtain
\begin{equation}
    \langle B_{Yi}(\vec p)B_{W^aj}(\vec q)\rangle=0
\end{equation}
at one-loop, where we have defined $B_{W^ai}\coloneq\epsilon^{ijk}\partial_jW^a_k$.

\subsection{Low-temperature phase}
\label{sec:1loop_broken}
In the $g_3^2 \ll g_3 v_3 $ limit, $B_{i}\coloneq\epsilon^{ijk}\partial_jA_k$ is the unconfined magnetic field, while magnetic fields associated with the other massive gauge fields are confined.
We here consider the one-loop corrections in the low-temperature regime where $g_3^2 \ll g_3 v_3 $ but still $3$d EFT is valid.

\begin{figure}[h]\centering
    \includegraphics[keepaspectratio, width=1.\textwidth]{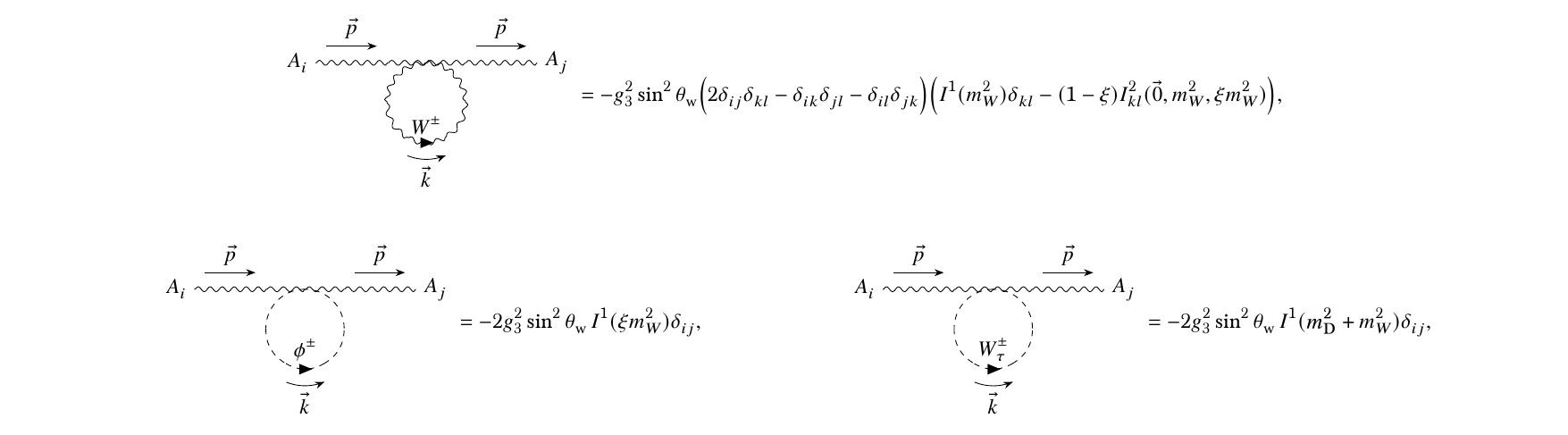}
    \caption{\label{fig:diagrams_broken_prop-4pt}One-loop corrections, which uses four-point interactions, to the $A$ propagator in the {\it low-temperature} phase. These diagrams add up to $g_3^2\sin^2\theta_{\rm w}\Bigl(\frac{4+3\sqrt\xi+2\xi\sqrt\xi}{6\pi}m_W\delta_{ij}+\frac{\sqrt{m_{\rm D}^2+m_W^2}}{2\pi}\delta_{ij}\Bigr)$.
    }
\end{figure}

\begin{figure}[h]\centering
    \includegraphics[keepaspectratio, width=1.\textwidth]{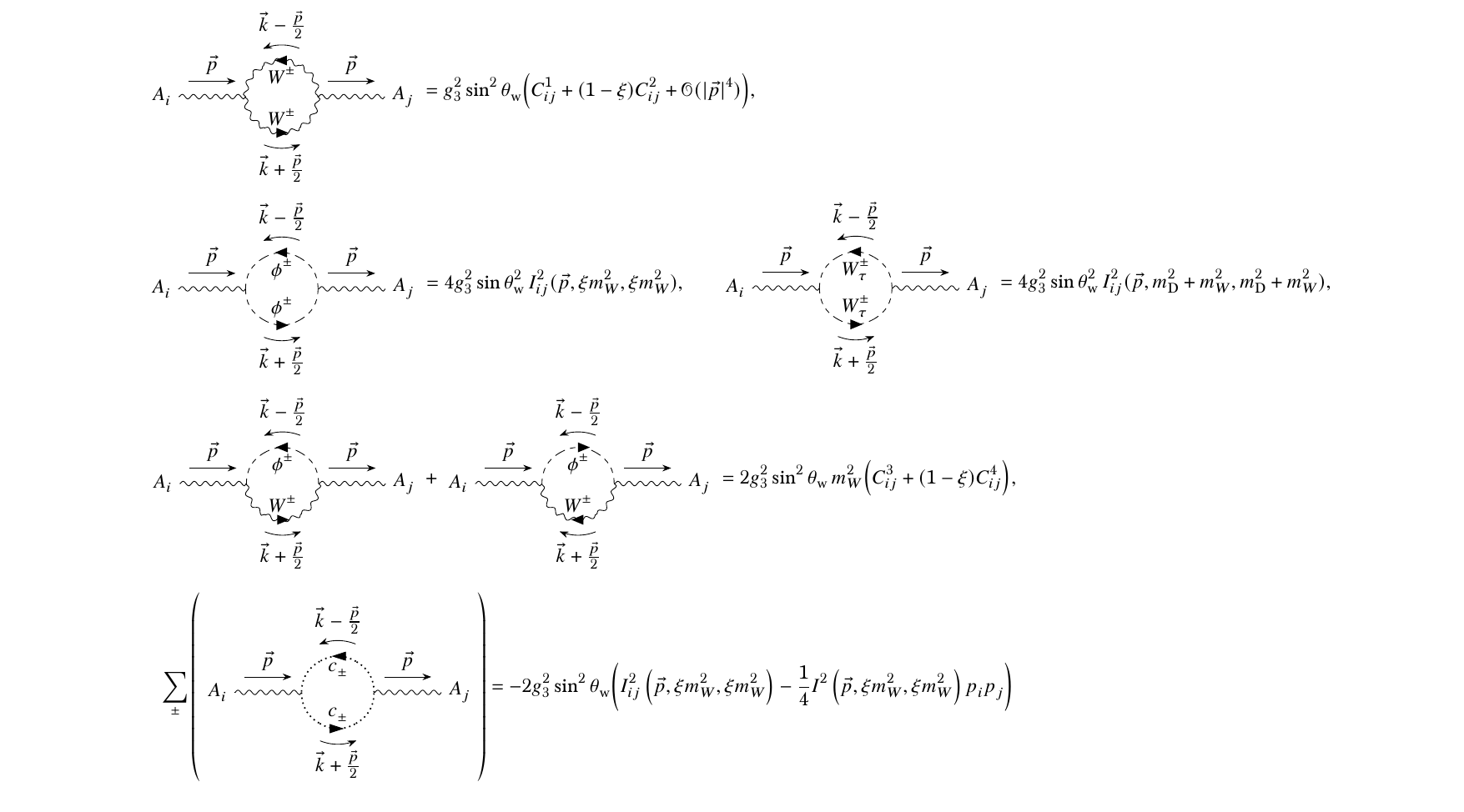}
    \caption{\label{fig:diagrams_broken_prop-3pt}One-loop corrections, which uses three-point interactions, to the $A$ propagator in the {\it low-temperature} phase. These diagrams add up to $g_3^2\sin^2\theta_{\rm w}\Bigl(-\frac{4+3\sqrt\xi+2\xi\sqrt\xi}{6\pi}m_W\delta_{ij}-\frac{\sqrt{m_{\rm D}^2+m_W^2}}{2\pi}\delta_{ij}+\frac{3+9\sqrt\xi-2\xi}{12\pi(1+\sqrt\xi)}\frac{\vert\vec p\vert^2P_{ij}(\hat p)}{m_W}-\frac{\vert\vec p\vert^2P_{ij}(\hat p)}{24\pi\sqrt{m_{\rm D}^2+m_W^2}}\Bigr)+\mathcal O(\vert\vec p\vert^4)$.}
\end{figure}

We first calculate the one-loop correction of the $\vec B$ correlation function.
The $A$ propagator obtains correction from internal $W^{\pm}_i,$ $\phi^{\pm},$ $W^{\pm}_\tau,$ and $c_{\pm}$ fields.
By using the short-hand notation of loop integrals defined in App.~\ref{sec:formulae}, we evaluate the relevant diagrams, which add up to
\begin{align}
    \sum\text{diagrams in Figs.~\ref{fig:diagrams_broken_prop-4pt} and \ref{fig:diagrams_broken_prop-3pt}}
        =\Pi_{AA}(\vert\vec p\vert)P_{ij}(\hat p)+\mathcal O(\vert\vec p\vert^4),\notag\\
    \Pi_{AA}(\vert\vec p\vert)
        \coloneq
        g_3^2\sin^2\theta_{\rm w}\qty(\dfrac{3+9\sqrt\xi-2\xi}{12\pi(1+\sqrt\xi)}
    -\dfrac{m_W}{24\pi\sqrt{m_{\rm D}^2+m_W^2}})\dfrac{\vert\vec p\vert^2}{m_W}.
    \label{eq:Pi_AA}
\end{align}

\begin{figure}[H]\centering
    \includegraphics[keepaspectratio, width=1.\textwidth]{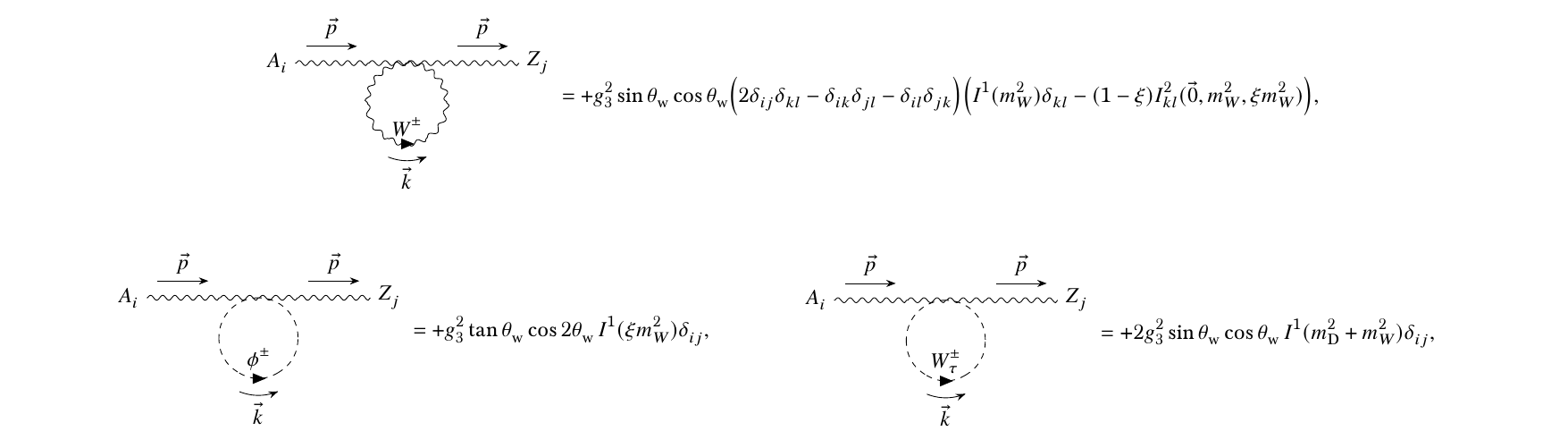}
    \caption{\label{fig:diagrams_broken_ver-4pt}One-loop corrections, which uses four-point interactions, to the $A$$-$$Z$ vertex in the {\it low-temperature} phase. These diagrams add up to $g_3^2\sin\theta_{\rm w}\cos\theta_{\rm w}\Bigl(-\frac{8+3\sqrt\xi+4\xi\sqrt\xi}{12\pi}m_W+\tan^2\theta_{\rm w}\frac{\sqrt\xi}{4\pi}m_W-\frac{1}{2\pi}\sqrt{m_{\rm D}^2+m_W^2}\Bigr)\delta_{ij}+\mathcal O(\vert\vec p\vert^2)$.}
\end{figure}

\begin{figure}[H]\centering
    \includegraphics[keepaspectratio, width=1.\textwidth]{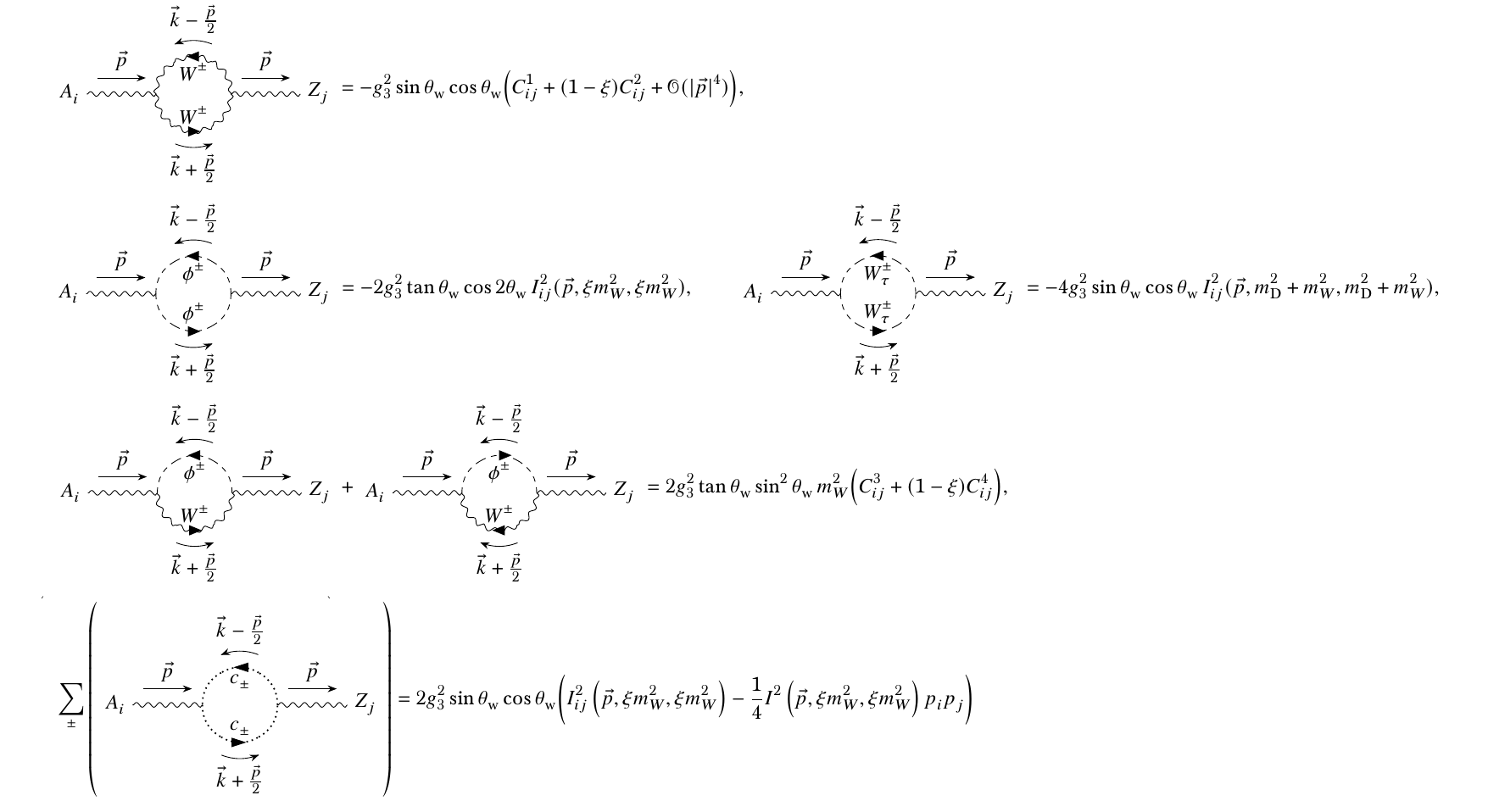}
    \caption{\label{fig:diagrams_broken_ver-3pt}One-loop corrections, which uses three-point interactions, to the $A$$-$$Z$ vertex in the {\it low-temperature} phase. These diagrams add up to $g_3^2\sin\theta_{\rm w}\cos\theta_{\rm w}\Bigl(\frac{3+3\sqrt\xi+\xi+\xi\sqrt\xi+\xi^2}{3\pi(1+\sqrt\xi)}m_W+\tan^2\theta_{\rm w}\frac{2-\sqrt\xi-\xi}{6\pi(1+\sqrt\xi)}m_W+\frac{1}{2\pi}\sqrt{m_{\rm D}^2+m_W^2}\Bigr)+\mathcal O(\vert\vec p\vert^2)$.}
\end{figure}

Next, we discuss the one-loop correction of the $A$$-$$Z$ mixing vertex.
It obtains correction from similar diagrams to the ones in Figs.~\ref{fig:diagrams_broken_prop-4pt} and \ref{fig:diagrams_broken_prop-3pt}.
They add up to
\begin{align}
    \sum\text{diagrams in Figs.~\ref{fig:diagrams_broken_ver-4pt} and \ref{fig:diagrams_broken_ver-3pt}}
        =V_{AZ}\delta_{ij}+\mathcal O(\vert\vec p\vert^2),\qquad
    V_{AZ}
        \coloneq 
        g_3^2\tan\theta_{\rm w}\dfrac{4+\sqrt\xi+\xi}{12\pi(1+\sqrt\xi)}m_W.
    \label{eq:V_AZ}
\end{align}

By combining Eqs.~\eqref{eq:Pi_AA} and \eqref{eq:V_AZ}, we obtain an expression for the $\vec B_Y$ correlation function at the one-loop order in the low-temperature phase,
\begin{align}
    \left\langle B_{Yi}(\vec p)B_{Yj}(-\vec p)\right\rangle'
    &=\epsilon^{ikl}\epsilon^{jmn}ip_ki(-p_m)\Bigl(
        \cos^2\theta_{\rm w}\left\langle A_l(\vec p)A_n(-\vec p)\right\rangle'+\sin\theta_{\rm w}\cos\theta_{\rm w}\left\langle A_l(\vec p)Z_n(-\vec p)\right\rangle'\notag\\
        &\hspace{30mm}+\sin\theta_{\rm w}\cos\theta_{\rm w}\left\langle Z_l(\vec p)A_n(-\vec p)\right\rangle'+\sin^2\theta_{\rm w}\left\langle Z_l(\vec p)Z_n(-\vec p)\right\rangle'
    \Bigr)
    \notag\\
    &=\epsilon^{ikl}\epsilon^{jmn}ip_ki(-p_m)
    \qty[
        \cos^2\theta_{\rm w}\qty(1+\dfrac{\Pi_{AA}(\vert\vec p\vert)}{\vert \vec p\vert^2})
        +2\sin\theta_{\rm w}\cos\theta_{\rm w}\dfrac{V_{AZ}}{m_Z^2}
        +\mathcal O(\vert\vec p\vert^2)
    ]\dfrac{\delta_{ln}}{\vert\vec p\vert^2}
    \notag\\
    &=
    \cos^2\theta_{\rm w}\qty(
        1+\dfrac{11g_3^2\sin^2\theta_{\rm w}}{12\pi m_W}-\dfrac{g_3^2\sin^2\theta_{\rm w}}{24\pi\sqrt{m_{\rm D}^2+m_W^2}}
    )\,P_{ij}(\vert\vec p\vert)+\mathcal O(\vert\vec p\vert^2),
\end{align}
which reproduces the result in Ref.~\cite{Kajantie:1996qd} when we take the magnetostatic limit, $m_{\rm D}\to\infty$.

\subsection{Gauge-independent corrections in the low-temperature regime}
\label{appendix:1loop_g-indep}
In the previous section, we have seen that the one-loop corrections to the photon propagator, $\Pi_{AA}$, and to the photon$-$$Z$ boson mixing vertex, $V_{AZ}$, are gauge-dependent.
The gauge-dependence is troublesome because, naively, $\Pi_{AA}$ would imply the photon wavefunction renormalization and is unphysical, while $V_{AZ}$ would modify the effective weak mixing angle and have physical consequences.

To disentangle the wavefunction renormalization and the physical mixing in a gauge-independent way, we have introduced $\mathcal W$ \cite{Nambu:1977ag} in Eq.~\eqref{eq:calW_def}, a gauge-invariant generalization of $W^3$.
Accordingly, we generalize magnetic fields in the low-temperature regime to define
\begin{align}
    \vec B_{\mathcal A}
        \coloneq\cos\theta_{\rm w}\vec B_{Y}-\sin\theta_{\rm w}\vec B_{\mathcal W},\qquad
    \vec B_{\mathcal Z}
        \coloneq\sin\theta_{\rm w}\vec B_{Y}+\cos\theta_{\rm w}\vec B_{\mathcal W}.
    \label{eq:cal_magneticfields_def}
\end{align}

\begin{figure}[H]\centering
    \includegraphics[keepaspectratio, width=1.\textwidth]{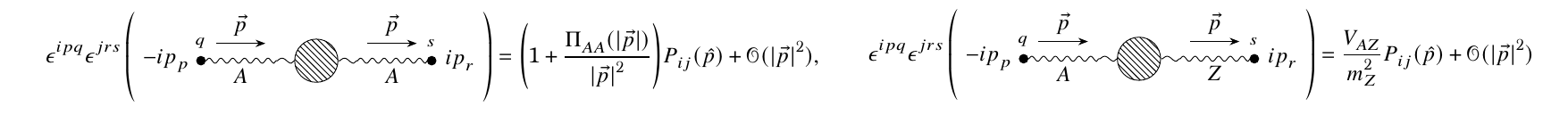}
    \caption{\label{fig:diagrams_GI_0-0}One-loop corrections in the {\it broken} phase derived in Sec.~\ref{sec:1loop_broken}.}
\end{figure}

\begin{figure}[H]\centering
    \includegraphics[keepaspectratio, width=1.\textwidth]{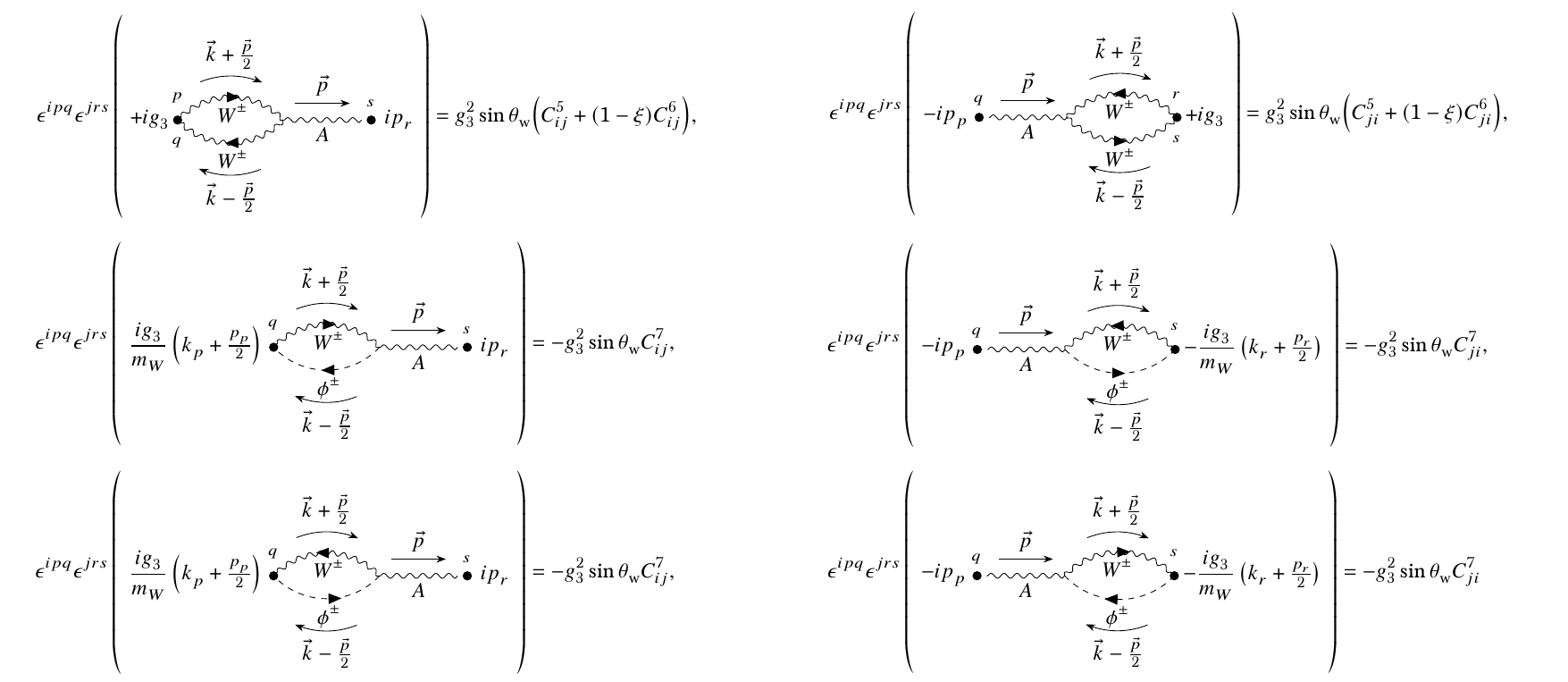}
    \caption{\label{fig:diagrams_GI_1-A}One-loop contributions from the correlation between the $\mathcal O(g_3)$ terms in Eq.~\eqref{eq:GI_W} and the photon field. Those diagrams in each column (Left and Right) add up to $g_3^2\sin\theta_{\rm w}\frac{-1+5\sqrt\xi+2\xi}{12\pi(1+\sqrt\xi)}\frac{P_{ij}(\hat p)}{m_W}+\mathcal O(\vert\vec p\vert^2)$.}
\end{figure}

\begin{figure}[H]\centering
    \includegraphics[keepaspectratio, width=1.\textwidth]{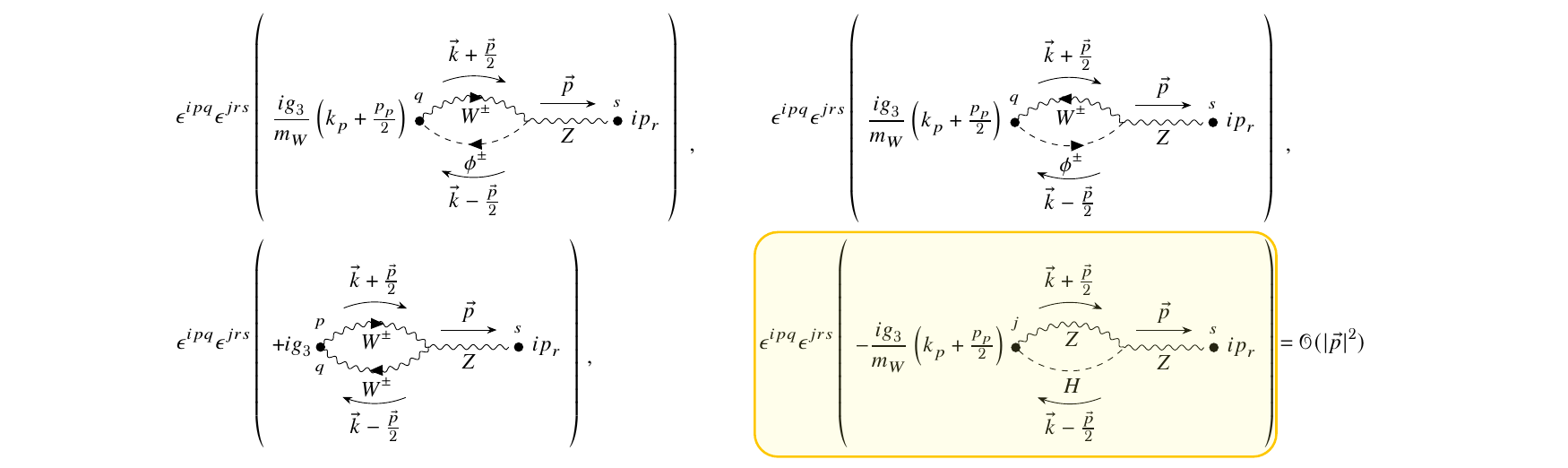}
    \caption{\label{fig:diagrams_GI_1-Z}Unshaded: one-loop contributions from the correlation between the $\mathcal O(g_3)$ terms in Eq.~\eqref{eq:GI_W} and $Z$ boson. With the alternative definition in Eq.~\eqref{eq:calWalt_def}, the yellow-shaded diagram should also be included.}
\end{figure}

\begin{figure}[H]\centering\vspace{-5mm}
    \includegraphics[keepaspectratio, width=.93\textwidth]{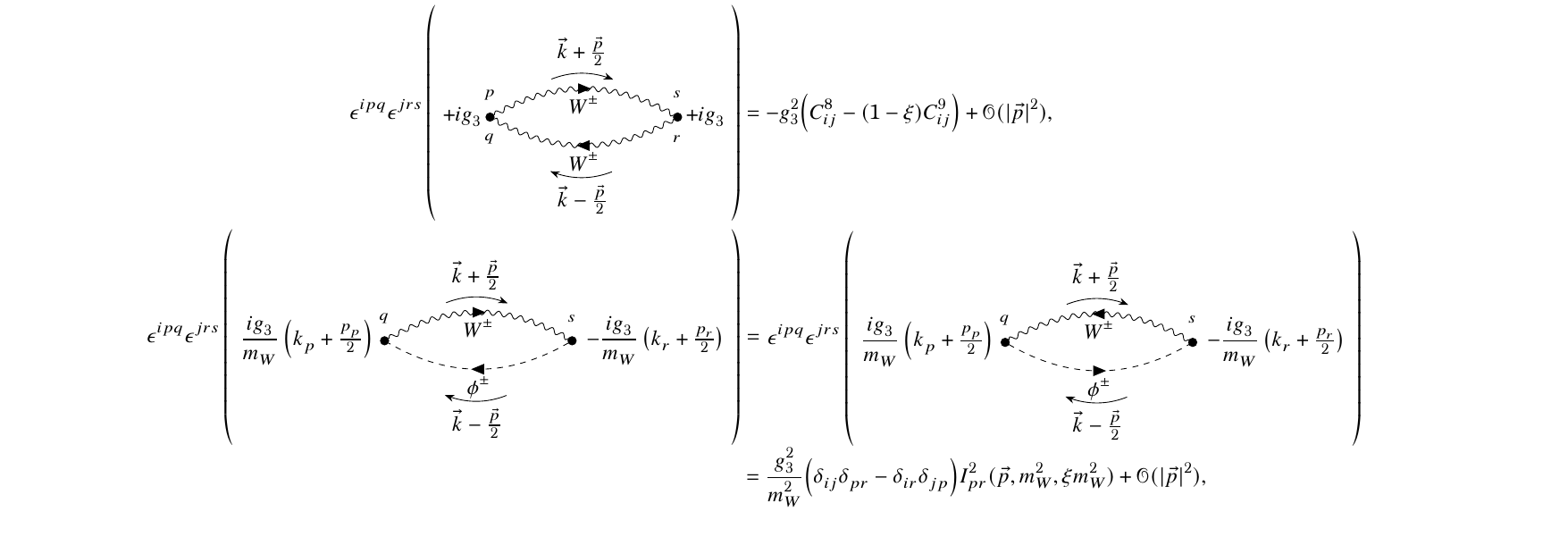}\vspace{-2.mm}
    \includegraphics[keepaspectratio, width=.93\textwidth]{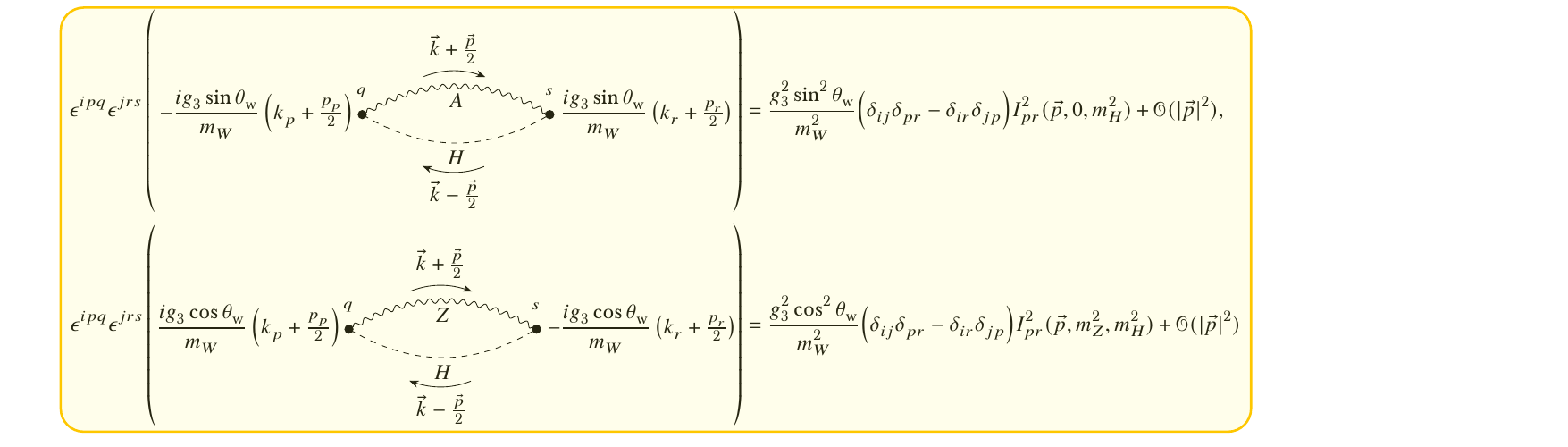}
    \caption{\label{fig:diagrams_GI_1-1}Unshaded: one-loop contributions from the correlation between pairs of the $\mathcal O(g_3)$ terms in Eq.~\eqref{eq:GI_W}, which add up to $-\frac{g_3^2\delta_{ij}}{4\pi m_W}+\mathcal O(\vert\vec p\vert^2)$. With the alternative definition in Eq.~\eqref{eq:calWalt_def}, the yellow-shaded diagrams are also included, and the total contribution in this set of diagrams becomes $-\frac{g_3^2\delta_{ij}}{12\pi m_W}\qty(3+\frac{2m_H}{m_W}+\frac{2\cos\theta_{\rm w}}{1+\cos\theta_{\rm w}\frac{m_H}{m_W}})+\mathcal O(\vert\vec p\vert^2)$.}
\end{figure}

\begin{figure}[H]\centering
    \includegraphics[keepaspectratio, width=.93\textwidth]{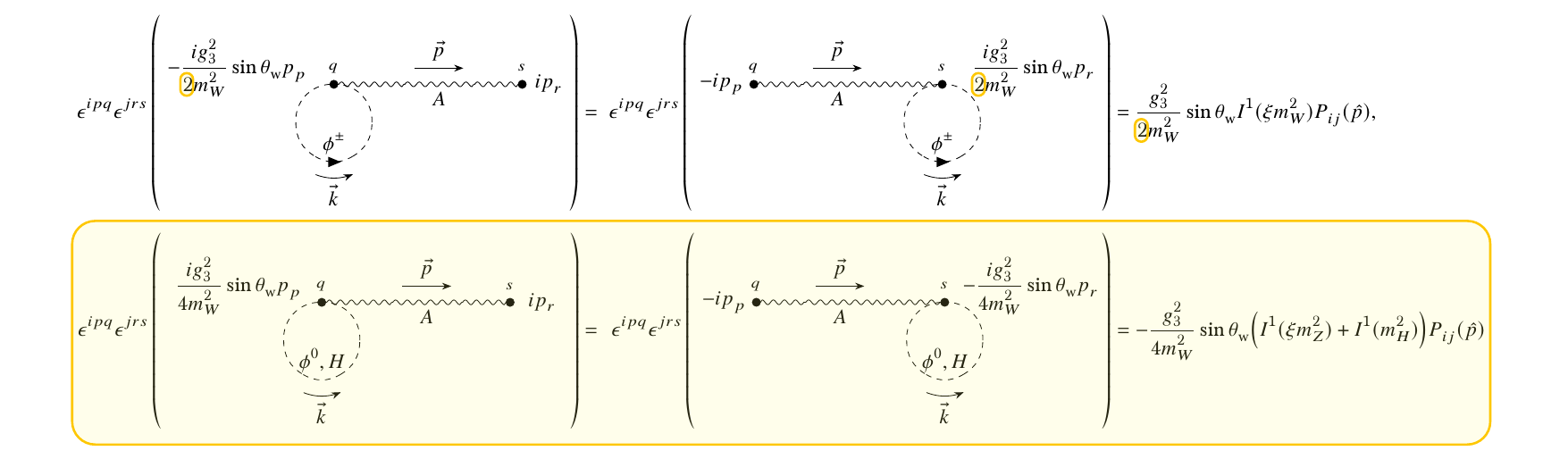}
    \caption{\label{fig:diagrams_GI_2-A}Unshaded: one-loop contributions from the correlation between the $\mathcal O(g_3^2)$ terms in Eq.~\eqref{eq:GI_W} and the photon field. The diagram in each column (Left and Right) is evaluated $-\frac{g_3^2\sin\theta_{\rm w}\sqrt\xi}{4\pi m_W}P_{ij}(\hat p)$. With the alternative definition in Eq.~\eqref{eq:calWalt_def}, the yellow-shaded numbers and diagrams are also included, and the total contribution of the diagrams in each column becomes $-\frac{g_3^2\sin\theta_{\rm w}}{16\pi}\qty(2\sqrt\xi-\frac{\sqrt\xi}{\cos\theta_{\rm w}}-\frac{m_H}{m_W})\frac{P_{ij}(\hat p)}{m_W}+\mathcal O(\vert\vec p\vert^2)$.}
\end{figure}

\begin{figure}[H]\centering
    \includegraphics[keepaspectratio, width=.93\textwidth]{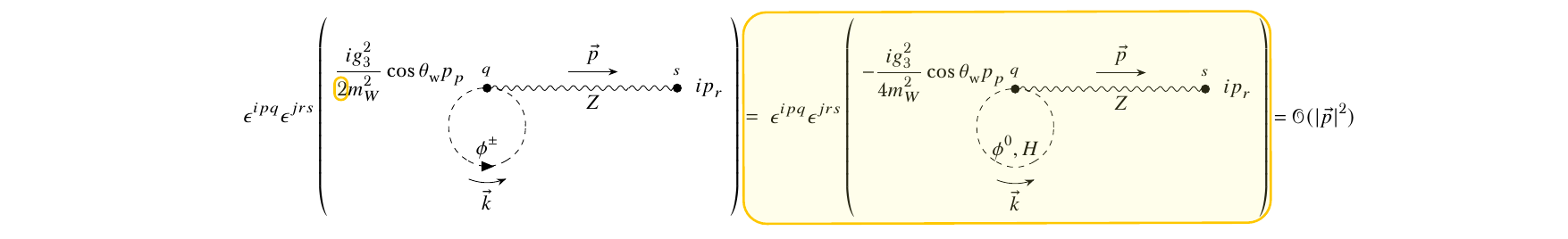}
    \caption{\label{fig:diagrams_GI_2-Z}Unshaded: one-loop contributions from the correlation between the $\mathcal O(g_3^2)$ terms in Eq.~\eqref{eq:GI_W} and $Z$ boson. With the alternative definition in Eq.~\eqref{eq:calWalt_def}, the yellow-shaded numbers and diagrams are also included.}
\end{figure}

\begin{figure}[H]\centering
    \includegraphics[keepaspectratio, width=1.\textwidth]{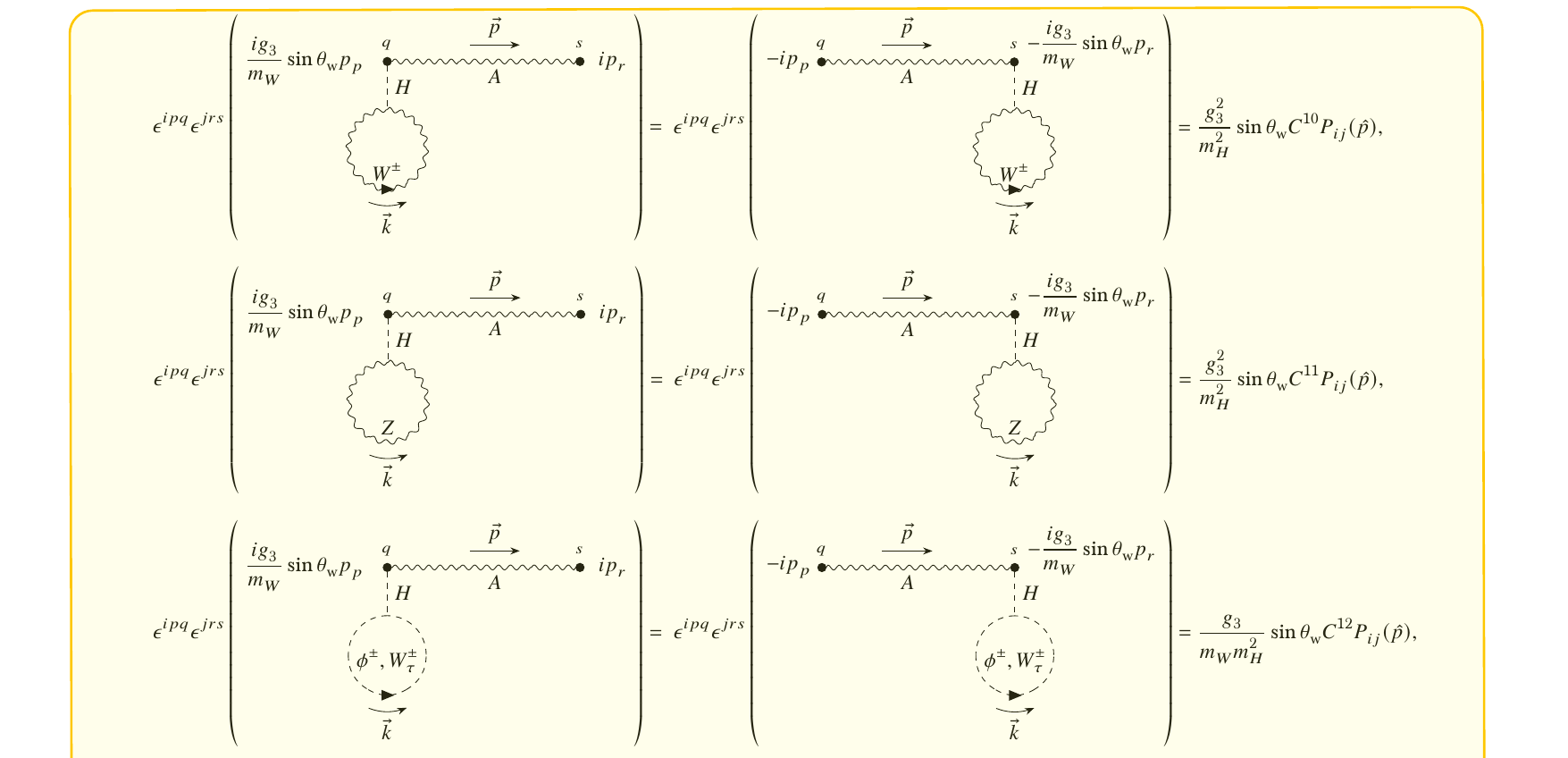}\vspace{-0.3mm}
    \includegraphics[keepaspectratio, width=1.\textwidth]{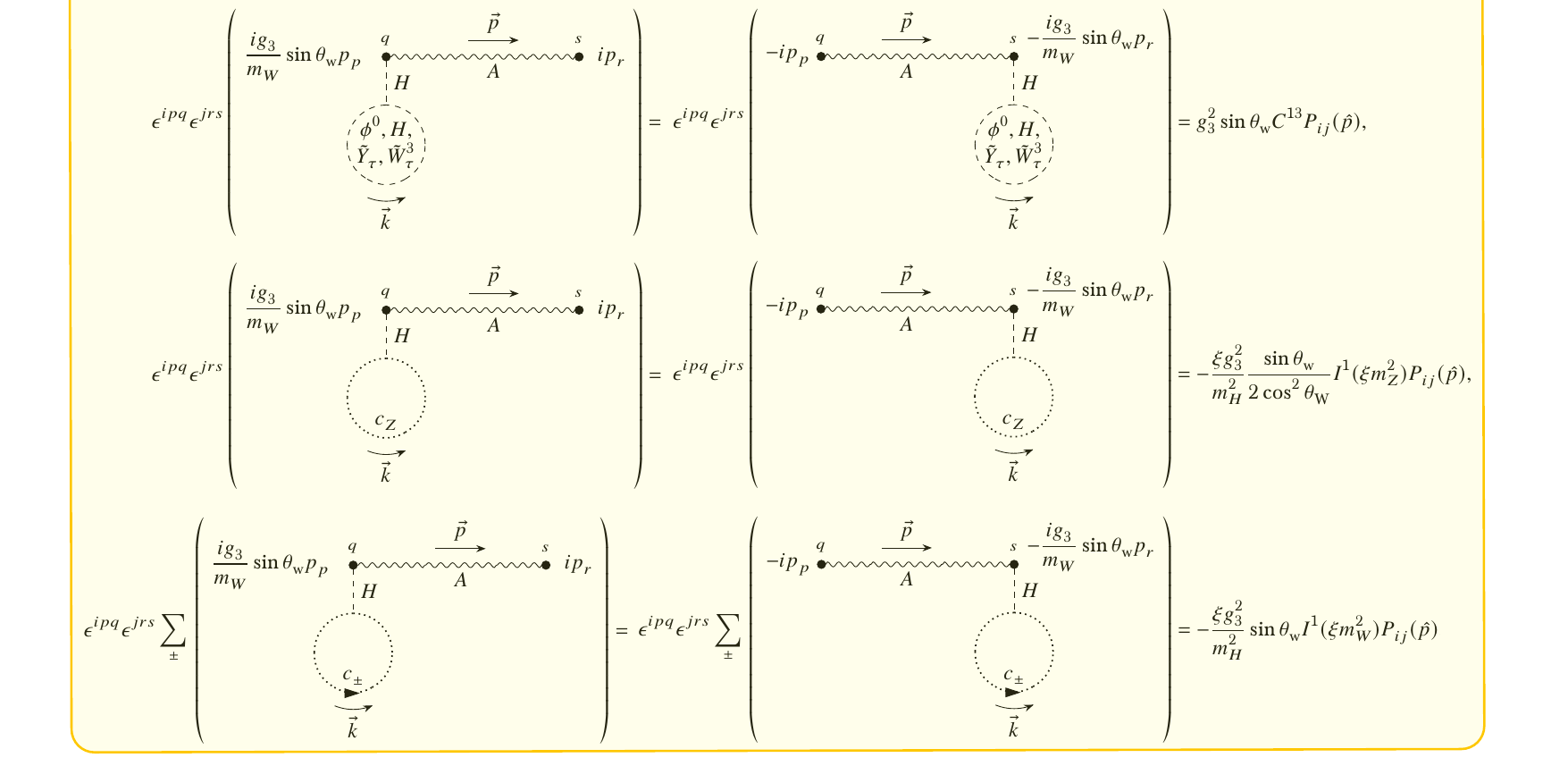}
    \caption{\label{fig:diagrams_GI_tp-A}One-loop tadpole contributions in the correlation between the $\mathcal O(g_3)$ terms and the photon field with the alternative definition in Eq.~\eqref{eq:calWalt_def}. Those diagrams in each column (Left and Right) add up to $-\frac{g_3^2\sin\theta_{\rm w}}{16\pi}\biggl(\qty(8+\frac{4}{\cos^3\theta_{\rm w}}+4\sqrt{1+\frac{m_{\rm D}^2}{m_W^2}})\frac{m_W^2}{m_H^2}+\frac{3m_H}{m_W}+2\sqrt\xi+\frac{\sqrt\xi}{\cos\theta_{\rm w}}$ $+2(\cos\theta_\tau \tan\theta_{\rm w}-\sin\theta_\tau)^2\frac{m_W\tilde m_-}{m_H^2}+2(\sin\theta_\tau \tan\theta_{\rm w}+\cos\theta_\tau)^2\frac{m_W\tilde m_+}{m_H^2}\biggr)\frac{P_{ij}(\hat p)}{m_W}+\mathcal O(g_3^4)$.}
\end{figure}

\begin{figure}[H]\centering
    \includegraphics[keepaspectratio, width=1.\textwidth]{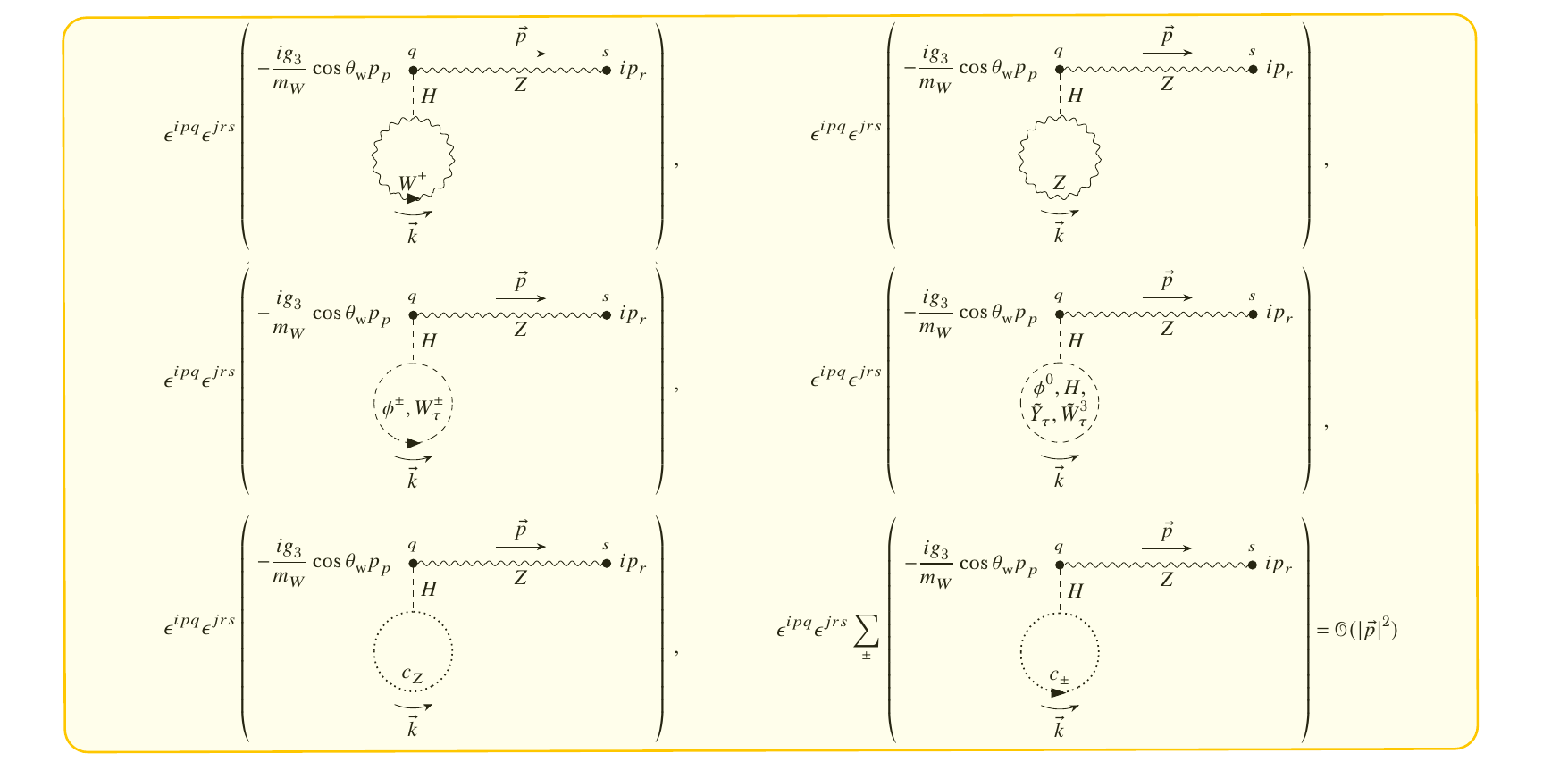}
    \caption{\label{fig:diagrams_GI_tp-Z}One-loop tadpole contributions in the correlation between the $\mathcal O(g_3)$ terms and $Z$ boson with the alternative definition in Eq.~\eqref{eq:calWalt_def}.}
\end{figure}

Our task in this section is to compute $\langle B_{\mathcal Ai}(\vec p)B_{\mathcal Aj}(\vec q)\rangle$ and $\langle B_{\mathcal Ai}(\vec p)B_{\mathcal Zj}(\vec q)\rangle$ at the one-loop level.
The gauge-independent magnetic fields, $B_{\mathcal A}$ and $B_{\mathcal Z}$, include contributions from
\begin{align}
    \mathcal W_{ij}
        =&\;\partial_iW_j^3-\partial_jW_i^3\notag\\
        &+g_3\Biggl[i\left(W^-_iW^+_j-W^+_iW^-_j\right)
            -\dfrac{\phi^-}{m_W}\left(\partial_i W_j^+-\partial_j W_i^+\right)
            -\dfrac{\phi^+}{m_W}\left(\partial_i W_j^--\partial_j W_i^-\right)\Biggr]\notag\\
        &+g_3^2\Biggl[\dfrac{i\phi^-}{m_W}\left(W_i^3W_j^+-W_i^+W_j^3\right)
            -\dfrac{i\phi^+}{m_W}\left(W_i^3W_j^--W_i^-W_j^3\right)
            -\dfrac{\phi^-\phi^+}{m_W^2}\left(\partial_i W_j^3-\partial_j W_i^3\right)\notag\\
        &\hspace{10mm}+\left(\dfrac{H}{2m_W^2}-\dfrac{i\phi^0}{2m_W^2}\right)\phi^-\left(\partial_i W_j^+-\partial_j W_i^+\right)
            +\left(\dfrac{H}{2m_W^2}+\dfrac{i\phi^0}{2m_W^2}\right)\phi^+\left(\partial_i W_j^--\partial_j W_i^-\right)\Biggr]
        +\mathcal O(g_3^3),
    \label{eq:GI_W}
\end{align}
where our discussion in Sec.~\ref{sec:1loop_broken} (summarized in Fig.~\ref{fig:diagrams_GI_0-0}) corresponds to considering only the first two terms.
The correction from the other terms includes diagrams in Figs.~\ref{fig:diagrams_GI_1-A}, \ref{fig:diagrams_GI_1-Z}, \ref{fig:diagrams_GI_1-1}, \ref{fig:diagrams_GI_2-A}, \ref{fig:diagrams_GI_2-Z}, \ref{fig:diagrams_GI_tp-A}, and \ref{fig:diagrams_GI_tp-Z}.
\newpage
\vspace{-5mm}
\vspace{-5mm}
\newpage
\newpage

By combining the evaluation of those diagrams in Figs.~\ref{fig:diagrams_GI_0-0}, \ref{fig:diagrams_GI_1-A}, \ref{fig:diagrams_GI_1-Z}, \ref{fig:diagrams_GI_1-1}, \ref{fig:diagrams_GI_2-A}, and \ref{fig:diagrams_GI_2-Z}, we obtain
\begin{align}
    \left\langle B_{\mathcal Ai}(\vec p)B_{\mathcal Aj}(-\vec p)\right\rangle'
        &=
        (\text{Left diagram in Fig.~\ref{fig:diagrams_GI_0-0}})
        -\sin\theta_{\rm w}(\text{Fig.~\ref{fig:diagrams_GI_1-A}}+\text{Fig.~\ref{fig:diagrams_GI_2-A}})
        +\sin^2\theta_{\rm w}(\text{Fig.~\ref{fig:diagrams_GI_1-1}})
        \notag\\
        &=
        R_{\mathcal{AA}}P_{ij}(\hat p)+S_{\mathcal{AA}}\delta_{ij}+\mathcal O(\vert \vec p\vert^2),\\
    \left\langle B_{\mathcal Ai}(\vec p)B_{\mathcal Zj}(-\vec p)\right\rangle'
        &=
        (\text{Right diagram in Fig.~\ref{fig:diagrams_GI_0-0}})
        -\sin\theta_{\rm w}(\text{Fig.~\ref{fig:diagrams_GI_1-Z}}+\text{Fig.~\ref{fig:diagrams_GI_2-Z}})\notag\\
        &\hspace{5mm}+\cos\theta_{\rm w}(\text{diagrams in the right columns in Figs.~\ref{fig:diagrams_GI_1-A} and \ref{fig:diagrams_GI_2-A}})\notag\\
        &\hspace{5mm}
        -\sin\theta_{\rm w}\cos\theta_{\rm w}(\text{Fig.~\ref{fig:diagrams_GI_1-1}})
        \notag\\
        &=
        R_{\mathcal{AZ}}P_{ij}(\hat p)+S_{\mathcal{AZ}}\delta_{ij}+\mathcal O(\vert \vec p\vert^2)
        ,
\end{align}
where
\begin{align}
    R_{\mathcal{AA}}
        &=1+\qty(\dfrac{5}{12}-\frac{1}{24\sqrt{1+m_{\rm D}^2/m_W^2}})\dfrac{g_3^2\sin^2\theta_{\rm w}}{\pi m_W},\\[1.em]
    R_{\mathcal{AZ}}
        &=\dfrac{g_3^2\sin\theta_{\rm w}\cos\theta_{\rm w}}{4\pi m_W},\\[1.em]
    S_{\mathcal{AA}}
        &=-\dfrac{g_3^2\sin^2\theta_{\rm w}}{4\pi m_W},\\[1.em]
    S_{\mathcal{AZ}}
        &=\dfrac{g_3^2\sin\theta_{\rm w}\cos\theta_{\rm w}}{4\pi m_W}.
\end{align}

\subsection{Ambiguity of the definition of $\mathcal W$}
\label{appendix:ambiguity}
In Sec.~\ref{sec:unconfined}, we have introduced the gauge-independent ${\rm SU}(2)_{\mathrm L}$ field strength $\mathcal W$ and defined the unconfined magnetic field as a mixture of $\mathcal W$ and $Y$ operators.
We believe that our definition is minimal and the most natural one, but there still remains an ambiguity to define $\mathcal W$.
Note, however, that different definitions should be consistent if one could rigorously calculate physical quantities such as the baryon number generated from decaying magnetic helicity.
Here, we discuss how the ambiguity of the definition of $\mathcal W$ affects the estimate of its mixing with $Y$ during the crossover.

First, we introduce an alternative definition of $\mathcal W$ \cite{Nambu:1977ag},
\begin{align}
    \mathcal W^{\rm alt}_{ij}
        \coloneq&-\dfrac{\Phi^\dagger\sigma^a\Phi}{v_3^2/2}W_{ij}^a\\
        =&\;\partial_iW_j^3-\partial_jW_i^3\notag\\
        &+g_3\Biggl[i\!\left(W^-_iW^+_j-W^+_iW^-_j\right)
            +\dfrac{H}{m_W}\!\left(\partial_i W_j^3-\partial_j W_i^3\right)
            -\dfrac{\phi^-}{m_W}\!\left(\partial_i W_j^+-\partial_j W_i^+\right)
            -\dfrac{\phi^+}{m_W}\!\left(\partial_i W_j^--\partial_j W_i^-\right)\Biggr]\notag\\
        &+g_3^2\Biggl[\dfrac{iH}{m_W}\left(W^-_iW^+_j-W^+_iW^-_j\right)
            +\dfrac{i\phi^-}{m_W}\left(W_i^3W_j^+-W_i^+W_j^3\right)
            -\dfrac{i\phi^+}{m_W}\left(W_i^3W_j^--W_i^-W_j^3\right)\notag\\
        &\hspace{10mm}+\left(\dfrac{H^2}{4m_W^2}+\dfrac{(\phi^0)^2}{4m_W^2}-\dfrac{\phi^-\phi^+}{2m_W^2}\right)\left(\partial_i W_j^3-\partial_j W_i^3\right)\notag\\
        &\hspace{10mm}-\left(\dfrac{H}{2m_W^2}+\dfrac{i\phi^0}{2m_W^2}\right)\phi^-\left(\partial_i W_j^+-\partial_j W_i^+\right)
            -\left(\dfrac{H}{2m_W^2}-\dfrac{i\phi^0}{2m_W^2}\right)\phi^+\left(\partial_i W_j^--\partial_j W_i^-\right)\Biggr]
        +\mathcal O(g_3^3),
    \label{eq:calWalt_def}
\end{align}
which is equivalent to $\mathcal W_{ij}$ and $W^3_{ij}$ at the tree-level in the low-temperature phase.
We then modify the definition of magnetic fields in the low-temperature regime, Eq.~\eqref{eq:cal_magneticfields_def}, as $\vec B_{\mathcal A^{\rm alt}}\coloneq\cos\theta_{\rm w}\vec B_{Y}-\sin\theta_{\rm w}\vec B_{\mathcal W^{\rm alt}}$ and $\vec B_{\mathcal Z^{\rm alt}}\coloneq\sin\theta_{\rm w}\vec B_{Y}+\cos\theta_{\rm w}\vec B_{\mathcal W^{\rm alt}}$.
With these definitions, we compute the mixing between $\mathcal W^{\rm alt}$ and $Y$ at the one-loop level.
The relevant diagrams are shown in Figs.~\ref{fig:diagrams_GI_0-0}--\ref{fig:diagrams_GI_tp-Z}, and we obtain
\begin{align}
    \left\langle B_{\mathcal A^{\rm alt}i}(\vec p)B_{\mathcal A^{\rm alt}j}(-\vec p)\right\rangle'
        &=
        (\text{Left diagram in Fig.~\ref{fig:diagrams_GI_0-0}})
        -\sin\theta_{\rm w}(\text{Fig.~\ref{fig:diagrams_GI_1-A}}+\text{Fig.~\ref{fig:diagrams_GI_2-A}}+\text{Fig.~\ref{fig:diagrams_GI_tp-A}})
        +\sin^2\theta_{\rm w}(\text{Fig.~\ref{fig:diagrams_GI_1-1}})
        \notag\\
        &=
        R_{\mathcal A^{\rm alt}\mathcal A^{\rm alt}}P_{ij}(\hat p)+S_{\mathcal A^{\rm alt}\mathcal A^{\rm alt}}\delta_{ij}+\mathcal O(\vert \vec p\vert^2)
        ,\\
    \left\langle B_{\mathcal A^{\rm alt}i}(\vec p)B_{\mathcal Z^{\rm alt}j}(-\vec p)\right\rangle'
        &=
        (\text{Right diagram in Fig.~\ref{fig:diagrams_GI_0-0}})
        -\sin\theta_{\rm w}(\text{Fig.~\ref{fig:diagrams_GI_1-Z}}+\text{Fig.~\ref{fig:diagrams_GI_2-Z}}+\text{Fig.~\ref{fig:diagrams_GI_tp-Z}})\notag\\
        &\hspace{5mm}+\cos\theta_{\rm w}(\text{diagrams in the right columns in Figs.~\ref{fig:diagrams_GI_1-A}, \ref{fig:diagrams_GI_2-A}, and \ref{fig:diagrams_GI_tp-A}})\notag\\
        &\hspace{5mm}
        -\sin\theta_{\rm w}\cos\theta_{\rm w}(\text{Fig.~\ref{fig:diagrams_GI_1-1}})
        \notag\\
        &=
        R_{\mathcal A^{\rm alt}\mathcal Z^{\rm alt}}P_{ij}(\hat p)+S_{\mathcal A^{\rm alt}\mathcal Z^{\rm alt}}\delta_{ij}+\mathcal O(\vert \vec p\vert^2)
        ,
\end{align}
where
\begin{align}
    R_{\mathcal A^{\rm alt}\mathcal A^{\rm alt}}
        &=1+\Biggl(\dfrac{5}{12}-\frac{1}{24\sqrt{1+m_{\rm D}^2/m_W^2}}+\dfrac{m_H}{4m_W}+\Biggl(1+\frac{1}{2\cos^3\theta_{\rm w}}+\dfrac{1}{2}\sqrt{1+m_{\rm D}^2/m_W^2}\,\Biggr)\frac{m_W^2}{m_H^2}\notag\\
        &\hspace{12mm}+\qty(\cos\theta_\tau \tan\theta_{\rm w}-\sin\theta_\tau)^2\frac{m_W\tilde m_-}{4m_H^2}+\qty(\sin\theta_\tau \tan\theta_{\rm w}+\cos\theta_\tau)^2\frac{m_W\tilde m_+}{4m_H^2}\Biggr)\dfrac{g_3^2\sin^2\theta_{\rm w}}{\pi m_W},
    \label{eq:RAA_alt}\\[1.em]
    R_{\mathcal A^{\rm alt}\mathcal Z^{\rm alt}}
        &=\Biggl(\dfrac{1}{4}-\dfrac{m_H}{8m_W}-\qty(\frac{1}{2}+\frac{1}{4\cos^3\theta_{\rm w}}+\frac{1}{4}\sqrt{1+m_{\rm D}^2/m_W^2}\,)\frac{m_W^2}{m_H^2}\notag\\
        &\hspace{6mm}-(\cos\theta_\tau \tan\theta_{\rm w}-\sin\theta_\tau)^2\frac{m_W\tilde m_-}{8m_H^2}-(\sin\theta_\tau \tan\theta_{\rm w}+\cos\theta_\tau)^2\frac{m_W\tilde m_+}{8m_H^2}\Biggr)\dfrac{g_3^2\sin\theta_{\rm w}\cos\theta_{\rm w}}{\pi m_W},
    \label{eq:RAZ_alt}\\[1.em]
    S_{\mathcal A^{\rm alt}\mathcal A^{\rm alt}}
        &=-\qty(\dfrac{1}{4}+\dfrac{m_H}{6m_W}+\dfrac{\cos\theta_{\rm w}}{6+6\cos\theta_{\rm w}\frac{m_H}{m_W}})\dfrac{g_3^2\sin^2\theta_{\rm w}}{\pi m_W},
    \label{eq:SAA_alt}\\[1.em]
    S_{\mathcal A^{\rm alt}\mathcal Z^{\rm alt}}
        &=\qty(\dfrac{1}{4}+\dfrac{m_H}{6m_W}+\dfrac{\cos\theta_{\rm w}}{6+6\cos\theta_{\rm w}\frac{m_H}{m_W}})\dfrac{g_3^2\sin\theta_{\rm w}\cos\theta_{\rm w}}{\pi m_W}.
    \label{eq:SAZ_alt}
\end{align}

Generalizing these results to the ones with a more general definition of $\mathcal W$, namely
\begin{align}
    \mathcal W^f_{ij}
        \coloneq f\qty(2\Phi^\dagger\Phi/v_3^2)\mathcal W_{ij}
\end{align}
for a given function $f$ such that $f(1)=1$, is rather an easy task.
Since we have
\begin{align}
    \dfrac{\Phi^\dagger\Phi}{v_3^2/2}
        =1+\Delta+\mathcal O(g_3^3),\qquad
    \Delta
        \coloneq \dfrac{g_3}{m_W}H+\dfrac{g_3^2}{4m_W^2}\qty(
            H^2+(\phi^0)^2+2\phi^-\phi^+
        ),
\end{align}
we obtain relations
\begin{align}
    \mathcal W^f_{ij}
        =&\qty(1+f'(1)\Delta+\dfrac{g_3^2f''(1)}{2m_W^2}H^2)
        \mathcal W_{ij}+\mathcal O(g_3^3),\\
    \vec B_{\mathcal A^f}
        \coloneq&\cos\theta_{\rm w}\vec B_{Y}-\sin\theta_{\rm w}\vec B_{\mathcal W^f}
        =\vec B_{\mathcal A}-\sin\theta_{\rm w}\qty(f'(1)\Delta+\dfrac{g_3^2f''(1)}{2m_W^2}H^2)\vec B_{\mathcal W}+\mathcal O(g_3^3),\\
    \vec B_{\mathcal Z^f}
        \coloneq&\sin\theta_{\rm w}\vec B_{Y}+\cos\theta_{\rm w}\vec B_{\mathcal W^f}
        =\vec B_{\mathcal Z}+\cos\theta_{\rm w}\qty(f'(1)\Delta+\dfrac{g_3^2f''(1)}{2m_W^2}H^2)\vec B_{\mathcal W}+\mathcal O(g_3^3).
\end{align}
As for the modification proportional to $f'(1)$, the consequence is almost explicit in Eqs.~\eqref{eq:RAA_alt}--\eqref{eq:SAZ_alt}. 
Namely, the yellow-shaded contribution in the diagrams in Figs.~\ref{fig:diagrams_GI_0-0}, \ref{fig:diagrams_GI_1-A}, \ref{fig:diagrams_GI_1-Z}, \ref{fig:diagrams_GI_2-A}, \ref{fig:diagrams_GI_2-Z}, \ref{fig:diagrams_GI_tp-A}, and \ref{fig:diagrams_GI_tp-Z} is multiplied by a factor $f'(1)$, and the one from the diagrams in Fig.~\ref{fig:diagrams_GI_1-1} is multiplied by $(f'(1))^2$.
As for the remaining modification, we need to include the correction from only the diagrams in Figs.~\ref{fig:diagrams_GI_2-A} and \ref{fig:diagrams_GI_2-Z} where $H$ runs in the loop.
Consequently, we obtain
\begin{align}
    \left\langle B_{\mathcal A^{f}i}(\vec p)B_{\mathcal A^{f}j}(-\vec p)\right\rangle'
        &=
        R_{\mathcal A^{f}\mathcal A^{f}}P_{ij}(\hat p)+S_{\mathcal A^{f}\mathcal A^{f}}\delta_{ij}+\mathcal O(\vert \vec p\vert^2)
        ,\\
    \left\langle B_{\mathcal A^{f}i}(\vec p)B_{\mathcal Z^{f}j}(-\vec p)\right\rangle'
        &=
        R_{\mathcal A^{f}\mathcal Z^{f}}P_{ij}(\hat p)+S_{\mathcal A^{f}\mathcal Z^{f}}\delta_{ij}+\mathcal O(\vert \vec p\vert^2)
        ,
\end{align}
where
\begin{align}
    R_{\mathcal A^{f}\mathcal A^{f}}
        &=R_{\mathcal AA}+f'(1)\Delta^1_{R_{\mathcal AA}}+(f'(1))^2\Delta^2_{R_{\mathcal AA}}+f''(1)\Delta^3_{R_{\mathcal AA}},
    \label{eq:RAA_f}\\[1.em]
    R_{\mathcal A^{f}\mathcal Z^{f}}
        &=R_{\mathcal AZ}+f'(1)\Delta^1_{R_{\mathcal AZ}}+(f'(1))^2\Delta^2_{R_{\mathcal AZ}}+f''(1)\Delta^3_{R_{\mathcal AZ}},
    \label{eq:RAZ_f}\\[1.em]
    S_{\mathcal A^{f}\mathcal A^{f}}
        &=S_{\mathcal AA}+f'(1)\Delta^1_{S_{\mathcal AA}}+(f'(1))^2\Delta^2_{S_{\mathcal AA}}+f''(1)\Delta^3_{S_{\mathcal AA}},
    \label{eq:SAA_f}\\[1.em]
    S_{\mathcal A^{f}\mathcal Z^{f}}
        &=S_{\mathcal AZ}+f'(1)\Delta^1_{S_{\mathcal AZ}}+(f'(1))^2\Delta^2_{S_{\mathcal AZ}}+f''(1)\Delta^3_{S_{\mathcal AZ}},
    \label{eq:SAZ_f}
\end{align}
and
\begin{align}
    \Delta^1_{R_{\mathcal AA}}
        &=\Biggl(\dfrac{m_H}{4m_W}+\Biggl(1+\frac{1}{2\cos^3\theta_{\rm w}}+\dfrac{1}{2}\sqrt{1+m_{\rm D}^2/m_W^2}\,\Biggr)\frac{m_W^2}{m_H^2}\notag\\
        &\hspace{7mm}+\qty(\cos\theta_\tau \tan\theta_{\rm w}-\sin\theta_\tau)^2\frac{m_W\tilde m_-}{4m_H^2}+\qty(\sin\theta_\tau \tan\theta_{\rm w}+\cos\theta_\tau)^2\frac{m_W\tilde m_+}{4m_H^2}\Biggr)\dfrac{g_3^2\sin^2\theta_{\rm w}}{\pi m_W},\\
    \Delta^2_{R_{\mathcal AA}}
        &=0,\\
    \Delta^3_{R_{\mathcal AA}}
        &=-\dfrac{g_3^2\sin^2\theta_{\rm w}m_H}{4\pi m_W^2},\\
    \Delta^1_{R_{\mathcal AZ}}
        &=\Biggl(-\dfrac{m_H}{8m_W}-\qty(\frac{1}{2}+\frac{1}{4\cos^3\theta_{\rm w}}+\frac{1}{4}\sqrt{1+m_{\rm D}^2/m_W^2}\,)\frac{m_W^2}{m_H^2}\notag\\
        &\hspace{7mm}-(\cos\theta_\tau \tan\theta_{\rm w}-\sin\theta_\tau)^2\frac{m_W\tilde m_-}{8m_H^2}-(\sin\theta_\tau \tan\theta_{\rm w}+\cos\theta_\tau)^2\frac{m_W\tilde m_+}{8m_H^2}\Biggr)\dfrac{g_3^2\sin\theta_{\rm w}\cos\theta_{\rm w}}{\pi m_W},\\
    \Delta^2_{R_{\mathcal AZ}}
        &=0,\\
    \Delta^3_{R_{\mathcal AZ}}
        &=\dfrac{g_3^2\sin\theta_{\rm w}\cos\theta_{\rm w}m_H}{8\pi m_W^2},\\
    \Delta^1_{S_{\mathcal AA}}
        &=0,\\
    \Delta^2_{S_{\mathcal AA}}
        &=-\qty(\dfrac{m_H}{m_W}+\dfrac{\cos\theta_{\rm w}}{1+\cos\theta_{\rm w}\frac{m_H}{m_W}})\dfrac{g_3^2\sin^2\theta_{\rm w}}{6\pi m_W},\\
    \Delta^3_{S_{\mathcal AA}}
        &=0,\\
    \Delta^1_{S_{\mathcal AZ}}
        &=0,\\
    \Delta^2_{S_{\mathcal AZ}}
        &=\qty(\dfrac{m_H}{m_W}+\dfrac{\cos\theta_{\rm w}}{1+\cos\theta_{\rm w}\frac{m_H}{m_W}})\dfrac{g_3^2\sin\theta_{\rm w}\cos\theta_{\rm w}}{6\pi m_W},\\
    \Delta^3_{S_{\mathcal AZ}}
        &=0.
\end{align}

\section{Formulae for the loop integrals}
\label{sec:formulae}
In this appendix, we list useful formulae for computing the momentum integration of loop diagrams.

For $a,b,c\geq0$, we have
\begin{align}\hspace{-1mm}
    \int_0^1\dd x\qty(\dfrac{1}{2}-x)\!\sqrt{x(1-x)a^2+b^2x+(1-x)c^2}
        &=\begin{cases}
            -\dfrac{(b-c)(b^2+3bc+c^2)}{15(b+c)^2}+\mathcal O(a^2),\\
            0,\hspace{23.5mm}\text{when $b=c=0$},
        \end{cases}\\
    \int_0^1\dd x\sqrt{x(1-x)a^2+b^2x+(1-x)c^2}
        &=\begin{cases}
            \dfrac{2(b^2+bc+c^2)}{3(b+c)}+\dfrac{2(b^2+3bc+c^2)}{15(b+c)^3}a^2+{\mathcal O}(a^4),\\[10pt]
            \dfrac{\pi a}{8},\hspace{40.5mm}\text{when $b=c=0$},
        \end{cases}\\
    \int_0^1\dd x\dfrac{1}{\sqrt{x(1-x)a^2+b^2x+(1-x)c^2}}
        &=\begin{cases}
            \dfrac{2}{b+c}-\dfrac{2a^2}{3(b+c)^3}+{\mathcal O}(a^4),\\[10pt]
            \dfrac{\pi}{a},\hspace{12mm}\text{when $b=c=0$},
        \end{cases}\\
    \int_0^1\dd x\dfrac{\frac{1}{2}-x}{\sqrt{x(1-x)a^2+b^2x+(1-x)c^2}}
        &=\begin{cases}
            \dfrac{b-c}{3(b+c)^2}-\dfrac{b-c}{5(b+c)^4}a^2+{\mathcal O}(a^4),\\[10pt]
            0,\hspace{23mm}\text{when $b=c=0$},
        \end{cases}\\
    \int_0^1\dd x\dfrac{\left(\frac{1}{2}-x\right)^2}{\sqrt{x(1-x)a^2+b^2x+(1-x)c^2}}
        &=\begin{cases}
            \dfrac{7b^2+6bc+7c^2}{30(b+c)^3}-\dfrac{19b^2-10bc+19c^2}{210(b+c)^5}a^2+{\mathcal O}(a^4),\\[10pt]
            \dfrac{\pi}{8a},\hspace{47mm}\text{when $b=c=0$}.
        \end{cases}
\end{align}
By using these formulae together with 
\begin{align}
    \int\dfrac{\dd^3k}{(2\pi)^3}\dfrac{1}{(\vert\vec k\vert^2+\mu^2)^n}
        &=\dfrac{\Gamma\left(n-\frac{3}{2}\right)}{(4\pi)^{\frac{3}{2}}\Gamma(n)}\mu^{3-2n},\hspace{13mm} n=1,2,\cdots,
\end{align}
and the Feynman parametrization \cite{Peskin:1995ev}, we obtain
\begin{align}
    I^1(m_1^2)
        &\coloneq\int\dfrac{\dd^3k}{(2\pi)^3}\dfrac{1}{\vert\vec k\vert^2+m_1^2}
        =-\dfrac{m_1}{4\pi},\\[1.em]
    I^2(\vec p,m_1^2,m_2^2)
        &\coloneq\int\dfrac{\dd^3k}{(2\pi)^3}\dfrac{1}{\vert\vec k-\frac{\vec p}{2}\vert^2+m_1^2}\dfrac{1}{\vert\vec k+\frac{\vec p}{2}\vert^2+m_2^2}\notag\\
        &=\begin{cases}
            \dfrac{1}{4\pi}\left(\dfrac{1}{m_1+m_2}-\dfrac{\vert\vec p\vert^2}{3(m_1+m_2)^3}\right)+{\mathcal O}(\vert\vec p\vert^4),\\[10pt]
            \dfrac{1}{8\vert\vec p\vert},\hspace{26.5mm}\text{when $m_1=m_2=0$},
        \end{cases}\\[1.em]
    I^2_{i}(\vec p,m_1^2,m_2^2)
        &\coloneq\int\dfrac{\dd^3k}{(2\pi)^3}\dfrac{k_i}{\vert\vec k-\frac{\vec p}{2}\vert^2+m_1^2}\dfrac{1}{\vert\vec k+\frac{\vec p}{2}\vert^2+m_2^2}\notag\\
        &=\begin{cases}-\dfrac{m_1-m_2}{24\pi(m_1+m_2)^2}p_i+\mathcal O(\vert\vec p\vert^3),\\[10pt]
        0,\hspace{12mm}\text{when $m_1=m_2=0$},\end{cases}\\[1.em]
    I^2_{ij}(\vec p,m_1^2,m_2^2)
        &\coloneq\int\dfrac{\dd^3k}{(2\pi)^3}\dfrac{k_ik_j}{\vert\vec k-\frac{\vec p}{2}\vert^2+m_1^2}\dfrac{1}{\vert\vec k+\frac{\vec p}{2}\vert^2+m_2^2}\notag\\
        &=-\dfrac{\delta_{ij}}{8\pi}\begin{cases}
            \dfrac{2}{3}\dfrac{m_1^2+m_1m_2+m_2^2}{m_1+m_2}+\dfrac{2}{15}\dfrac{m_1^2+3m_1m_2+m_2^2}{(m_1+m_2)^3}\vert\vec p\vert^2+{\mathcal O}(\vert\vec p\vert^4)\\[10pt]
            \dfrac{\pi}{8}\vert\vec p\vert
        \end{cases}\notag\\
        &\quad\;+\dfrac{p_ip_j}{8\pi}\begin{cases}
            \dfrac{7m_1^2+6m_1m_2+7m_2^2}{30(m_1+m_2)^3}+{\mathcal O}(\vert\vec p\vert^4),\\[10pt]
            \dfrac{\pi}{8\vert\vec p\vert},\hspace{9.5mm}\text{when $m_1=m_2=0$},
        \end{cases}\\[1.em]
    I^3(\vec p,m_1^2,m_2^2,m_3^2)
        &\coloneq\int\dfrac{\dd^3k}{(2\pi)^3}\dfrac{1}{\vert\vec k-\frac{\vec p}{2}\vert^2+m_1^2}\dfrac{1}{\vert\vec k+\frac{\vec p}{2}\vert^2+m_2^2}\dfrac{1}{\vert\vec k+\frac{\vec p}{2}\vert^2+m_3^2}\notag\\
        &=\begin{cases}
            \dfrac{1}{4\pi(m_1+m_2)(m_1+m_3)(m_2+m_3)}+{\mathcal O}(\vert\vec p\vert^2),\\[10pt]
            \dfrac{1}{8m_2^2\vert\vec p\vert}-\dfrac{1}{4\pi m_2^3}+{\mathcal O}(\vert\vec p\vert^2),\hspace{7mm}\text{when $m_1=m_3=0$},
        \end{cases}\\[1.em]
    I^3_{i}(\vec p,m_1^2,m_2^2,m_3^2)
        &\coloneq\int\dfrac{\dd^3k}{(2\pi)^3}\dfrac{k_i}{\vert\vec k-\frac{\vec p}{2}\vert^2+m_1^2}\dfrac{1}{\vert\vec k+\frac{\vec p}{2}\vert^2+m_2^2}\dfrac{1}{\vert\vec k+\frac{\vec p}{2}\vert^2+m_3^2}\notag\\
        &=\begin{cases}
            -\dfrac{3m_1^2+m_1m_2+m_1m_3-m_2m_3}{24\pi(m_2+m_3)(m_1+m_2)^2(m_1+m_3)^2}p_i+{\mathcal O}(\vert\vec p\vert^3),\\[10pt]
            -\dfrac{p_i}{24\pi m_2^3}+{\mathcal O}(\vert\vec p\vert^3),\hspace{17.5mm}\text{when $m_1=m_3=0$},
        \end{cases}\\[1.em]
    I^3_{ij}(\vec p,m_1^2,m_2^2,m_3^2)
        &\coloneq\int\dfrac{\dd^3k}{(2\pi)^3}\dfrac{k_ik_j}{\vert\vec k-\frac{\vec p}{2}\vert^2+m_1^2}\dfrac{1}{\vert\vec k+\frac{\vec p}{2}\vert^2+m_2^2}\dfrac{1}{\vert\vec k+\frac{\vec p}{2}\vert^2+m_3^2}\notag\\
        &\hspace{-27mm}=\begin{cases}
            \dfrac{m_1m_2+m_1m_3+m_2m_3}{12\pi(m_1+m_2)(m_1+m_3)(m_2+m_3)}\delta_{ij}\\[9pt]
            \quad-\dfrac{m_2^2(m_1^2+3m_1m_3+m_3^2)+m_1m_2(2m_1^2+7m_1m_3+3m_3^2)+m_1^2m_3(2m_1+m_3)}{60\pi(m_2+m_3)(m_1+m_2)^3(m_1+m_3)^3}\vert\vec p\vert^2\delta_{ij}\\[9pt]
            \quad+\dfrac{m_2^2(7m_1^2+6m_1m_3+7m_3^2)+2m_1m_2(7m_1^2+2m_1m_3+3m_3^2)+m_1^2(15m_1^2+14m_1m_3+7m_3^2)}{240\pi(m_2+m_3)(m_1+m_2)^3(m_1+m_3)^3}p_ip_j\\[10pt]
            \quad+{\mathcal O}(\vert\vec p\vert^4),\\[10pt]
            \left(\dfrac{1}{12\pi m_2}-\dfrac{\vert\vec p\vert}{64m_2^2}+\dfrac{\vert\vec p\vert^2}{60\pi m_2^3}\right)\delta_{ij}+\left(\dfrac{1}{64m_2^2\vert\vec p\vert}-\dfrac{7}{240\pi m_2^3}\right)p_ip_j+{\mathcal O}(\vert\vec p\vert^4),\hspace{1.5mm}\text{when $m_1=m_3=0$},
        \end{cases}\\[1.em]
    I^3_{ijk}(\vec p,m_1^2,m_2^2,m_3^2)
        &\coloneq\int\dfrac{\dd^3k}{(2\pi)^3}\dfrac{k_ik_jk_k}{\vert\vec k-\frac{\vec p}{2}\vert^2+m_1^2}\dfrac{1}{\vert\vec k+\frac{\vec p}{2}\vert^2+m_2^2}\dfrac{1}{\vert\vec k+\frac{\vec p}{2}\vert^2+m_3^2}\notag\\
         &=\dfrac{p_i\delta_{jk}+p_j\delta_{ik}+p_k\delta_{ij}}{120\pi(m_2+m_3)}\left(\dfrac{m_1^2(m_1^2-m_1(m_2+m_3)-3m_2m_3)}{(m_1+m_2)^2(m_1+m_3)^2}-1\right)+\mathcal O(\vert\vec p\vert^3),\\[1.em]
    I^4_{ij}(\vec p,m_1^2,m_2^2)
        &\coloneq\int\dfrac{\dd^3k}{(2\pi)^3}\dfrac{k_ik_j}{\vert\vec k-\frac{\vec p}{2}\vert^2+m_1^2}\dfrac{1}{\vert\vec k+\frac{\vec p}{2}\vert^2+m_1^2}\dfrac{1}{\vert\vec k-\frac{\vec p}{2}\vert^2+m_2^2}\dfrac{1}{\vert\vec k+\frac{\vec p}{2}\vert^2+m_2^2}\notag\\
        &=\dfrac{1}{24\pi(m_1+m_2)^3}\delta_{ij}+{\mathcal O}(\vert\vec p\vert^2),\\[1.em]
    I^4_{llij}(\vec p,m_1^2,m_2^2)
        &\coloneq\int\dfrac{\dd^3k}{(2\pi)^3}\dfrac{\vert\vec k\vert^2k_ik_j}{\vert\vec k-\frac{\vec p}{2}\vert^2+m_1^2}\dfrac{1}{\vert\vec k+\frac{\vec p}{2}\vert^2+m_1^2}\dfrac{1}{\vert\vec k-\frac{\vec p}{2}\vert^2+m_2^2}\dfrac{1}{\vert\vec k+\frac{\vec p}{2}\vert^2+m_2^2}\notag\\
        &=\dfrac{m_1^2+3m_1m_2+m_2^2}{24\pi (m_1+m_2)^3}\delta_{ij}
        -\dfrac{2m_1^2+5m_1m_2+2m_2^2}{120\pi(m_1+m_2)^5}\vert\vec p\vert^2\delta_{ij}\notag\\
        &\quad+\dfrac{9m_1^2+10m_1m_2+9m_2^2}{480\pi(m_1+m_2)^5}p_ip_j
        +{\mathcal O}(\vert\vec p\vert^4),\\[1.em]
    I^4_{mnij}(\vec p,m_1^2,m_2^2)p_mp_n
        &\coloneq\int\dfrac{\dd^3k}{(2\pi)^3}\dfrac{(\vec k\cdot\vec p)^2k_ik_j}{\vert\vec k-\frac{\vec p}{2}\vert^2+m_1^2}\dfrac{1}{\vert\vec k+\frac{\vec p}{2}\vert^2+m_1^2}\dfrac{1}{\vert\vec k-\frac{\vec p}{2}\vert^2+m_2^2}\dfrac{1}{\vert\vec k+\frac{\vec p}{2}\vert^2+m_2^2}\notag\\
        &=\dfrac{m_1^2+3m_1m_2+m_2^2}{120\pi (m_1+m_2)^3}\left(\vert\vec p\vert^2\delta_{ij}+2p_ip_j\right)
        +{\mathcal O}(\vert\vec p\vert^4),\\[1.em]
    I^4_{llmmij}(\vec p,m_1^2,m_2^2)
        &\coloneq\int\dfrac{\dd^3k}{(2\pi)^3}\dfrac{\vert\vec k\vert^4k_ik_j}{\vert\vec k-\frac{\vec p}{2}\vert^2+m_1^2}\dfrac{1}{\vert\vec k+\frac{\vec p}{2}\vert^2+m_1^2}\dfrac{1}{\vert\vec k-\frac{\vec p}{2}\vert^2+m_2^2}\dfrac{1}{\vert\vec k+\frac{\vec p}{2}\vert^2+m_2^2}\notag\\
        &=-\dfrac{3m_1^4+9m_1^3m_2+11m_1^2m_2^2+9m_1m_2^3+3m_2^4}{24\pi(m_1+m_2)^3}\delta_{ij}\notag\\
        &\quad\;-\dfrac{11m_1^4+55m_1^3m_2+92m_1^2m_2^2+55m_1m_2^3+11m_2^4}{480\pi(m_1+m_2)^5}\vert\vec p\vert^2\delta_{ij}\notag\\
        &\quad\;+\dfrac{13m_1^4+65m_1^3m_2+96m_1^2m_2^2+65m_1m_2^3+13m_2^4}{480\pi(m_1+m_2)^5}p_ip_j
        +{\mathcal O}(\vert\vec p\vert^4).
\end{align}

To evaluate diagrams in App.~\ref{appendix:1loop_g-indep}, we define several particular combinations of these integrals:
\begin{align}\hspace{2mm}
    C_{ij}^1\coloneq&\,
            2I^2_{ll}(\vec{p},m_W^2,m_W^2)\delta_{ij}+6I^2_{ij}(\vec{p},m_W^2,m_W^2)+\dfrac{9}{2}I^2(\vec{p},m_W^2,m_W^2)\qty(\vert\vec p\vert^2\delta_{ij}-p_ip_j)\notag\\
            =&\,
            -\dfrac{3}{2\pi}m_W\delta_{ij}
            +\dfrac{11}{24\pi}\dfrac{\vert\vec p\vert^2\delta_{ij}}{m_W}
            -\dfrac{1}{2\pi}\dfrac{p_ip_j}{m_W}
            +\mathcal O(\vert\vec p\vert^4),\\[1.em]
        C_{ij}^2\coloneq&\,
            -2\qty(I^4_{llmmnn}+\xi m_W^2I^4_{llmm})\delta_{ij}+2I^4_{llmmij}+2\xi m_W^2I^4_{llij}\notag\\
            &\,
            +\qty(\qty(\biggl(\dfrac{5}{2}I^4_{llkk}+3\xi m_W^2I^4_{ll}\biggr)\delta_{mn}
            -6I^4_{llmn}-2\xi m_W^2I^4_{mn})\delta_{ij}
            -\qty(3I^4_{llij}+\dfrac{7}{2}\xi m_W^2I^4_{ij})\delta_{mn}
            +2I^4_{mnij})p_mp_n\notag\\
            &\,
            +\qty(4I^4_{llmi}+3\xi m_W^2I^4_{mi})p_mp_j
            +\qty(4I^4_{llmj}+3\xi m_W^2I^4_{mj})p_mp_i
            -\dfrac{7}{2}\qty(I^4_{llmm}+\xi m_W^2I^4_{ll})p_ip_j\notag\\
            =&\,
            \dfrac{3+6\sqrt\xi+4\xi+2\xi\sqrt\xi}{6\pi(1+\sqrt\xi)^2}m_W\delta_{ij}
            +\dfrac{-1-4\sqrt\xi+25\xi+20\xi\sqrt\xi}{120\pi(1+\sqrt\xi)^4}\dfrac{\vert\vec p\vert^2\delta_{ij}}{m_W}\notag\\
            &\,-\dfrac{1+4\sqrt\xi+15\xi+10\xi\sqrt\xi}{60\pi(1+\sqrt\xi)^4}\dfrac{p_ip_j}{m_W}
            +\mathcal O(\vert\vec p\vert^4),
        \label{eq:C2}\\[1.em]
        C_{ij}^3\coloneq&\,
            I^2(\vec{p},\xi m_W^2,m_W^2)\delta_{ij}\notag\\
            =&\,
            \dfrac{1}{4\pi(1+\sqrt\xi)}\dfrac{\delta_{ij}}{m_W}
            -\dfrac{1}{12\pi(1+\sqrt\xi)^3}\dfrac{\vert\vec p\vert^2\delta_{ij}}{m_W^3}
            +\mathcal O(\vert\vec p\vert^4),\\[1.em]
        C_{ij}^4\coloneq&\,
            -I^3_{ij}(\vec{p},\vec{p},\xi m_W^2,m_W^2,\xi m_W^2)
            -\dfrac{1}{2}I^3_i(\vec{p},\vec{p},\xi m_W^2,m_W^2,\xi m_W^2)p_j
            -\dfrac{1}{2}I^3_j(\vec{p},\vec{p},\xi m_W^2,m_W^2,\xi m_W^2)p_i\notag\\
            &\,
            -\dfrac{1}{4}I^3(\vec{p},\vec{p},\xi m_W^2,m_W^2,\xi m_W^2)p_ip_j\notag\\
            =&\begin{cases}
                -\dfrac{2+\sqrt\xi}{24\pi(1+\sqrt\xi)^2}\dfrac{\delta_{ij}}{m_W}
                +\dfrac{5+12\sqrt\xi+3\xi}{480\pi(1+\sqrt\xi)^4\sqrt\xi}\dfrac{\vert\vec p\vert^2\delta_{ij}}{m_W^3}
                -\dfrac{5+4\sqrt\xi+\xi}{120\pi(1+\sqrt\xi)^4\sqrt\xi}\dfrac{p_ip_j}{m_W^3}
                +\mathcal O(\vert\vec p\vert^4),\\[1.em]
                \qty(-\dfrac{1}{12\pi m_W}+\dfrac{\vert\vec p\vert}{64 m_W^2}-\dfrac{\vert\vec p\vert^2}{60\pi m_W^3})\delta_{ij}
                +\qty(-\dfrac{3}{64 m_W^2\vert\vec p\vert}+\dfrac{1}{16\pi m_W^3})p_ip_j
                +\mathcal O(\vert\vec p\vert^4)
                \;\text{for}\;\xi=0,
            \end{cases}
            \\[1.em]
        C^5_{ij}\coloneq&\,
            3I^2(\vec p,m_W^2,m_W^2)P_{ij}(\hat p)\notag\\
            =&\dfrac{3}{8\pi}\dfrac{P_{ij}(\hat p)}{m_W}+\mathcal O(\vert\vec p\vert^2),\\[1.em]
        C^6_{ij}\coloneq&\,
            \epsilon^{ipq}\epsilon^{jrs}\dfrac{p_r}{\vert\vec p\vert^2}\Bigl(2\delta_{qs}I^3_{ttp}(\vec p, m_W^2,m_W^2,\xi m_W^2)
            -2p_pI^3_{qs}(\vec p, m_W^2,m_W^2,\xi m_W^2)\notag\\
            &\hspace{22mm}
            -2\delta_{qs}p_tI^3_{pt}(\vec p, m_W^2,m_W^2,\xi m_W^2)
            +\delta_{qs}p_pI^3_{tt}(\vec p, m_W^2,m_W^2,\xi m_W^2)\Bigr)
            +\mathcal O(\vert\vec p\vert^2)\notag\\
            =&-\dfrac{3+9\sqrt\xi+4\xi}{24\pi(1+\sqrt\xi)^3}\dfrac{P_{ij}(\hat p)}{m_W}+\mathcal O(\vert\vec p\vert^2),\\[1.em]
        C^7_{ij}\coloneq&\,
            \epsilon^{ipq}\epsilon^{jrs}
            \delta_{qs}\dfrac{p^r}{\vert\vec p\vert^2}\qty(-I_p^2(\vec p,m_W^2,\xi m_W^2)+\dfrac{p_p}{2}I^2(\vec p,m_W^2,\xi m_W^2))\notag\\
            =& \dfrac{2+\sqrt\xi}{12\pi(1+\sqrt\xi)^2}\dfrac{P_{ij}(\hat p)}{m_W}
            +\mathcal O(\vert\vec p\vert^2),\\[1.em]
        C^8_{ij}\coloneq&\,
            -2I^2(\vec p,m_W^2,m_W^2)\delta_{ij}\notag\\
            =&-\dfrac{\delta_{ij}}{4\pi m_W}
            +\mathcal O(\vert\vec p\vert^2),\\[1.em]
        C^9_{ij}\coloneq&\,
            \epsilon^{ipq}\epsilon^{jrs}\qty(I^3_{ps}(\vec p,m_W^2,m_W^2,\xi m_W^2)\delta_{qr}+I^3_{qr}(\vec p,m_W^2,m_W^2,\xi m_W^2)\delta_{ps})\notag\\
            =&-\dfrac{1+2\sqrt\xi}{6\pi(1+\sqrt\xi)^2}\dfrac{\delta_{ij}}{m_W}
            +\mathcal O(\vert\vec p\vert^2),\\[1.em]
        C^{10}\coloneq&\,
            (2+\xi)I^1(m_W^2)+(1-\xi)\xi m_W^2 I^2(\vec 0, m_W^2,\xi m_W^2)\notag\\
            =&-\dfrac{2+\xi\sqrt\xi}{4\pi}m_W,\\[1.em]
        C^{11}\coloneq&\,
            \dfrac{1}{2\cos^2\theta_{\rm w}}\qty((2+\xi)I^1(m_Z^2)+(1-\xi)\xi m_Z^2 I^2(\vec 0, m_Z^2,\xi m_Z^2))\notag\\
            =&-\dfrac{2+\xi\sqrt\xi}{8\pi\cos^3\theta_{\rm w}}m_W,\\[1.em]
        C^{12}\coloneq&\,
            \dfrac{g_3m_H^2}{2m_W}I^1(\xi m_W^2)+2h_3v_3I^1(m_{\rm D}^2+h_3v_3^2)\notag\\
            =&-\dfrac{g_3\sqrt\xi}{8\pi}m_H^2-\dfrac{g_3}{4\pi}m_W\sqrt{m_{\rm D}^2+m_W^2}+\mathcal O(g_3^3),\\[1.em]
        C^{13}\coloneq&\,
            \dfrac{1}{4m_W^2}\qty(I^1(\xi m_Z^2)+3I^1(m_H^2))\notag\\
            &\,+\dfrac{1}{2m_H^2}\qty((\cos\theta_\tau \tan\theta_{\rm w}-\sin\theta_\tau)^2I^1(\tilde m_-^2)+(\sin\theta_\tau \tan\theta_{\rm w}+\cos\theta_\tau)^2I^1(\tilde m_+^2))\notag\\
            =&-\dfrac{1}{16\pi m_W}\qty(\dfrac{\sqrt\xi}{\cos\theta_{\rm w}}+\dfrac{3m_H}{m_W})\notag\\
            &\,-\dfrac{1}{8\pi m_H^2}\qty((\cos\theta_\tau \tan\theta_{\rm w}-\sin\theta_\tau)^2\tilde m_-+(\sin\theta_\tau \tan\theta_{\rm w}+\cos\theta_\tau)^2\tilde m_+),
\end{align}
where the argument of $I^4$s in Eq.~\eqref{eq:C2}, $(\vec p,m_W,\xi m_W^2)$, is implicit.

\section{Magnetic helicity as the Gauss' linking number of magnetic field lines}
\label{sec:linking}
In this appendix, we review the topological interpretation of magnetic helicity in terms of magnetic field lines \cite{1969JFM....35..117M,ricca2024knotted}.
We here consider a pure ${\rm U}(1)$ gauge theory in either three or four dimensional space(time).\footnote{
While we keep the context general, we intend to apply the discussion here to the confined and the unconfined magnetic fields in the Standard Model.
The corresponding vector potentials in the Coulomb gauge in Eq.~\eqref{eq:def_unconfined_helicity} may be defined by the the Biot--Savart-like equation, Eq.~\eqref{eq:Biot--Savart}.}

To highlight the topological interpretation, we assume that the magnetic field $\vec B=\vec\nabla\times\vec A$ associated with the ${\rm U}(1)$ is confined so that magnetic field lines are quantized as
\begin{align}
    \int_\sigma\vec B\cdot\dd\vec s
        =\dfrac{4\pi}{g}n,\qquad n\in\mathbb Z,
\end{align}
where $\sigma$ is the cross section of a magnetic flux tube.
By taking $n=1$ and $\sigma\to0$ limit, we specify a unit magnetic field flux along an oriented one-dimensional contour $\gamma$ given as
\begin{align}
    \vec B^{\gamma}(\vec x)
        =\dfrac{4\pi}{g}\int_\gamma \dd\vec l\,\delta^3(\vec x-\vec l).
\end{align}

\begin{figure}\centering
    \includegraphics[keepaspectratio, width=.35\textwidth]{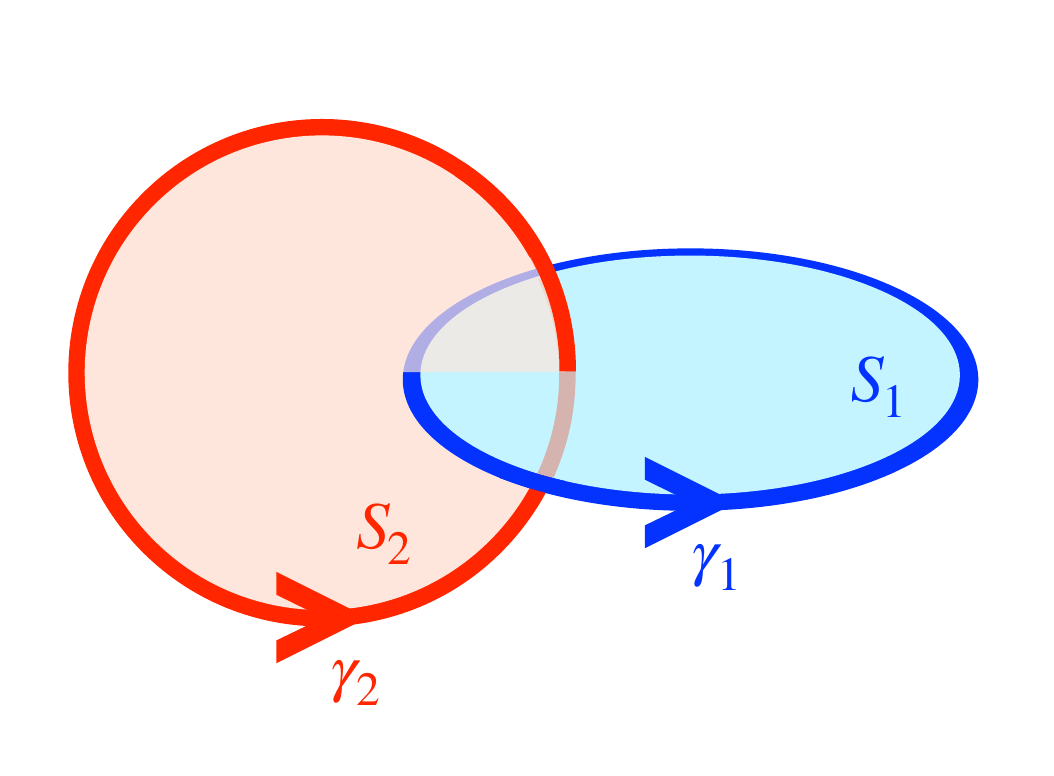}
    \caption{\label{fig:Hopf} The Hopf link $\gamma_1\sqcup\gamma_2$, illustrating the same topology as the one in the leftmost panel of Fig.~\ref{fig:Hopf_link}.}
\end{figure}

Suppose that we have a Hopf link, where two closed loops $\gamma_1$ and $\gamma_2$ are interlinked (Fig.~\ref{fig:Hopf}).
In what follows, we shall interpret the quantized magnetic helicity,
\begin{align}
    N_{\rm M}
        \coloneq\dfrac{g^2}{16\pi^2}\int \dd^3x\,\vec A(\vec x)\cdot\vec B(\vec x),
    \label{eq:def_NM}
\end{align}
by evaluating its value for a magnetic field configuration, $\vec B^{\gamma_1\sqcup\gamma_2}=\vec B^{\gamma_1}+\vec B^{\gamma_2}$.
Note that $N_{\mathrm M}$ is gauge-independent if we take  an appropriate integration volume, {\it i.e.}, the absence of magnetic monopoles, $\vec\nabla\cdot\vec B=0$, inside and $\vec B\cdot\dd\vec s=0$ on the boundary of the volume are sufficient to ensure the gauge-invariance.
We then proceed in the Coulomb gauge, where $\vec\nabla\cdot\vec A=0$.
For a given $\vec B^\gamma(\vec x)$, the corresponding vector potential $\vec A^\gamma(\vec x)$ is given by a Biot--Savart-like equation,
\begin{align}
    \vec A^\gamma(\vec x)
        =\dfrac{1}{4\pi}\int \dd^3y\,\dfrac{\vec B^\gamma(\vec y)\times(\vec x-\vec y)}{\vert\vec x-\vec y\vert^3}\qquad
    \text{in the Coulomb gauge}.
    \label{eq:Biot--Savart}
\end{align}
Therefore, we obtain an expression for the mutual part of the quantized magnetic helicity for $\gamma_1$ and $\gamma_2$,\footnote{For a self-knotted configuration such as a trefoil, we should take {\it self-linking} into account as well \cite{ricca2024knotted}.}
\begin{align}
    N_{\rm {M}}^{\rm Hopf}
        &\coloneq
        \dfrac{g^2}{16\pi^2}\int \dd^3x\qty(\vec A^{\gamma_1}(\vec x)\cdot\vec B^{\gamma_2}(\vec x)+A^{\gamma_2}(\vec x)\cdot\vec B^{\gamma_1}(\vec x))\notag\\
        &=\dfrac{g^2}{32\pi^3}\oint_{\gamma_1}\!\dd^3x\oint_{\gamma_2}\!\dd^3y\,\qty(\vec B^{\gamma_1}(\vec x)\times\vec B^{\gamma_2}(\vec y))\cdot\dfrac{(\vec x-\vec y)}{\vert\vec x-\vec y\vert^3}\notag\\
        &=\dfrac{1}{2\pi}\!\oint_{\gamma_1}\oint_{\gamma_2}\,\qty(\dd\vec l_1\times\dd\vec l_2)\cdot\dfrac{\vec l_1-\vec l_2}{\vert\vec l_1-\vec l_2\vert^3},
\end{align}
which is nothing but twice the Gauss' linking number of the two loops $\gamma_1$ and $\gamma_2$.
As understood from a direct calculation using the definition \eqref{eq:def_NM},
\begin{align}
    N_{\rm M}^{\rm Hopf}
        &=\dfrac{g^2}{16\pi^2}\int_{\rm Hopf} \dd^3x\,\vec A(\vec x)\cdot\vec B(\vec x)
        =\dfrac{g^2}{16\pi^2}\int_{\rm Hopf} B\,\dd^2s\,\vec A(\vec x)\cdot\dd\vec l
        =\dfrac{g}{4\pi}\int_{\rm Hopf}\vec A(\vec x)\cdot\dd\vec l\notag\\
        &=\dfrac{g}{4\pi}\int_{\rm S_1}\dd\vec s\cdot\vec B(\vec x)+\dfrac{g}{4\pi}\int_{\rm S_2}\dd\vec s\cdot\vec B(\vec x)
        \hspace{23mm}\text{where}\quad\partial S_1=\gamma_1,\;\partial S_2=\gamma_2\notag\\
        &=+2,
\end{align}
the Hopf link has the magnetic helicity $N_{\rm M}^{\rm Hopf}=+2$ and correspondingly the Gauss' linking number $+1$.

\small
\bibliographystyle{utphys}
\bibliography{ref}

\end{document}